\documentclass[journal=jacs,manuscript=article]{achemso}

\usepackage[colorlinks=true,allcolors=black]{hyperref}

\usepackage[table]{xcolor} 

\usepackage{soul} 

\usepackage{textgreek}

\usepackage{graphicx}			
\DeclareGraphicsExtensions{.pdf,.png,.jpg} 

\usepackage{siunitx}			
\DeclareSIUnit\photon{photon}
\DeclareSIUnit\molar{M}
\DeclareSIUnit\Molar{M}
\DeclareSIUnit\rev{rev} 
\DeclareSIUnit\OD{OD}
\DeclareSIUnit\cal{cal} 
\DeclareSIUnit\calorie{cal} 
\DeclareSIUnit\lines{lines}

\usepackage{bm} 

\usepackage{tikz} 

\SectionNumbersOn 

\usepackage[version=3]{mhchem} 


\makeatletter
\newcommand{\textsubsuperscript}[2]{%
  \begingroup
    \settowidth{\@tempdima}{\textsubscript{#1}}%
    \settowidth{\@tempdimb}{\textsuperscript{#2}}%
    \ifdim\@tempdima<\@tempdimb
      \setlength{\@tempdima}{\@tempdimb}%
    \fi
    \makebox[\@tempdima][l]{%
      \rlap{\textsubscript{#1}}\textsuperscript{#2}}%
  \endgroup}
\makeatother

\author{James P. Pidgeon} 
\affiliation{School of Mathematical and Physical Sciences, University of Sheffield, Sheffield, UK}
\email{j.pidgeon@sheffield.ac.uk}
\author{George A. Sutherland}
\affiliation{School of Biosciences, University of Sheffield, Sheffield, UK}
\author{Matthew S. Proctor}
\affiliation{School of Biosciences, University of Sheffield, Sheffield, UK}
\author{Shuangqing Wang} 
\affiliation{School of Mathematical and Physical Sciences, University of Sheffield, Sheffield, UK}
\alsoaffiliation{Department of Chemistry and Biochemistry, University of California San Diego, La Jolla, California, US}
\author{Dimitri Chekulaev} 
\affiliation{School of Mathematical and Physical Sciences, University of Sheffield, Sheffield, UK}
\author{Sayantan Bhattacharya}
\affiliation{School of Mathematical and Physical Sciences, University of Sheffield, Sheffield, UK}
\alsoaffiliation{Central Laser Facility, Research Complex at Harwell, Science and Technology Facilities Council, Didcot, UK}
\author{Rahul Jayaprakash} 
\affiliation{School of Mathematical and Physical Sciences, University of Sheffield, Sheffield, UK}
\author{Andrew Hitchcock}
\affiliation{School of Biosciences, University of Sheffield, Sheffield, UK}
\author{Ravi Kumar Venkatraman} 
\affiliation{School of Mathematical and Physical Sciences, University of Sheffield, Sheffield, UK}
\alsoaffiliation{Department of Chemistry, SRM Institute of Science and Technology, Kattankulathur, Chengalpattu, Tamil Nadu, India}
\author{Matthew P. Johnson}
\affiliation{School of Biosciences, University of Sheffield, Sheffield, UK}
\author{C. Neil Hunter}
\affiliation{School of Biosciences, University of Sheffield, Sheffield, UK}
\author{Jenny Clark}
\email{jenny.clark@sheffield.ac.uk}
\affiliation{School of Mathematical and Physical Sciences, University of Sheffield, Sheffield, UK}

\title{Assessment of S* in the Orange\\Carotenoid Protein}

\begin{document}

\begin{abstract}
	
\noindent
The orange carotenoid protein (OCP) is the water-soluble mediator of non-photo\-chemical quenching in cyanobacteria, a crucial photoprotective mechanism in response to excess illumination. OCP converts from a dark-adapted inactive state (OCPo) to an active quenching conformation (OCPr) under high-light conditions, resulting in a concomitant redshift in the absorption of the bound carotenoid. Here, we test whether a long-lived carotenoid singlet excited state (S*) is required for this photoconversion. We measured pump wavelength-dependent transient absorption of OCPo trapped in trehalose-sucrose glass films. We found that initial OCP photoproducts are still formed despite the glass preventing completion to OCPr, and that S* is only apparent for $<$\SI{495}{\nano\metre} pumps. By comparison to the pump wavelength-dependence of the OCPo to OCPr conversion in buffer, we show that S* is not required for photoconversion, and that S* likely arises from ground-state heterogeneity within OCPo.
\begin{center}
	\textbf{TOC Graphic}
	
	\vspace{6pt}
	\fbox{\includegraphics[scale=1.0]{"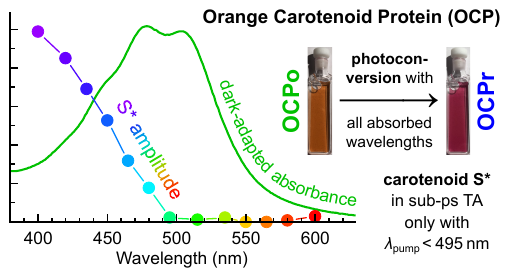"}} \end{center}
\end{abstract}

\clearpage

\tableofcontents

\clearpage

\section*{Introduction}

\begin{figure}[h]
	\vspace{10pt}
	\centering
	\includegraphics[scale=1.0]{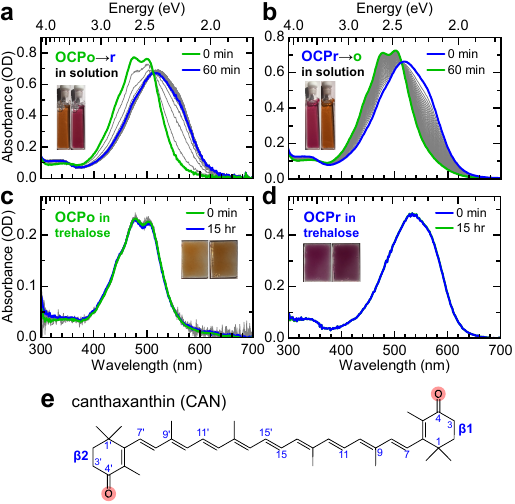}
	\caption{\textbf{OCP conversion and absorbance change in solution (a,b), trapping in trehalose glass (c,d), and canthaxanthin structure (e).} Absorbance spectra of OCPo (a,c) taken in \SI{1}{\minute} intervals (uo to 1 hour) under constant white-light illumination (\SI{1600}{\micro\mol\photon\per\square\m\per\s}), or OCPr (b,d) taken in \SI{1}{\minute} intervals in darkness, both at \SI{22}{\degreeCelsius}. For trehalose samples (c,d), the spectra at 15 hours are also displayed. Noise contributions in (a,c) are due to scatter of the white-light illumination. The optical path length for solution measurements was \SI{1}{\milli\m}. Data are reproduced from our previous work.\cite{Sutherland2023} \textbf{(e)} Chemical structure of canthaxanthin (CAN), bound in our OCP for this study. Conventional numeration of carbons and \textbeta-rings are shown as blue text. Carbonyl substitutions relative to \textbeta-carotene are highlighted in red.}
	\label{fig:OCP_steady_state}
\end{figure}

Oxygenic photosynthetic organisms have evolved mechanisms to protect themselves from excess light conditions and the consequent formation of damaging reactive oxygen species. In cyanobacteria, this photoprotective response occurs over two different timescales. On long and medium timescales (hours/days), photoprotection is achieved by attenuation of protein expression and state transitions.\cite{vanThor1998, Mullineaux2005} On shorter timescales (seconds/minutes) photoprotection relies on a small (\SI{35}{\kilo\dalton}), water-soluble, single carotenoid-binding protein known as the `orange carotenoid protein' (OCP). This photoprotection process, sometimes referred to as `non-photochemical quenching' (NPQ), is initiated by the absorption of light by the OCP. Under these conditions, OCP undergoes a conformational change from a globular, inactive conformation (OCPo) to an extended, active state (OCPr) \cite{Kerfeld2003, Leverenz2015, Gupta2015, Dominguez-Martin2022} allowing it to bind to and dissipate excess energy in the photosynthetic light-harvesting phycobilisome.\cite{Kirilovsky2007, Tian2012, Kirilovsky2016} 
OCP's photoactivated conformational change is accompanied by a redshift of its absorption spectrum, changing colour from orange (OCPo) to red (OCPr) (Figure \ref{fig:OCP_steady_state}a,b) as the effective conjugation length of the bound carotenoid extends.\cite{Wilson2008, Niedzwiedzki2014, Kish2015, Bondanza2020}

While several groups have recently focused on understanding the mechanism of OCPr-induced phycobilisome quenching,\cite{Tian2011, Tian2012, Berera2012, Berera2013, Niyogi1997, Pascal2005, Harris2016, Squires2019, Lou2020, Maksimov2019, Krasilnikov2020, Wilson2022, Sauer2024, Liguori2024, Yang2024} here we focus instead on understanding the initial step of the OCP photocycle to help determine how excitation leads to the dramatic conformational change from OCPo to OCPr. The answer to this question is of interest because the photoswitch in OCP appears to be unique,\cite{Bandara2017, Wilson2008}  and unlike other molecular and pigment-protein photoswitches such as rhodopsin\cite{Lutz2001} or phytochrome.\cite{Yang2011} This OCP photoswitch is, nevertheless, relevant to an entire class of primary producers in many ecosystems,\cite{Kerfeld2017} and understanding the mechanism in greater depth will not only help our understanding of these key organisms, but may also help develop biomimetic optoelectronic or photonic technologies for light adaptation\cite{Piccinini2022} or energy storage.\cite{Dominguez-Martin2019}

The significant work reported to date on this protein suggests that in the orange form (OCPo), the carotenoid is bound to the protein in a twisted, strained geometry.\cite{Kerfeld2003, Bandara2017, Kish2015} It is held in this strained geometry mainly by hydrogen-bonding between its \textbeta1-ring carbonyl group (C=O) and two conserved amino acid residues in the C-terminal domain of the protein: tryptophan 288 and tyrosine 201.\cite{Kerfeld2003, Bandara2017, Wilson2008} Upon photoexcitation, at some time both hydrogen-bonds are believed to break,\cite{Bandara2017} either releasing the strain allowing the carotenoid to planarise and translocate into the N-terminal domain,\cite{Konold2019a, Maksimov2020, Bandara2017, Pigni2020, Bondanza2020, Bondanza2020a, Gupta2019} or simply as a result of N- and C-terminal domain separation.\cite{Chukhutsina2022b}

While there is consensus that the initial trigger of the OCP photocycle is absorption of light by the carotenoid, the processes immediately following this photoexcitation and the causes of hydrogen-bond breakage between the carbonyl group and protein Trp288/Tyr201 are an active topic of debate.\cite{Maksimov2015, Maksimov2017, Bandara2017, Chukhutsina2022b} A comprehensive study by Konold \textit{et al.}~using time-resolved UV-vis and polarised mid-infrared spectroscopy was able to follow the electronic and vibrational degrees of freedom through the OCP photocycle from femtoseconds to \SI{0.5}{\milli\s}.\cite{Konold2019a} A key hypothesis of this study is that a relatively low-yield and long-lived carotenoid singled excited state, dubbed S*,  triggers the photocycle.

The presence of a putative S* feature in transient absorption spectroscopy studies of carotenoids has been discussed in the literature for several decades.  Although debate remains, as described further in our literature review in the Supplementary Information (SI), the prevailing hypothesis seems to be that S* is due to ground state heterogeneity, giving rise to small changes in spectral features with different excitation wavelengths.\cite{Ostroumov2011, Polak2019, Nizinski2022a} This description of the S* feature as an artefact of ground-state heterogeneity, rather than a distinct excited state, suggests that the role of S* in OCP photoconversion merits further investigation.

To determine whether there is a correlation between the S* feature and OCP photoswitching, we measure the relative yield of both the S* feature and the OCPo$\rightarrow$OCPr photoconversion process as a function of pump wavelength. To immobilise the carotenoid for spectroscopic analysis, we trapped OCP in either its OCPo or OCPr form at room temperature using a trehalose-sucrose glass \cite{Jain2009, Kurashov2018} (herein referred to simply as `trehalose'), as previously described.\cite{Sutherland2020,Sutherland2023}

We conclude that the S* spectral feature cannot be directly correlated with the photoconversion yield and that this spectral feature merely arises from ground-state OCPo heterogeneity, comparable with solution-phase carotenoid studies.\cite{Ostroumov2011, Polak2019, Polgar1942, Zechmeister1944, Pesek1990}

\section*{Results}

\subsection*{OCP binding $\sim$100\% canthaxanthin prevents chromophore heterogeneity} When extracted from \textit{Synechocystis}, OCP binds a mixture of carotenoids that proportionally vary depending on the method used for production, leading to heterogeneous samples that complicate spectroscopic analysis. Here we employed a dual-plasmid system, comprised of pAC-CANTHipi to generate carotenoid ($\sim$100\% CAN)\cite{Cunningham2007} and pET28a::OCP to produce OCP. Maintenance of both plasmids in BL21(DE3) \textit{E.~coli} allows incorporation of the carotenoid \textit{in vivo} producing OCP binding near-100\% CAN. See SI Section \ref{sec:methods_sample_prep} for further details on the method. Unless stated otherwise, all experimentation was conducted with OCP containing near-100\% CAN.

\subsection*{OCP trapping does not affect initial photophysics}

\begin{figure*}[h!]
\centering
\includegraphics[scale=0.9]{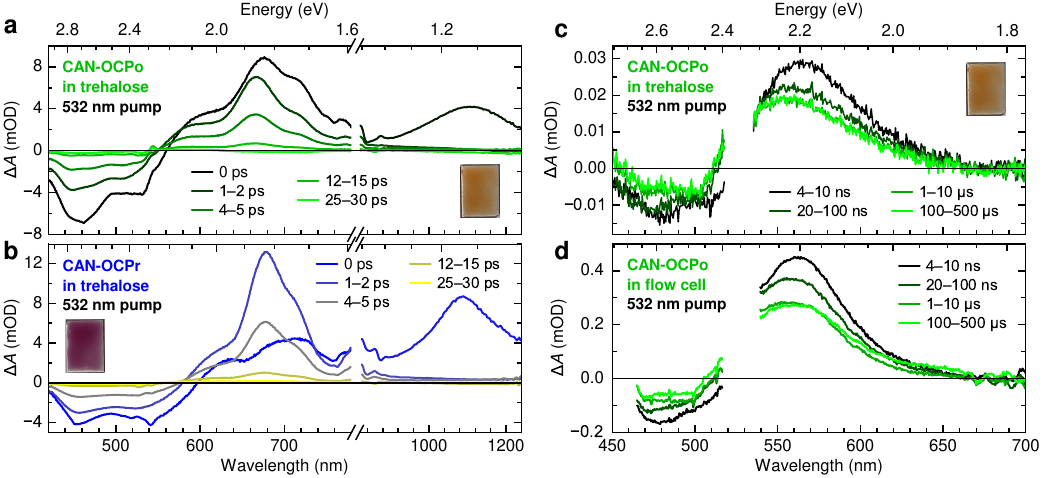}
\caption{\textbf{Picosecond (a,b) and \SI{}{\nano\s}--\SI{}{\milli\s} (c,d) transient absorption spectra of CAN-binding OCP in trehalose glass (a--c) and in buffer (d) with pump wavelength \SI{532}{\nano\m}.} Singlet decay occurs over $\sim$\SI{5}{\pico\s} (a,b), and long-lived and static features are seen on the long-time (c,d), consistent with the literature on OCPo and OCPr in buffer.\cite{Slouf2017, Khan2020, Konold2019a} Spectra have been averaged between the times indicated. In (d), buffer was constantly refreshed by use of a flow cell. Pump fluence was set to \SI{200}{\micro\J\per\square\centi\m} for (a--c), and to \SI{1600}{\micro\J\per\square\centi\m} for (d). For (a,b), the nominal delay of \SI{0}{\pico\s} is heavily impacted by a coherent artefact, and data are reproduced from our previous work.\cite{Sutherland2023} For (c,d), the noisy $\sim$\SI{532}{\nano\m} region (due to pump scatter and a notch filter) has been omitted.}
\label{fig:STTA_LTTA_OCPo_r}
\end{figure*}

We aim to study the early-time photophysics of OCP without problems associated with full photoconversion of the sample during measurement. We therefore trapped OCPo and OCPr in a trehalose matrix,\cite{Sutherland2020} as we had done previously,\cite{Sutherland2023} but performed extra experimentation to assure us the photophysics are comparable to that in buffered solution before our key pump wavelength-dependent measurement here. We reproduce some of that data from our previous work\cite{Sutherland2023} for clarity.

\begin{sloppypar}
The lack of complete photoconversion in trehalose is demonstrated in Figure \ref{fig:OCP_steady_state}. Here we monitored the absorbance spectra of OCPo/OCPr in buffer (a,b) and in trehalose (c,d) as a function of time either under illumination with intense white-light (\SI{1600}{\micro\mol\photon\per\square\m\per\s}; Figure \ref{fig:OCP_steady_state}a,c) or in the dark (Figure \ref{fig:OCP_steady_state}b,d) at room temperature. Full photoconversion of OCPo in solution occurred within \SI{6}{\minute} of illumination and the reverse conversion completed within \SI{60}{\minute} in darkness. No conversion was observed in either sample encapsulated in trehalose during the 15 hours measured, with the absorbance remaining the same.
\end{sloppypar}

Transient absorption spectra of OCP in trehalose are shown in Figure \ref{fig:STTA_LTTA_OCPo_r}a--c (see SI Section \ref{sec:methods_TA} for setup details\cite{Berera2009}). To confirm that the early-time photophysics of OCP are not impacted by being trapped in the trehalose, or by the presence of only one carotenoid type (near-100\% CAN), we compare these spectra to solution measurements (Figure \ref{fig:STTA_LTTA_OCPo_r}d) and those reported in the literature, \textit{e.g.}~Ref.\cite{Konold2019a, Slouf2017, Khan2020}. We find that the spectra of OCP measured on these timescales in trehalose show no obvious differences to those measured in flowed solution over the same timescales. 
For example, the picosecond spectra and dynamics (Figure \ref{fig:STTA_LTTA_OCPo_r}a,b) are similar to those observed in CAN-binding OCP,\cite{Slouf2017} RCP\cite{Slouf2017} and N-terminal domain helical carotenoid proteins (HCP) HCP2 and HCP3 in solution.\cite{Khan2020} This includes the presence of significant excited-state absorption (ESA) in the \SI{700}{\nano\m} to \SI{900}{\nano\m} spectral range due to intramolecular charge transfer (ICT) state absorption.\cite{Slouf2017, Khan2020}
The longer-time behaviour (Figure \ref{fig:STTA_LTTA_OCPo_r}c) is also near-identical  to solution measurements in buffer (Figure \ref{fig:STTA_LTTA_OCPo_r}d), up to at least \SI{0.7}{\milli\s}. These measurements are similar to results on OCP binding 3$'$hydroxyechinenone (3$'$hECN),\cite{Konold2019a} although ours have relatively small-magnitude $\Delta A$ with $<\SI{510}{\nano\metre}$ probe, likely attributable to weak probe intensity in this region in these \SI{}{\nano\s}--\SI{}{\milli\s} experimental runs.
Overall, our results demonstrate that the early-time photophysics of CAN-OCPo are not affected by the trehalose environment. 

Importantly, we also observe no photoconversion during the transient measurements on trehalose films: the transient absorption spectra of individual sweeps from Figure \ref{fig:STTA_LTTA_OCPo_r}c did not change over the course of the 18-hour experiment (see SI Figure \ref{fig:LTTA_OCPo_50sweeps}) and never resemble those seen for a comparable measurement on OCPr in trehalose (see SI Figure \ref{fig:LTTA_OCPo_r_tr}). Also, visual inspection of the films after all experiments with pulsed lasers show no detectable colour change (apart from bleaching in measurements using high pump fluences; see SI Figure \ref{fig:degredation_CAN-OCPo_trehalose}).

To summarise, trapping the OCPo and OCPr protein conformations in trehalose glass prevents the full OCPo$\rightarrow$OCPr photoconversion. However, the initial photophysics remains unchanged in the trehalose compared with buffer and OCP remains in its native form within the trehalose. Trehalose-OCP films therefore provide us with stable solid-state samples that can help elucidate the mechanism of OCPo$\rightarrow$OCPr photoconversion. 

\subsection*[Transient absorption spectroscopy shows different OCPo ground-state\\conformations]{Transient absorption spectroscopy shows different OCPo ground-state conformations}
\begin{figure}[h!]
\centering
\includegraphics[scale=0.95]{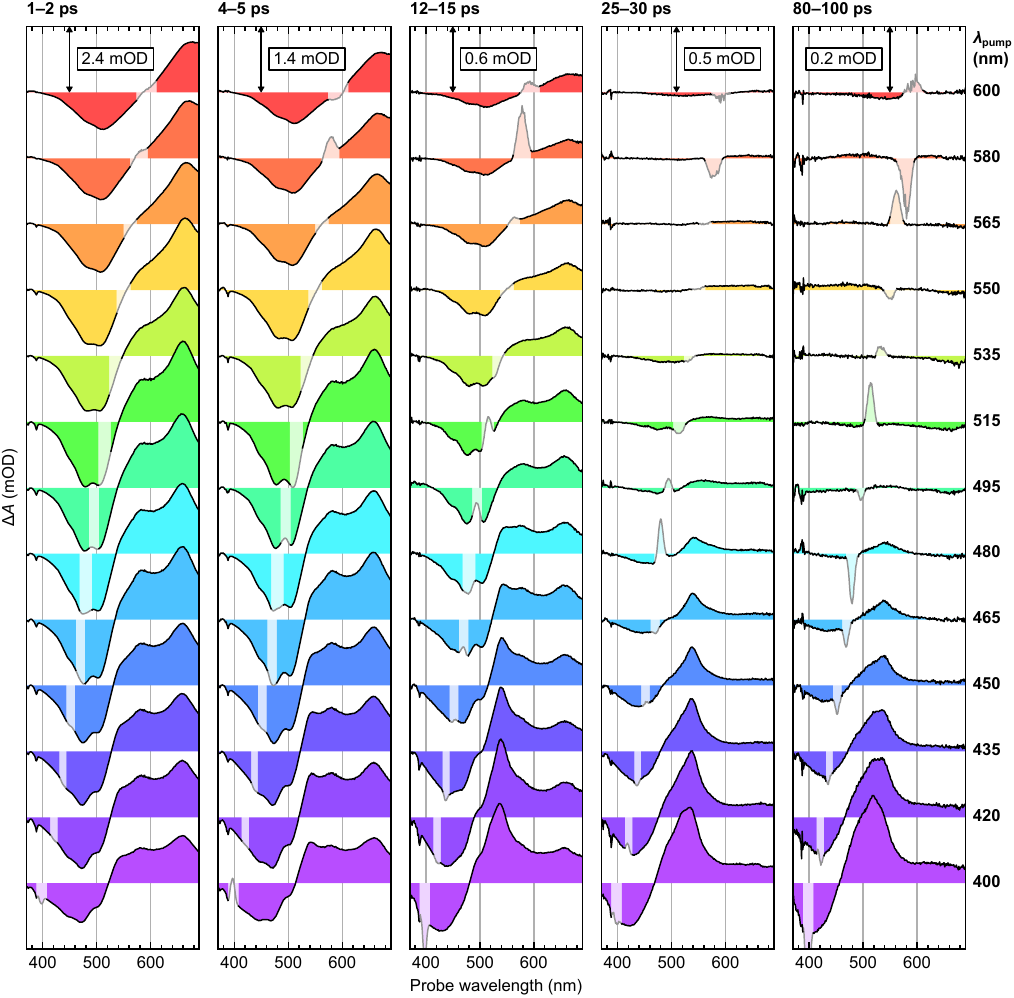}
\caption{\textbf{Transient absorption spectra of CAN-binding OCPo in trehalose glass with pump wavelengths in the range \SI{400}{\nano\m} to \SI{600}{\nano\m} (specified on the right).} Spectra have been averaged between the times specified on the top, with arrows denoting the size of the $y$-axis scale ($\Delta A$ in mOD) for that column of spectra. Pump powers were tuned to give an initial $\sim$\SI{3}{\milli\OD} peak GSB response at \SI{1}{\pico\s}. We observe that higher-energy (lower wavelength) pump wavelengths (400-\SI{480}{\nano\m}: blue/purple) give rise to long-lived features previously assigned to S*-like features.\cite{Konold2019a} As the pump wavelength is tuned to lower energies (longer wavelengths), the spectra continuously shift spectrally, demonstrating ground-state heterogeneity. Note that probe regions affected by pump scatter and pre-signal background correction are in lighter shades of colour.
}
\label{fig:STTA_OCPo_ex_dep}
\end{figure}

To test the hypothesis that a long-lived singlet (often dubbed S*) is an initial trigger for the photoconversion mechanism,\cite{Konold2019a, Yaroshevich2021, Maksimov2020} we first turn to pump wavelength-dependent transient absorption measurements on OCPo in trehalose, shown in Figure \ref{fig:STTA_OCPo_ex_dep}. Here, to ensure a consistent excitation density between each pump wavelength, we tuned the pump powers to target an initial $\sim$\SI{3}{\milli\OD} peak ground-state bleach (GSB) response at \SI{1}{\pico\s}. A different sample spot was pumped and probed per $\lambda_\textrm{pump}$. For most pump wavelengths, this resulted in a relatively low pump power (see SI Figure \ref{fig:STTA_ex_wav_dep_pump_scatter} for pump profiles and powers used). We chose a UV-vis probe spanning \SI{370}{\nano\m} to \SI{690}{\nano\m} to better resolve the GSB region. This OCPo sample was additionally encapsulated with a microscope cover slip (see image in Figure \ref{fig:STTA_OCPo_ex_dep} and SI Section \ref{sec:methods_sample_prep} for preparation details) to avoid atmospheric rehydration. 

We start by describing spectra with \SI{495}{\nano\m} pump, close to the peak absorption of OCPo and similar to pump wavelengths used in previous reports.\cite{Slouf2017, Khan2020, Kish2015, Konold2019a} As expected, the spectra show signatures of S\textsubscript{1} decay ($\sim$\SI{5}{\pico\s} visible ESA decay) and associated effects due to intramolecular vibrational redistribution (IVR) and vibrational cooling (VC) (\textit{e.g.}~narrowing, shifting, changes in vibronic lineshapes, see Refs.~\cite{Balevicius2015,Balevicius2016,Balevicius2018, Balevicius2019,Sebelik2022} for details). We note that carotenoids, particularly keto-carotenoids such as CAN, demonstrate a wealth of different diabatic states and that any S\textsubscript{1}-like features we observe likely originate from S\textsubscript{1}/ICT (and other) admixtures, with confounding effects due to IVR/VC. In this work to simplify the discussion we describe the states as S\textsubscript{1}, S\textsubscript{2}, \textit{etc.}~with the implicit assumption of them being mixed states.

We now discuss the impact of redder pump wavelengths (from \SI{495}{\nano\m} to \SI{600}{\nano\m}) on the OCPo photophysics. For all times $>$\SI{1}{\pico\s}, as $\lambda_\textrm{pump}$ increases from \SI{495}{\nano\m} to \SI{600}{\nano\m}, we observe a continuous spectral evolution, consisting of two main factors: (1) a decrease in the S\textsubscript{1}-S\textsubscript{n} ESA vibronic replica around \SI{585}{\nano\m}, together with an apparent narrowing of the entire ESA feature, and (2) a continuous change in the GSB vibronic structure, with the peak around \SI{507}{\nano\m} increasing relative to the $\sim$\SI{481}{\nano\m} peak. While (1) could be partially explained in terms of effects associated with IVR and VC, as modelled by Balevičius \textit{et al.}\cite{Balevicius2018}, the effect we observe is significantly more pronounced and long-lived, and therefore cannot be solely due to IVR and VC. In addition, considerations of IVR and VC do not explain (2). We therefore assign the $\lambda_\textrm{pump}$-dependence of the spectra as the result of heterogeneity in `CAN-OCPo' with a presence of multiple ground-state species, as others have done for OCP.\cite{Kish2015, Slouf2017, Polivka2013, Khan2020, Kuznetsova2020a, Nizinski2022a, Pishchalnikov2019} These species have similar lifetimes associated with their respective S\textsubscript{2} ($\sim$\SI{100}{\femto\s} time constant) and S\textsubscript{1} ($\sim$\SI{5}{\pico\s}) states with implicit admixtures (see global target analysis\cite{Snellenburg2012, Mullen2007} in SI Sections \ref{sec:GTA_SADS_OCPo_summary} and \ref{sec:GTA_SADS_OCPo}).

Pumping with wavelengths $\lambda_\textrm{pump}=\SI{480}{\nano\m}$ and bluer results in the formation of additional and distinct S\textsubscript{1}-like excited-state features with significantly longer decay time constants than the expected $\sim$\SI{5}{\pico\s}. For $\lambda_\textrm{pump}=\SI{480}{\nano\m}$ to \SI{400}{\nano\m}, new features appear in the 25--\SI{30}{\pico\s} and 80--\SI{100}{\pico\s} spectra, with increasing intensity with decreasing pump wavelength. These have a comparatively unstructured GSB, and an ESA consisting of broad peak(s) around \SI{520}{\nano\m} to \SI{540}{\nano\m} and a long, unstructured red tail. These features are consistent with additional ground-state species of OCPo that are only excited at $\sim$\SI{480}{\nano\m} and below, very similar to the contribution reported as S* and assigned to ground-state non-all-\emph{trans} conformations in studies of carotenoids in solution.\cite{Ostroumov2011, Polak2019} Its spectrum is similar to that described as S* by Konold \textit{et al.}\cite{Konold2019a}~for 3$'$hECN-OCP and thought to be the precursor of OCP switching.\cite{Konold2019a, Yaroshevich2021, Maksimov2020} It is also similar to the S\textsuperscript{$\sim$} state reported by Niziński \textit{et al.}~in ECN-OCP, who assign it to a non-photoactive carotenoid `impurity' (\textit{e.g.}~\textbeta-carotene) in their OCP sample.\cite{Nizinski2022a} We note that while largely obscured by this new S*-like feature, the aforementioned spectral evolution behaviour described above as (1) and (2) are still present with $\lambda_\textrm{pump}=\SI{400}{\nano\m}$ to \SI{480}{\nano\m}.

We note that a different sample spot was pumped and probed per $\lambda_\textrm{pump}$ to avoid any degradation-associated affects. The sample spots chosen were practically random. We rule out heterogeneity across the OCPo sample volume as the source of the phenomena seen for the following reasons: (1) the transient absorption spectra and dynamics for different sample spots probed with near-identical pumps (the same nominal $\lambda_\textrm{pump}$ between different generation methods, or `bands'), shown in SI Figures \ref{fig:STTA_480nm_comp} and \ref{fig:STTA_535nm_comp}, appear to be near-identical (see SI Section \ref{sec:UV-ps-TA_replicates_same_wavelength} for details), (2) the variations seen with decreasing $\lambda_\textrm{pump}$ are directed trends, as opposed to random variations (as would be expected with practically random choices of the pump/probed sample spot). We therefore assign heterogeneity in OCPo itself. We note a $\lambda_\textrm{pump}$-dependence is observed in OCPr, but only \SI{532}{\nano\m} and \SI{485}{\nano\m} pumps were tested, with further study beyond the scope of this work; see SI Section \ref{sec:STTA_OCPo_OCPr} for details and figures.

To summarise, from inspection of the spectra of pump wavelength-dependent transient absorption on OCPo in trehalose (Figure \ref{fig:STTA_OCPo_ex_dep}), we have identified ground-state heterogeneity within OCPo, including forms yielding a long-lived ($\sim$\SI{65}{\pico\s}) decay time constant that we associate with S*.

\begin{figure}[h!]
\centering
\includegraphics[scale=1.0]{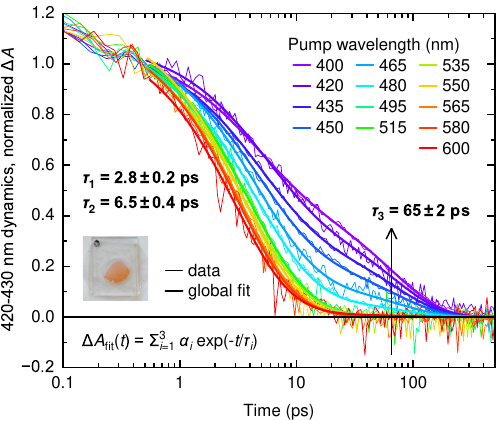}
\caption{\textbf{Normalised dynamics and a triexponential global fit (equation inset bottom-left) in the long-lived GSB region for transient absorption experiments using different pump wavelengths on CAN-binding OCPo.} The S* amplitude clearly increases as pump wavelength is reduced from \SI{480}{\nano\m} to \SI{400}{\nano\m}. Dynamics are taken as the 420--\SI{430}{\nano\m} probe range average, and subsequently normalised to average one in the \SI{0.35}{\pico\s} to \SI{0.65}{\pico\s} range. Errors denote fit parameter standard errors.}
\label{fig:STTA_OCPo_kinetics_superfigure}
\end{figure}

\begin{figure}[h!]
\centering
\includegraphics[scale=1.0]{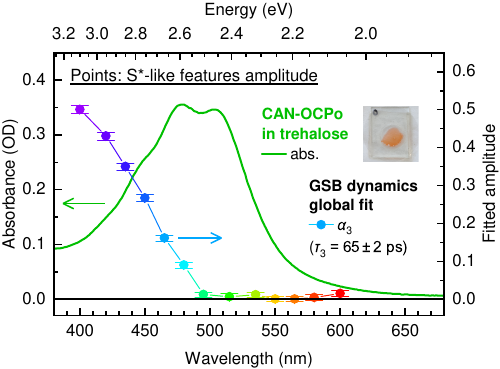}
\caption{\textbf{Fitted amplitudes from the global fit of transient absorption dynamics with absorbance of CAN-binding OCPo.} Absorbance of OCPo in trehalose (line, left axis), with fitted amplitude $\alpha_3$ from a global triexponential fit of picosecond transient absorption dynamics (points, right axis). Error bars denote fit parameter standard errors. Dynamic fits are shown and described in Figure \ref{fig:STTA_OCPo_kinetics_superfigure} and SI Table \ref{tab:STTA_OCP_fit_params}.}
\label{fig:OCP_fitted_kinetics_comparison}
\end{figure}

To quantify the long-lived S* yield, we perform a global triexponential fit of normalised dynamics in the $>$\SI{500}{\femto\s} time-region (Figure \ref{fig:STTA_OCPo_kinetics_superfigure}). We take the average $\Delta A$ in the 420--\SI{430}{\nano\m} probe range (close to the peak GSB of the long-lived singlet features -- see Figure \ref{fig:STTA_OCPo_ex_dep}), normalise to unity at $\sim$0.35--\SI{0.65}{\pico\s}, plot against time, and fit for times $>$\SI{500}{\femto\s} (beyond any features associated with the coherent artefact and decay of S\textsubscript{2}-like states) with a global triexponential fit with the following equation:
\begin{equation}
\Delta A_\textrm{fit}(t) = \alpha_1\exp(-t/\tau_1) + \alpha_2\exp(-t/\tau_2) + \alpha_3\exp(-t/\tau_3)
\end{equation}
Here $t$ is the delay time in the transient absorption, $\alpha_i$ are fitted amplitudes with the applied lower-bound condition $\alpha_i \geq 0$, and $\tau_i$ are fitted time constants that are shared between the different pump wavelengths in the global fit. The dynamics and fit are shown in Figure \ref{fig:STTA_OCPo_kinetics_superfigure}, and the fit parameters are printed in SI Table \ref{tab:STTA_OCP_fit_params}. We find that $\alpha_3$, the amplitude associated with S* ($\tau_3 = 65\pm\SI{2}{\pico\s}$), shows a significant wavelength dependence, increasing from zero at $\lambda_\textrm{pump} \geq \SI{495}{\nano\m}$ to $\sim$0.5 at $\lambda_\textrm{pump} = \SI{400}{\nano\m}$, see Figure \ref{fig:OCP_fitted_kinetics_comparison}. This dependence matches with supplementary global target analysis; an additional component is required for fitting $\lambda_\textrm{pump} < \SI{495}{\nano\m}$ (see SI Sections \ref{sec:GTA_SADS_OCPo_summary} and \ref{sec:GTA_SADS_OCPo}).

In short, using pump wavelength-dependent transient absorption measurements on OCPo in trehalose, we have found evidence of heterogeneity in OCPo with populations of distinct OCPo `forms', including S*-yielding forms. We have quantified the proportion of the S* feature as a function of $\lambda_\textrm{pump}$. If, as reported,\cite{Konold2019a, Yaroshevich2021, Maksimov2020} S* is a trigger for the OCPo to OCPr photocycle, the OCPr yield should show a similar excitation wavelength-dependence; in particular, it should show an onset with $<$\SI{495}{\nano\m} pumps. Therefore, we next assessed the excitation wavelength-dependence of the solution OCPo$\rightarrow$OCPr photoconversion.

\subsection*{Photoconversion occurs for all wavelengths absorbed by OCPo}
\begin{figure}[h!]
\centering
\includegraphics[scale=1.0]{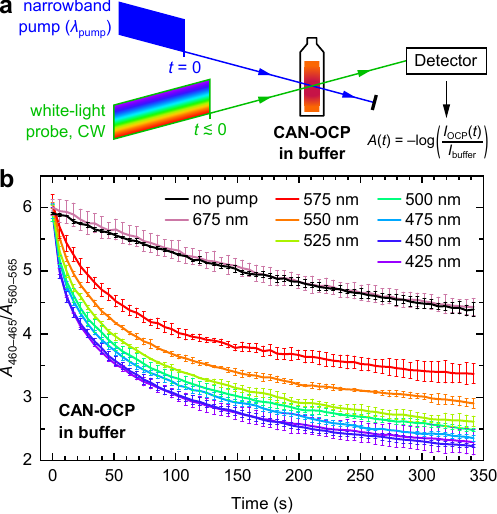}
\caption{\textbf{Experimental setup for absorbance measurements of CAN-binding OCP in buffer that is continually photoconverted by narrowband pump light (a), and average absorbance ratios against time (b).} The pump power was tuned to give approximately the same initial OCPo excitation rate per-pump wavelength (except for \SI{675}{\nano\m}). All absorbed pump wavelengths give a conversion yield greater than the white-light probe alone, showing that all absorbed wavelengths cause photoconversion. The OCPr yield apparently increases as pump wavelength is reduced from \SI{575}{\nano\m} to \SI{425}{\nano\m}, although as noted in the main text and SI, this could be due to experimental artefacts. Absorbance ratios are averaged from three experimental replicates, and the error bars show the standard deviation. The cuvette width was \SI{2}{\milli\m}.}
\label{fig:PC_OCP_Aratio}
\end{figure}

We assess the degree of OCPo$\rightarrow$OCPr photoconversion as a function of pump wavelength using absorbance measurements of dark-adapted OCPo in buffer as a function of time, Figure \ref{fig:PC_OCP_Aratio}. A simplified diagram of the experiment is shown in Figure \ref{fig:PC_OCP_Aratio}a and details of the method are outlined in SI Section \ref{sec:methods_PC_OCP}; importantly, the pump power was tuned to give approximately the same initial rate of OCPo excitations (\textit{i.e.} rate of photon absorptions) for each pump wavelength (except for \SI{675}{\nano\m}) based on the dark-adapted absorbance. The conversion from OCPo to OCPr can be tracked with a ratio of absorbance: 460--\SI{465}{\nano\m} (average) representing primarily OCPo absorption to 560--\SI{565}{\nano\m} representing primarily OCPr ($A_{460-465}/A_{560-565}$). The dynamics in Figure \ref{fig:PC_OCP_Aratio}b are the average of three experimental replicates.

Figure \ref{fig:PC_OCP_Aratio}b shows that the OCP photocycle is triggered by excitation into the OCPo absorption band, right down to the band-edge ($\sim$\SI{575}{\nano\m}). Note that the white-light probe alone triggers some OCPo$\rightarrow$OCPr photoconversion; see the average absorbance-ratio dynamics obtained without the pump (probe light only; black line) and \SI{675}{\nano\m} pump (non-absorbed pump; pink line). Significantly, we observe photoconversion with pump wavelengths that do not generate S* features in transient absorption spectroscopy measurements ($\geq$\SI{495}{\nano\m}, see Figures \ref{fig:OCP_fitted_kinetics_comparison}). This suggests that S* is not required for photoconversion.

We note that the traces in Figure \ref{fig:PC_OCP_Aratio}b appear to suggest a pump wavelength-dependence on the photoconversion, with OCPr yield apparently increasing with decreasing pump wavelength. It is possible that this dependence is the result of artefacts arising from experimental oversights; these are discussed in SI Section \ref{sec:PC_OCP_oversights}. We refrain from further analysis of this dataset, and take this experimentation as a preliminary suggestion of a pump wavelength-dependence on the photoconversion yield. Further study with an optimised experiment, such as the well-considered ones recently employed by others,\cite{Rose2023,Nizinski2022,Nizinski2025} is beyond the scope of this work. We note that Niziński \textit{et al.}\cite{Nizinski2025}~exclude a difference in the photoconversion mechanism with \SI{512}{\nano\m} against \SI{488}{\nano\m} and obtain comparable photoproduct yields (although they controlled for the pump energy density rather than number of absorbed photons), while Rose \textit{et al.}\cite{Rose2023}~mention a possible pump wavelength-dependence on their OCP photoconversion kinetics but focus on \SI{430}{\nano\m} pump in their study.

To summarise, measurements of OCP in buffer continually photoconverting under narrowband pumps have shown that all wavelengths absorbed by OCPo cause the transition to OCPr, which does not match the $<$\SI{495}{\nano\m} onset of S* in transient absorption spectroscopy. This leads us to reject the hypotheses \cite{Konold2019a, Yaroshevich2021, Maksimov2020} that S* is the sole trigger for the photocycle.

\section*{Discussion}

\subsection*{Long-lived singlets do not appear to be required for photoconversion}
Long-lived singlet excited states have been suggested to play a pivotal role in the initial steps of OCPo to OCPr photoconversion.\cite{Konold2019a, Yaroshevich2021, Maksimov2020} This argument was based on sequential-model global lifetime analysis on transient absorption data with UV-vis and mid-IR probes, where the `S* state' was assigned a \SI{24}{\pico\s} lifetime,\cite{Konold2019a} within the 20--\SI{100}{\pico\s} range typically assigned to the S* feature in free carotenoids\cite{Ostroumov2011, Polak2019} and other OCP studies.\cite{Konold2019a, Maksimov2020, Kuznetsova2020a, Nizinski2022a, Khan2020} Here we show that photoconversion does not correlate with the long-lived S* feature in CAN-OCPo; there is no apparent onset pump wavelength for the photoconversion in buffer (Figure \ref{fig:PC_OCP_Aratio}) like that seen for the pump wavelength-dependence of the `S*' yield (Figure \ref{fig:OCP_fitted_kinetics_comparison}). Note our rejection of the hypothesis is independent of the physical identity of S*, which we argue below arises from ground-state heterogeneity originating from CAN-associated chromophores, but is superfluous to our argument here.

\subsection*{OCPo photophysics show ground-state heterogeneity} Our pump wavelength-dependent transient absorption spectra (Figure \ref{fig:STTA_OCPo_ex_dep}) demonstrate a surprisingly rich excitation dependence, given our attempts to restrict the chromophore heterogeneity (all samples initially contain $\sim$100\% CAN). We note, in particular, that the observed changes in ground-state bleach (GSB) as a function of pump wavelength indicate that many ground-state OCPo species are present. The clearest are those yielding S* features, discussed above. They are only apparent with pump wavelengths $<$\SI{495}{\nano\m} and yield comparatively blue GSBs (high S\textsubscript{2}-like energy) with a relatively long decay time constant ($\sim$\SI{65}{\pico\s}). However, further distinct forms are present, made apparent by redshifts of the GSB contribution (lowering of the S\textsubscript{2} energy) with tuning of the pump wavelength from 495 to \SI{600}{\nano\m}.

We principally consider the possible origin of the S* feature. The photophysical properties of carotenoids strongly depend on their effective conjugation length; as the length increases, the S\textsubscript{1} lifetime and energy decrease, the S\textsubscript{0}$\rightarrow$S\textsubscript{2} transition energy decreases, and the C=C vibration energy (\textnu\textsubscript{1}-peak in Raman spectroscopy) decreases.\cite{Llansola-Portoles2017, Kloz2016, Fuciman2015} Such behaviour might explain the origin of the long-lived `S*'. From the non-radiative energy gap law,\cite{Chynwat1995} we estimate the S*-associated S\textsubscript{1} energy to be $\sim$\SI{16400}{\per\centi\meter}, corresponding to a conjugation length of $N\approx8.2$.\cite{Fujii1998, Koyama1999} Similarly, in a resonance Raman spectroscopy study, Kish \textit{et al.}\cite{Kish2015}~noted an excitation wavelength-dependent blue \SI{1527}{\per\centi\m} shoulder of the main \textnu\textsubscript{1} peak. This \textnu\textsubscript{1} energy corresponds to an effective conjugation length of $N\approx9$.\cite{Kish2015} These values are shorter than the expected CAN conjugation length $N\approx13$ and would require complete non-participation of the \textbeta1 and \textbeta2 rings to the effective conjugation length, which has not been observed to our knowledge. It is also unlikely that out-of-plane distortions of the terminal rings could change the effective conjugation length to such an extent.\cite{Slouf2017,Fuciman2015}

Studies of carotenoids in solution, where signatures extremely similar to our S* features have been identified in samples that are nominally all-\textit{trans}, may contain impurities. A transient absorption study of \textbeta-carotene in solution by Ostroumov \textit{et al.}~showed that a long-lived contribution with lifetime 20--\SI{100}{\pico\s} disappears if the samples are purified to all-\textit{trans}-\textbeta-carotene immediately prior to a measurement. A 3-level carotenoid model with consideration of VC in the S\textsubscript{1} state is then sufficient to explain the photophysics in these purified samples.\cite{Ostroumov2011} It was suggested that the `impurities' removed through purification might be due to chromophore products from \textbeta-carotene, such as shorter conjugation-length carotenoids. Work by Polak \textit{et al.}~undertaking pump wavelength-dependent transient absorption measurements on \textbeta-carotene and astaxanthin (a keto-carotenoid) in toluene confirmed the pump wavelength-dependence of the long-lived contribution and showed the dependence in turn matches Ostroumov \textit{et al.}'s impurity absorbance.\cite{Polak2019} The wavelength dependence of Ostroumov \textit{et al.}'s impurity absorbance and Polak \textit{et al.}'s long-lived signal amplitude both follow a similar profile to our own pump wavelength-dependence on S* in OCPo (\textit{e.g.}~the dependence of $\alpha_3$ on pump wavelength; Figure \ref{fig:OCP_fitted_kinetics_comparison}). 

From the observations and assignments made regarding the ECN-OCPo resonance Raman spectra by Kish \textit{et al.}\cite{Kish2015},~the similarity to an `impurity' contribution identified for carotenoid-in-solution studies,\cite{Ostroumov2011, Polak2019} and the similarity to a 60--\SI{100}{\pico\s} feature in transient absorption spectroscopy, assigned to a non-photoconverting `impurity'-OCP fraction,\cite{Nizinski2022a} we hypothesise that the S*-yielding species in OCPo are due to ground-state contributions of short conjugation length carotenoids or carotenoid-like chromophores. The precise identity of these chromophores remains to be determined, although similarities with Ostroumov's work \cite{Ostroumov2011} suggests they may be CAN-associated products.

We now turn to the identities of the further ground-state forms mentioned above. We infer the presence of these due to significant changes in the excited state absorption (ESA) vibronic replica intensities and the ground state bleach (GSB) vibrational structure in transient absorption experiments scanning pump wavelengths from \SI{495}{\nano\m} to \SI{600}{\nano\m} (Figure \ref{fig:STTA_OCPo_ex_dep}). We observe IVR and VC signatures at all pump wavelengths, but the pump wavelength-dependence of these cannot explain the extent of the changes in the transient absorption spectra.\cite{Balevicius2015, Balevicius2016, Balevicius2018, Balevicius2019, Sebelik2022} As explained in SI Section \ref{sec:GTA_SADS_OCPo_summary}, the effects of IVR and VC make application of multi-exponential dynamic fitting and global lifetime/target analysis poorly suited for this system, particularly with both forms decaying with similar lifetimes.\cite{Ruckebusch2012, vanStokkum2004, Fernandez-Teran2022} We recall our global triexponential fit of the normalised dynamics in the 420--\SI{430}{\nano\m} GSB region (Figure \ref{fig:STTA_OCPo_kinetics_superfigure}); while $\tau_3 = \SI{65}{\pico\s}$ can safely be associated with the S*-yielding forms of OCPo due to its 1-order time difference between that and the earlier-time dynamics, the extracted $\tau_1 = \SI{2.8}{\pico\s}$ and $\tau_2 = \SI{6.5}{\pico\s}$ have a mixed and unknown physical correspondence, likely mixed S\textsubscript{1}/IVR/VC for multiple ground-state OCPo forms. Meanwhile, the SADS2 and SADS3 (Figure \ref{fig:STTA_OCPo_SADS_superfigure}b,c) extracted from global target analysis and associated with fitted time constants $\sim$\SI{5}{\pico\s} also have a muddled correspondence, and vary significantly with pump wavelength. Applying a model accounting for IVR and VC such as those designed by Balevičius \textit{et al.}\cite{Balevicius2015, Balevicius2016, Balevicius2018, Balevicius2019, Sebelik2022}~is beyond the scope of this paper, considering the ground-state heterogeneity within the OCPo system.

We hypothesise that these `non-S*-yielding forms' of dark-adapted OCP arise from heterogeneities in the protein environment, associated with only minor CAN conformational differences.\cite{Mendes-Pinto2013,Llansola-Portoles2017} We reach this because, while the S\textsubscript{1} profiles and decay times of these forms are similar, the S\textsubscript{2}-associated GSB is significantly different in the two forms. S\textsubscript{1} and S\textsubscript{2} transition lifetimes/energies are affected differently by changes in conformation and environment; S\textsubscript{1} lifetime and energy are known to be insensitive to environment (namely its polarity)\cite{Bondarev1994} but highly sensitive to conjugation length,\cite{Chynwat1995}~while the S\textsubscript{0}$\rightarrow$S\textsubscript{2} transition energy is sensitive to the polarisability of the environment.\cite{Llansola-Portoles2017, Pishchalnikov2019} Our results therefore suggest that these non-S* forms result from different CAN-protein microenvironments, rather than differences in CAN conjugation length. We do not exclude the possibility that these forms are better recognised as OCPr (acting as a contaminant within our trehalose films) rather than OCPo.

This hypothesis can also be reconciled with the resonance Raman spectroscopy of Kish \textit{et al}.\cite{Kish2015} We previously assigned the reported \SI{1527}{\per\centi\m} `shoulder' as due to the S*-associated forms in CAN-OCPo, so forms not yielding S* would have a \textnu\textsubscript{1} wavenumber of \SI{1518}{\per\centi\m}, within experimental tolerances. We note that the \textnu\textsubscript{1} wavenumber for carotenoids has only minor dependence on the microenvironment.\cite{Kish2015, Llansola-Portoles2017, Mendes-Pinto2013}

In the literature, similar heterogeneities to the ones we observe here have occasionally been assigned as due to differences in keto-carotenoid hydrogen-bonding to Tyr201 and Trp288,\cite{Polivka2005,Maksimov2020,Tsoraev2023, Sluchanko2024,Tsoraev2025} but as zeaxanthin (ZEA) has no carbonyl groups yet ZEA-OCP still shows similar heterogeneity,\cite{Slouf2017} a pure hydrogen-bond-heterogeneity origin to these forms appears unlikely.\cite{Kuznetsova2020a} This is supported by a theoretical study by Pigni \textit{et al.}~who modelled the CAN-OCP photoconversion. In their `redshifted OCPo' form, the CAN \textbeta1-ring still forms hydrogen-bonds to Tyr201 and Typ288, but the protein environment is reorganised, giving a relatively planar CAN configuration with a different position for the \textbeta2-ring within the NTD.\cite{Pigni2020} Similarly, the theoretical work by Arcidiacono and co-workers assign heterogeneous CAN populations that all maintain hydrogen-bonds to both Tyr201 and Typ288.\cite{Arcidiacono2023} 

It is worth highlighting that multiple structural studies on dark-adapted OCP using X-ray crystallography have not observed  inhomogeneity in carotenoid conformation. Such structural studies have been conducted on wild-type OCPo,\cite{Bandara2017,Chukhutsina2022b,Leverenz2015,Wilson2010,Kuznetsova2020a,Kerfeld2003, Lopez-Igual2016, Wilson2022a, Chukhutsina2022b, Han2023, Sluchanko2024, Yang2023} mutant OCP,\cite{Yaroshevich2021,Wilson2010, Slonimskiy2022, Han2023, Moldenhauer2023} and various OCP homologues.\cite{Dominguez-Martin2019a, Melnicki2016, Sklyar2024} It appears to counter the hypotheses above regarding the presence of multiple ground-state forms within CAN-OCPo due to distortions of the bound carotenoid. We can reconcile this by speculating that crystal packing and/or cryogenic temperatures used during data collection could artificially be responsible for preferentially adopting one
ground-state OCPo geometry. This contrasts with our spectroscopic studies on dilute OCPo and OCPr in trehalose glass probed at room temperature, along with the aforementioned solution-based studies at room temperature, where heterogeneity is apparent.\cite{Kish2015,Slouf2017, Polivka2013, Khan2020, Kuznetsova2020a, Nizinski2022a}

In short, we have identified a continuum of ground-state forms within CAN-OCPo. By comparison to the OCP and carotenoid literature, we hypothesise that the forms leading to a relatively long-lived S* features incorporate short-conjugation length molecules associated with CAN.\cite{Polak2019, Ostroumov2011} Meanwhile, other forms may arise due to minor distortions of CAN leading to different interactions between CAN and the protein.

\subsection*[Photoexcitation of OCPo in trehalose forms initial photoproducts, without\\completion of the cycle]{Photoexcitation of OCPo in trehalose forms initial photoproducts, without completion of the cycle}
We end the discussion by recalling that our transient absorption spectroscopy measurements were performed on OCPo in trehalose glass, a method used to prevent the full transition to OCPr (Figure \ref{fig:OCP_steady_state}c,d). Despite the lack of full photoconversion, the transient absorption spectroscopy (Figure \ref{fig:STTA_LTTA_OCPo_r}a,c) shows very similar features to those observed in OCPo buffer by us (Figure \ref{fig:STTA_LTTA_OCPo_r}d) and in the literature.\cite{Slouf2017, Konold2019a} Inspection of the films (Figure \ref{fig:degredation_CAN-OCPo_trehalose}) and individual transient absorption sweeps (Figure \ref{fig:LTTA_OCPo_50sweeps}) show that no permanent change is induced by laser excitation (aside from bleaching in high pump-fluence measurements). Therefore the trehalose prevents the overall transition to OCPr, yet the trehalose does not prevent the initial (pre-OCPr) photoproducts from forming.

We do not assign identities (protein conformation, interactions such as hydrogen-bonds, \textit{etc.}) of the $>$\SI{}{\nano\s} photoproducts seen in our spectroscopic measurements. These identities are under active debate in the literature, with experimental work\cite{Konold2019a, Gupta2019, Leverenz2015, Tsoraev2023, Tsoraev2025} and theoretical studies \cite{Bondanza2020,Pigni2020,Bondanza2020a} generally suggesting a process involving rupture of the hydrogen-bonds between the \textbeta1-ring carbonyl group and Typ201/Tyr288, subsequent translocation of the carotenoid into the NTD, and final separation of the CTD and NTD, giving OCPr. Some studies question whether the translocation occurs,\cite{Bandara2017} and a recent study suggested that carotenoid photoisomerisation and subsequent structural rearrangements are the first events that occur, with hydrogen-bond breakage merely a secondary effect accompanying domain separation.\cite{Chukhutsina2022b} Our findings are consistent with initial photo-induced formation of a red-shifted chromophore that sits within a `compact' OCPo-type protein structure, even with the full conversion to OCPr (with associated domain separation) prevented by the trehalose. 

Finally, we note that the use of trehalose to prevent photoconversion of OCP also offers easy and efficient use of material. We found that the sample could be stored at room temperature in the dark without degradation or photoconversion for weeks at a time, allowing us to perform multiple experiments on the same sample, thus saving both material and time (avoiding the need to prepare large volumes of (mutant) samples,\cite{Maksimov2015, Maksimov2017,Maksimov2020,Yaroshevich2021,Tsoraev2025} setting up and cleaning the flow cell\cite{Konold2019a, Slouf2017, Nizinski2022a,Tsoraev2025} or cryogenically freezing the solution \cite{Berera2013,Niedzwiedzki2014}). The trehalose glass system is therefore a relatively straightforward, material-efficient, and artefact-mitigated method for spectroscopic studies of photoactive proteins.

\section*{Conclusions}

We fabricated samples of canthaxanthin-binding inactive (OCPo) and active (OCPr) forms of orange carotenoid protein (OCP) in trehalose glass. We found that formation of OCPr is prevented in this matrix, enabling static and time-resolved photophysical characterisation using samples that are stable and material-efficient.

We use these samples to perform transient absorption spectroscopy employing narrowband pumps spanning the visible region. We show that S* features, seen only for pumps $<\SI{495}{\nano\m}$, are not required for photoconversion. Conversion to OCPr occurs for all visible wavelengths absorbed by the dark-adapted OCPo. We hypothesise that `S*' in OCP and wider carotenoid photophysics arises due to chromophore heterogeneity, with other effects attributable to heterogeneity in the protein environment.

Our results clear up some confusion from the literature regarding S* and its putative role in the photoconversion, but we leave questions unanswered. Our experiments on photoconverting OCP solution hint at a wavelength-dependent photoconversion efficiency, but the experimental design requires improvement before we can assert a real trend and whether it correlates with any of the features seen in sub-ps and ns--ms transient absorption spectroscopy. Pump wavelength-dependent behaviour in photoconversion can originate from two broad categories: chromophore/environmental heterogeneity, which appears to be present in this sample, or excitation-dependent processes such as intersystem crossing, which has previously been observed in carotenoids\cite{Nielsen1996} and implicated in OCP photoconversion.\cite{Khan2022} Again, with the available data, we cannot conclude whether either, or both, of these factors influence the photoconversion yield and suggest that more detailed pump wavelength-dependent studies using additional spectroscopic handles such as photoluminescence,\cite{Gurchiek2018, Rose2025} Raman,\cite{Chrupkova2024, Kish2015} and mid-IR absorption\cite{Konold2019a, Leccese2023}  are required on a range of timescales, namely `fast' (afforded by $\sim$\SI{}{\kilo\Hz} lasers) and `slow' (such as $\sim$s--min transient absorption\cite{Rose2023,Nizinski2022,Nizinski2025}). In the immediate instance, we highlight the importance, due to this ground-state heterogeneity, of pump wavelength-dependence on the OCP photophysics.

\section*{Author contributions}

\begin{sloppypar}
	J.P.P.~and G.A.S.~conceived the study. J.P.P., G.A.S., S.W., R.J.~and J.C.~designed the experiments. G.A.S., M.S.P.~and A.H.~prepared protein samples under the supervision of M.P.J.~and C.N.H. The absorbance measurements in Figure \ref{fig:OCP_steady_state} were performed by G.A.S. Transient absorption was conducted by J.P.P.~and S.W.~within the Lord Porter Laser Facility, with S.B.~assisting and D.C.~providing facility management. S.B.~finalised construction of the nanosecond-millisecond transient absorption system. J.P.P.~and R.J.~conducted the narrowband-pump photoconversion experiments. J.P.P.~and G.A.S.~analysed the data. R.K.V.~assisted with the literature review. J.P.P., G.A.S.~and J.C.~wrote the manuscript and prepared the figures with input from all authors.
\end{sloppypar}

\section*{Conflicts of interest}

There are no conflicts to declare.

\section*{Supplementary information (SI) available} Literature review clarifying our current understanding of S* in carotenoids and OCP; experimental methods, including canthaxanthin-binding OCP production, OCPo and OCPr in sugar glasses preparation, spectroscopic setups, beam characterisation methods, and data analysis procedures; supplementary visible/NIR picosecond transient absorption materials, including pump spectra, \SI{485}{\nano\m} pump measurements on OCPo and OCPr, and the results of global lifetime analysis; supplementary nanosecond-millisecond transient absorption materials, including 50-sweep averages of the OCPo spectra, measurement on OCPr, and results of global lifetime analysis on OCPo in trehalose and buffer; image of encapsulated OCPo with bleaching but no conversion after high-fluence measurements; supplementary UV-vis picosecond transient absorption materials, including pump spectra, replicate measurements at the same nominal pump fluence, measurements demonstrating pump fluence independence, fit parameters for the global fit of dynamics, and the results of global target analysis; further details on the narrowband-pump photoconverting solution experiment, in particular about the experimental oversights. Refs.\cite{Polivka2009,Niedzwiedzki2017,Kolano2006,Niedzwiedzki2007,LakowiczBook} are referenced in the SI and not additionally in the main text above.

\section*{Data availability}

Data will be available on ORDA (the University of Sheffield's repository, provided by Figshare) once the paper has been accepted for publication. The Python software used to run the nanosecond to millisecond transient absorption system is available on GitHub at \url{https://github.com/fast-spectroscopy-sheffield/TA-software}

\section*{Acknowledgements}

The authors thank James D.~Shipp and David G.~Bossanyi for assistance with transient absorption measurements. The authors thank Andrew J.~Musser, Daniel W.~Polak, and D.G.B.~for their contributions to the nanosecond-millisecond transient absorption system.

\begin{sloppypar}
	\textbf{Funding:} J.P.P.~thanks the Engineering and Physical Sciences Research Council (EPSRC) for support through a Doctoral Training Partnership Scholarship (EP/R513313/1) and through an Impact Acceleration Account (EP/X525790/1). G.A.S.~and C.N.H.~acknowledge European Research Council (ERC) Synergy Grant 854126. The authors thank the EPSRC for a Capital Equipment Award (EP/L022613/1 and EP/R042802/1) which funded the Lord Porter Ultrafast Laser Spectroscopy Laboratory used in this study. J.C., C.N.H, G.A.S., and S.W.~thank the EPSRC for funding through EP/S002103/1.  J.P.P., J.C., R.J., and R.K.V.~ thank the EPSRC for funding through EP/T012455/1. J.C.~and S.W.~also thank the EPSRC for funding through EP/N014022/1. S.B.~thanks the EPSRC for support through EP/R045305/1. M.S.P.~and M.P.J.~were supported by Leverhulme Trust award RPG-2019-045. A.H.~acknowledges support of a Royal Society University Research Fellowship (award URF{\textbackslash}R1{\textbackslash}191548).
\end{sloppypar}

\clearpage

\bibliography{library}

\clearpage

\renewcommand{\thepage}{S\arabic{page}}
\setcounter{page}{1}
\renewcommand{\thesection}{} 
\setcounter{section}{1}
\renewcommand{\thesubsection}{S\arabic{subsection}}
\renewcommand{\thetable}{S\arabic{table}}
\setcounter{table}{1}
\renewcommand{\thefigure}{S\arabic{figure}}
\setcounter{figure}{1}
\renewcommand{\theequation}{S\arabic{equation}}
\setcounter{equation}{1}


\section*{Supplementary information (SI)}

\subsection{Literature review on S* in carotenoids and OCP}
We start our discussion with a description of S* as a generic spectral feature without assigning it to any distinct excited-state or process at this point. S* is a catch-all term for a excited-state absorption (ESA) spectral feature located at the red edge of the S\textsubscript{0}$\rightarrow$S\textsubscript{2} ground-state bleach (GSB), and distinct from the main singlet S\textsubscript{1} ESA band \cite{Polivka2009}. The S* feature is typically weak ($\sim$1--5\% of the S\textsubscript{1}$\rightarrow$S\textsubscript{n} absorption intensity) with a long lifetime ($\sim$20--\SI{100}{\pico\s}) compared to the S\textsubscript{1} absorption features. The S* feature is sometimes also associated with narrowing of the GSB and often shows pump (excitation) wavelength-dependent behaviour (it is relatively more intense with higher-energy excitation) \cite{Ostroumov2011, Polak2019, Nizinski2022a}. This broad definition of the S* spectral feature and its location at the edge of the GSB, which can be prone to artefacts (because depositing energy into a system can lead to an increase in local temperature, resulting in changes in the sample's refractive index and absorption spectrum in the probe region), means that S* has been assigned to a variety of different phenomena \cite{Polivka2009, Sebelik2022, Ostroumov2011, Niedzwiedzki2017}. 

In OCP, Konold \emph{et al.}\cite{Konold2019a}~suggest that the S* feature represents a key intermediate electronic excited state that drives the switch from OCPo to OCPr. They hypothesise that S* is a structurally distorted form of the lowest singlet excited state in carotenoids \cite{Konold2019a,Kloz2016,Niedzwiedzki2007}, and that this distortion enables hydrogen-bond rupture between C=O and Trp288/Tyr201. 
A study by Yaroshevich \textit{et al.}~supports the  hypothesis that S* is a distinct excited state \cite{Yaroshevich2021}, but suggests that it enables accumulation of intramolecular charge transfer (ICT) states which are themselves responsible for hydrogen-bond breaking \cite{Yaroshevich2021} (though we note more recent work from the group questions this hypothesis\cite{Tsoraev2025}). In both of these pictures, hydrogen-bond rupture should occur during the S* lifetime, which was indeed reported in a tryptophan fluorescence study of an OCP mutant, where four Trp residues were mutated, allowing analysis of the crucial W288-carotenoid hydrogen bond interaction (the one involved in hydrogen-bonding to C=O) \cite{Maksimov2020}. However, hydrogen-bond breaking should also be observable through shifts in the transient mid-infrared or UV-vis absorption spectra; either as a blue-shifted ESA of the C=O band \cite{Kolano2006}, or as a ESA to the blue-edge of the GSB in UV-vis transient absorption spectroscopy \cite{Yaroshevich2021}. To our knowledge, neither of these features have been observed on the S* timescale ($\sim$20--\SI{100}{\pico\s}), although it is possible that these features are simply outside of the measured spectral ranges \cite{Konold2019a} or obscured by other spectral features. In addition, a more recent study by Chukhutsina \emph{et al.}, using time-resolved crystallography and optical spectroscopy suggests that the hydrogen-bond breakage occurs on 5--\SI{10}{\minute} timescales, and not during the S* lifetime, and is merely as a consequence of N- and C-terminal domain separation.\cite{Chukhutsina2022b}

Other research by Balevičius \textit{et al.}~supports the hypothesis that observation of an S* feature in visible transient absorption spectroscopy is key to understanding the trigger to photoswitching in OCP \cite{Balevicius2019}. However, Balevičius' study on carotenoids in solvent, together with earlier work \cite{Balevicius2015, Balevicius2016, Balevicius2018}, suggest that the S* spectral feature is not an excited-state signature, but instead arises due to transient heating of the carotenoid and solvent during and immediately following internal conversion \cite{Balevicius2019}. Internal conversion in carotenoids is incredibly rapid and most of the energy deposited into OCP by light ($\sim$\SI{2}{\electronvolt}) is converted into vibrational or kinetic energy within $\sim$\SI{100}{\femto\s} to \SI{20}{\pico\s} through intramolecular vibrational redistribution (IVR) and vibrational cooling (VC) to the surroundings. IVR and VC populate both higher-lying vibronic levels ($\nu\geq 1$) in the ground (S\textsubscript{0}) and excited (S\textsubscript{1}) electronic states \cite{Balevicius2015, Balevicius2016, Balevicius2018} and cause local heating in the form of population of low-energy molecular, solvent, or protein vibrational/rotational modes. Both of these result in transient spectral features in the GSB spectral region which could be assigned to S* \cite{Balevicius2015, Balevicius2016, Balevicius2018, Balevicius2019, Sebelik2022}. From these carotenoid-in-solution studies, Balevičius \textit{et al.}\cite{Balevicius2019}~suggested that the S* feature seen in OCP by Konold \textit{et al.}\cite{Konold2019a}~was consistent with a residual `hot ground state', associated with a non-equilibrium distribution of carotenoid vibronic populations as well as elevated local temperature (rather than a distinct excited state), that may provide enough energy to break the weak hydrogen-bonds \cite{Balevicius2019} (estimated to have a bond energy of $\sim$\SI{8}{\kilo\cal\per\mol} or \SI{0.35}{\electronvolt} per molecule \cite{Yaroshevich2021}). A recent study employing femtosecond stimulated Raman spectroscopy demonstrated signatures concomitant with S* in a simultaneous transient absorption experiment; these features were assigned to a hot ground state.\cite{Chrupkova2024}

Still others claim that the S* feature in transient absorption spectroscopy is not due to an intermediate excited state, or due to heating, but is instead due to inhomogeneity of the sample. The inhomogeneity is observed as a pump wavelength-dependent change in resonance Raman \cite{Kish2015} and transient absorption spectra \cite{Slouf2017, Polivka2013, Khan2020, Kuznetsova2020a, Nizinski2022a} and is usually attributed to carotenoid conformational inhomogeneity as the carotenoid adopts more than one conformation in the ground-state \cite{Kish2015, Slouf2017, Kuznetsova2020a,Pishchalnikov2019, Arcidiacono2023}, possibly due to spontaneous hydrogen-bond disruption \cite{Maksimov2020}. This spectral heterogeneity has been observed in OCPo binding 3$'$-hydroxyechinenone (3$'$hECN)\cite{Polivka2005,Polivka2013}, echinenone (ECN)\cite{Kish2015, Slouf2017, Nizinski2022a,Pishchalnikov2019}, canthaxanthin (CAN)\cite{Slouf2017, Arcidiacono2023} and zeaxanthin (ZEA)\cite{Slouf2017}, in N-terminal domain helical carotenoid proteins (HCP) HCP2 and HCP3 containing CAN,\cite{Khan2020} and in the non-canonical OCP2 clade.\cite{Kuznetsova2020a}

The hypothesis that the pump wavelength-dependence of OCP spectral features is due to conformational heterogeneity of the carotenoid has basis in studies of isolated carotenoid in solution \cite{Polgar1942, Zechmeister1944, Pesek1990, Ostroumov2011, Polak2019}. For example, an important study by Ostroumov \emph{et al.}\cite{Ostroumov2011}~demonstrated that the pump wavelength-dependence could be removed by purifying \textbeta-carotene immediately prior to measurement. The resulting transient absorption spectra of all-\textit{trans} \textbeta-carotene showed no sign of the putative S* spectral feature. While the aforementioned transient heating due to rapid internal conversion is also predicted to show some pump wavelength-dependence \cite{Balevicius2018}, the effect is small and short-lived, and cannot on its own explain the sometimes dramatic changes in transient absorption spectra in relation to pump wavelength \cite{Polak2019}.

The literature places great emphasis on the S* feature because it is widely believed to be directly correlated with the first steps of the photoconversion mechanism of OCP \cite{Konold2019a, Yaroshevich2021, Maksimov2020}. We find that the evidence for this key assumption is largely circumstantial and requires further scrutiny. Therefore, in the main text, we aim to test the hypothesis that the S* spectral feature is directly correlated with photoconversion.

\subsection{Methods}

\subsubsection{Sample preparation}
\label{sec:methods_sample_prep}

OCP binding near-100\% CAN was produced using the same method to that in previous work.\cite{Sutherland2023} CAN-OCP from BL21(DE3) \textit{Escherichia coli} (\textit{E.~coli}) using a dual-plasmid system comprised of pAC-CANTHipi \cite{Cunningham2007} and pET28a containing the gene encoding OCP (slr1963) from \textit{Synechocystis} sp.~PCC 6803. Briefly, \SI{500}{\milli\litre} cultures were grown at \SI{37}{\degreeCelsius} (\SI{200}{\rev\per\minute} agitation) in \SI{2}{\litre} baffled Erlenmeyer flasks using lysogeny broth medium containing the appropriate concentrations of antibiotics. When the absorbance of the medium at \SI{600}{\nano\m} had reached $A_{600}=0.6$ (\SI{1}{\centi\m} path length), protein production was induced by addition of \SI{0.5}{\milli\Molar} isopropyl \textbeta-D-1-thiogalatopyranoside and the cultures incubated for 16 hours at \SI{18}{\degreeCelsius}.

Cells were harvested by centrifugation (4,400$\times${}$g$, \SI{30}{\minute}, \SI{4}{\degreeCelsius}) and resuspended in binding buffer (\SI{50}{\milli\Molar} HEPES, pH 7.4, \SI{500}{\milli\Molar} NaCl, \SI{5}{\milli\Molar} imidazole). Cells were lysed by sonication and then centrifuged (53,000$\times${}$g$, \SI{30}{\minute}, \SI{4}{\degreeCelsius}). The supernatant was collected and filtered (\SI{0.22}{\micro\m} filter pores) and applied to a Chelating Sepharose Fast Flow column (GE Healthcare) pre-equilibrated with NiSO\textsubscript{4}. The column was washed with binding buffer, wash buffer (\SI{50}{\milli\Molar} HEPES, pH 7.4, \SI{500}{\milli\Molar} NaCl, \SI{50}{\milli\Molar} imidazole) and elution buffer (\SI{50}{\milli\Molar} HEPES, pH 7.4, \SI{100}{\milli\Molar} NaCl, \SI{400}{\milli\Molar} imidazole) with the elution pooled for further purification. The protein sample was buffer exchanged into buffer A (\SI{50}{\milli\Molar} HEPES, pH 7.4) loaded onto a Fast Flow Q-Sepharose column (GE Healthcare) and a linear gradient of 0--\SI{1}{\molar} NaCl was applied. Fractions were analysed by SDS-PAGE and appropriate samples taken forward for size exclusion chromatography on a Superdex 200 Increase column (GE Healthcare) in buffer B (\SI{50}{\milli\Molar} HEPES, pH 7.4, \SI{200}{\milli\Molar} NaCl). Where necessary OCP samples were concentrated using centrifugal dialysis (VivaSpin, Sartorius).

Trehalose glass encapsulation of OCPo and OCPr was conducted using a similar method to that in previous work.\cite{Sutherland2020, Sutherland2023} \SI{100}{\micro\litre} of concentrated protein solution ($A_\textrm{max}\sim2$, \SI{1}{\centi\m} path length) in aqueous buffer (\SI{50}{\milli\molar} HEPES, \SI{200}{\milli\molar} NaCl, pH 7.4) was mixed with \SI{100}{\micro\litre} of a trehalose-sucrose mixture (\SI{0.5}{\molar} trehalose, \SI{0.5}{\molar} sucrose). \SI{200}{\micro\litre} of the protein-trehalose mixture was drop-cast in the centre of a quartz-coated glass substrate (S151, Ossila; 15$\times$20$\times$\SI{1.1}{\milli\m}). The substrate was incubated under vacuum (\SI{-70}{\kilo\pascal}) with an excess of calcium sulphate desiccant (Drierite) at room temperature for at least 48 hours.

Some samples were additionally encapsulated with imaging spacers and a cover slip to protect the trehalose against atmospheric rehydration. For these samples, a stack of two imaging spacers (SecureSeal, Grace BioLabs; \SI{9}{\milli\m} diameter, \SI{0.12}{\milli\m} thickness) were attached to the quartz-coated glass substrate (S151, Ossila; 15$\times$20$\times$\SI{1.1}{\milli\m}) and \SI{40}{\micro\litre} of the protein-trehalose mixture drop-cast in the centre of the imaging spacer. The substrate was then placed in vacuum as above; pressure was released under a continuous flow of ultra-pure nitrogen gas, and a glass microscope cover slip (ThermoScientific; 22$\times$\SI{22}{\milli\m}, No.1 thickness) was attached to the upper imaging spacer.

For OCPo samples, all preparation steps and desiccation were conducted in the dark. For OCPr, preparation and desiccation were conducted under bright white light, with samples illuminated for \SI{30}{\minute} (\SI{1600}{\micro\mol\photon\per\square\m\per\s}) prior to the addition of the trehalose-sucrose solution and constant weaker illumination (\SI{500}{\micro\mol\photon\per\square\m\per\s}) for the duration of the desiccation. After encapsulation, samples were stored at room temperature in the dark.

\subsubsection{Steady-state absorbance spectroscopy}

Absorbance of samples were (unless specified otherwise) measured in a commercial Cary double-beam spectrometer (Cary 60 UV-Vis Spectrophotometer, Agilent Technologies). Both zero and baseline corrections were applied with a blank sample; in the case of trehalose-encapsulated samples, a trehalose blank was used.

\subsubsection{Laser beam power and diameter measurements}
\label{sec:methods_pump_fluence}

Laser beam spot diameters ($1/e^2$) were measured at the sample positions with a CCD beam profiler (BC106N-VIS/M, Thorlabs). Unless stated otherwise, laser beam powers were measured slightly before the sample position (off-focus) with a photodiode power sensor (S120VC, Thorlabs) and meter console (PM100A or PM100D, Thorlabs). Both measurements were used in subsequent calculations for the pump fluence using the formula
\begin{equation}
\label{eqn:pump_fluence}
\textrm{pump fluence} = \frac{P}{f \pi r_1 r_2}
\end{equation}
where $P$ is the measured power, $f$ is the laser repetition rate, and $r_1$, $r_2$ are the major/minor radii ($1/e^2$) of the beam at the sample position.

\subsubsection{Transient absorption spectroscopy}
\label{sec:methods_TA}

\begin{figure}[h]
	\centering
	\includegraphics[scale=1.0]{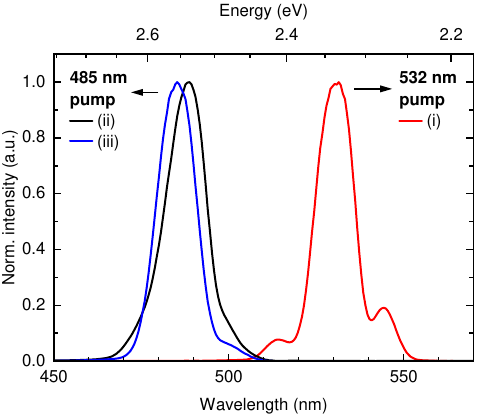}
	\caption{\textbf{Intensity spectra of the pumps used in the visible/NIR \SI{}{\pico\s} transient absorption experiments, normalised to the maximum intensity.} Small variations in the tuneable pump generation resulted in slightly different pump profiles on each experimental day. \SI{532}{\nano\m} pump (i) was used in taking the visible-probe data shown in Figures \ref{fig:STTA_LTTA_OCPo_r}a,b, \ref{fig:STTA_OCPo}a, and \ref{fig:STTA_OCPr}a. \SI{485}{\nano\m} pump (ii) was used in taking the visible-probe data shown in Figures \ref{fig:STTA_OCPo}b and \ref{fig:STTA_OCPr}b. \SI{485}{\nano\m} pump (iii) was used in taking the NIR-probe data shown in Figures \ref{fig:STTA_OCPo}b and \ref{fig:STTA_OCPr}b. The \SI{532}{\nano\m} pump used in taking the NIR-probe data of Figures \ref{fig:STTA_LTTA_OCPo_r}a,b, \ref{fig:STTA_OCPo}a, and \ref{fig:STTA_OCPr}a was not recorded, and may have had a slightly different spectrum to (i).}
	\label{fig:STTA_pump_scatter}
\end{figure}

Picosecond transient absorption spectroscopy was undertaken with a commercial spectrometer (Helios, Ultrafast Systems) outfitted with a Ti:Sapphire seed laser (MaiTai, Spectra-Physics) providing \SI{800}{\nano\m} pulses (\SI{84}{\mega\hertz}, \SI{25}{\femto\s} nominal FWHM) and a Ti:Sapphire chirped-pulse amplifier (Spitfire Ace PA-40, Spectra-Physics) amplifying \SI{800}{\nano\m} pulses (\SI{10}{\kilo\hertz}, \SI{12}{\W} average power, \SI{40}{\femto\s} nominal FWHM). Pump pulses were generated by seeding a part of the \SI{800}{\nano\m} beam into either a frequency doubler/tripler utilising \textbeta-barium borate crystals (TimePlate, Photop Technologies) for \SI{400}{\nano\m} excitation, or an optical parametric amplifier (TOPAS Prime, Light Conversion) and subsequent frequency mixer (NirUVis, Light Conversion) for \SI{420}{\nano\m} to \SI{600}{\nano\m} excitations. An optical chopper was used to modulate the pump frequency to \SI{5}{\kilo\hertz}. Pump spectra used for some of the visible/NIR probe experiments are shown in Figure \ref{fig:STTA_pump_scatter}, and pump spectra for the UV-vis probe experiments are shown in Figure \ref{fig:STTA_ex_wav_dep_pump_scatter}. Supercontinuum probes were generated with a part of the \SI{800}{\nano\m} pulse focused on either a continuously translating CaF\textsubscript{2} crystal for UV-vis probes (350--\SI{750}{\nano\m}; we note that the use of a hot mirror and filters typically restricted the range to 370--\SI{690}{\nano\m}, and with low probe in the 370--\SI{400}{\nano\m} sub-range), a sapphire crystal for visible probes (450--\SI{800}{\nano\m}), or a YAG crystal for NIR probes (800--\SI{1600}{\nano\m}). Pump-probe delay was controlled with a motorised delay stage with a random stepping order per-sweep, and the resulting sweeps are averaged. The signal was dispersed with a grating and detected with a CMOS sensor for UV-vis and visible probes, or an InGaAs sensor for NIR probe. The pump and probe polarisations were set to the magic angle. The room temperature was controlled at \SI{19}{\degreeCelsius}.

Nanosecond to millisecond transient absorption spectroscopy was undertaken with a home-built pump-probe-reference setup. A Q-switched Nd:YVO\textsubscript{4} laser (Picolo AOT 1, Innolas) outfitted with an integrated harmonic module for frequency-doubling (to \SI{532}{\nano\m}) provided pump pulses (\SI{500}{\hertz}, $<$\SI{800}{\pico\s} nominal FWHM). A Ti:Sapphire regenerative amplifier (Solstice, Spectra-Physics) provided \SI{800}{\nano\m} pulses (\SI{1}{\kilo\hertz}, \SI{4}{\watt} nominal power, \SI{90}{\femto\s} nominal FWHM) for supercontinuum generation. 450--\SI{700}{\nano\m} supercontinuum pulses were generated by focusing part of the \SI{800}{\nano\m} pulse on a sapphire crystal, and subsequently split into probe and reference beams with a 50:50 beamsplitter. The pump and probe were focused and overlapped on the sample, while the reference was focused on the sample $\sim$\SI{2}{\milli\m} away from the pump/probe overlap. Pump-probe/reference delay was controlled electronically with a digital delay generator (DG645, Stanford) with a linear stepping order per-sweep, and the resulting sweeps are averaged. To reduce \SI{532}{\nano\m} pump scatter, a \SI{532}{\nano\m} notch filter (NF533-17, Thorlabs) was placed after the sample. The probe and reference were dispersed with a volume phase holographic grating (Wasatch, \SI{360}{\lines\per\milli\metre} at \SI{550}{\nano\m} CW, \SI{30}{\milli\m} diameter, \SI{3}{\milli\metre} thick, BK7) and directed onto two linear image sensors (S7030, Hamamatsu) driven and read out at the probe/reference repetition rate (\SI{1}{\kilo\hertz}) by a custom-built board (Entwicklungsbuero Stresing). Transient absorption data was acquired with home-built Python software (GitHub link: https://github.com/fast-spectroscopy-sheffield/TA-software). The pump and probe polarisations were set to the magic angle. The room temperature was controlled at \SI{19}{\degreeCelsius}.

Surface Xplorer 4.3.0 (Ultrafast Systems) was used in processing the transient absorption datasets. Noisy edges of the spectra were trimmed, and the program's bad spectra replacement procedure was applied. To account for pump scatter, which typically appeared as a negative $\Delta A$ contribution at all times in the pre-processed datasets, background correction (`subtract scattered light') was applied, averaging the spectra before any apparent response from the sample. Note that this background correction resulted in apparently positive/negative pump scatter artefacts in some spectra shown. For \SI{}{\pico\s} transient absorption data using the visible and UV-vis probes, chirp correction was applied, choosing points at the first apparent signal for a given dynamic. Chirp was not discernible in the NIR-probe \SI{}{\pico\s} transient absorption data nor all \SI{}{\nano\s}--\SI{}{\milli\s} transient absorption data, so a chirp correction was not applied to these. Time zero was adjusted to the time of maximum initial signal in \SI{}{\pico\s} transient absorption data, while for \SI{}{\nano\s}--\SI{}{\milli\s} transient absorption data it was adjusted to the time of first signal. Further processing and some analysis was performed with home-built Python code. Glotaran 1.5.1 (http://glotaran.org)\cite{Snellenburg2012}, a GUI for the R package TIMP\cite{Mullen2007}, was used in global lifetime/target analysis.

\subsubsection{Pump wavelength-dependent photoconversion experimentation}
\label{sec:methods_PC_OCP}

Absorbance spectroscopy on continually-photoconverting OCP in solution under narrowband pump was undertaken using a home-built system. OCP in buffer was placed in a \SI{2}{\milli\m} path length (at normal incidence) quartz cuvette (Hellma 110-2-40). Conversion to OCPr was continuously triggered by narrow-band pump light (\SI{78.3}{\mega\hertz}, $<$\SI{10}{\nano\m} nominal FWHM, $\pm$\SI{5}{\nano\m} nominal centre wavelength accuracy) turned on at time $t=0$. This pump light was provided by a supercontinuum laser (SuperK EXTREME EXU-6 PP, NKT Photonics) outfitted with a tuneable filter (SuperK VARIA, NKT Photonics). Pump light was focused by a lens (LA4380, Thorlabs) through the cuvette at $\sim$\SI{40}{\degree} from the cuvette's normal to minimise transmission and scattering into the spectrometer. The power of the pump was controlled such that the initial number rate of photon absorptions (\textit{i.e.}~OCPo excitations) was approximately the same for each pump wavelength, achieved by keeping a constant product
\begin{equation}
P \lambda_{\textrm{pump}} (1-10^{A_\textrm{dark}})
\end{equation}
where $P$ is the steady-state power of the pump, $\lambda_{\textrm{pump}}$ is the pump wavelength, and $A_\textrm{dark}$ (shown as a green line in Figure \ref{fig:PC_OCP_absorbance}) is the absorbance of dark-adapted OCP in buffer determined at normal incidence using a separate spectrometer (FluoroMax-4, Horiba, fitted with a Xenon lamp). The exception to this pump power control was the non-absorbed \SI{675}{\nano\m} pump, where instead the photon rate was kept the same as that for \SI{550}{\nano\metre} pump (so controlling $P \lambda_{\textrm{pump}}$ between those two pumps). We note that pump powers $P$ were not measured close to the sample position to prevent pre-measurement photoconversion and also due to spatial limitations; they were instead measured and controlled at a position such that to reach the sample, reflection by three UV-enhanced Al mirrors (PF10-03-F01, Thorlabs) and focusing through the aforementioned lens is required. This results in a slightly wavelength-dependent fraction of power difference between what was measured as $P$ and what was incident on the sample. After measurements were completed, this was checked by positioning the power meter at the sample position and at the far position, and summarised in Table \ref{tab:PC_OCP_pump_powers_calc}. We note also that $A_\textrm{dark}$ was determined at normal incidence, but the pump light was at $\sim$\SI{40}{\degree} from the cuvette's normal, so that the fraction of pump photons absorbed differs from $(1-10^{A_\textrm{dark}})$. We give full details of the pump power control, including an explanation and correction of these noted oversights, in Section \ref{sec:PC_OCP_oversights}; in short, these particular oversights did not have a substantial effect on our results. The yield of OCPr was monitored by measuring the absorbance spectrum of OCP using a weak white-light probe from a fibre-coupled halogen-tungsten/deuterium lamp  (DH-2000-BAL, Ocean Optics), turned on just before the pump; turning both on at $t=0$ was not possible. This white-light was focused onto the sample close to normal incidence using collimating and focusing lenses. The pump and white light were overlapped in the cuvette close to their focuses. The sample-attenuated white light was collimated and focused by subsequent lenses into an optical fibre, in turn coupled to a CCD spectrometer (Andor Shamrock SR-303i-A, Oxford Instruments), measuring white-light transmission through the sample. Absorbance $A(t)$ was calculated using the measured transmission through a cuvette containing solvent (buffer), with the equation
\begin{equation}
A(t) = -\log\left(\frac{I_\textrm{OCP}(t)}{I_\textrm{buffer}}\right)
\end{equation}
where $I_\textrm{OCP}$ is the transmission in counts measured by the CCD for OCP in solution, and $I_\textrm{buffer}$ is that for buffer only. The pump spot diameter was $\sim$\SI{30}{\micro\m} and the white light spot diameter was $\sim$\SI{1}{\milli\metre} at the overlap/sample position. The pump spot diameter could not be determined precisely using a beam profiler due to the large angle of incidence of the pump beam, and potentially varies in position and size per-wavelength due to the dispersive optics. The larger white-light spot size also potentially yields an artefact; see Section \ref{sec:PC_OCP_oversights} for further detail. Experimental consistency was verified by repeating the experiment twice on the same OCP in buffer sample. No apparent degradation of the OCP occurred, and back-conversion from OCPr to OCPo in the dark was successful in the $\geq$1 hour between experimental runs. The room temperature was controlled at \SI{18}{\degreeCelsius}.

\subsubsection{Figure preparation}

OriginPro 9.6.0.172 (OriginLab), home-built Python code, and Inkscape 1.1.2 (\url{https://inkscape.org/}) were used to prepare the plots.

\clearpage

\subsection{Supplementary visible/NIR \SI{}{\pico\s} transient absorption materials}
\label{sec:STTA_OCPo_OCPr}

\subsubsection{Pump spectra}

\clearpage

\subsubsection{Transient absorption on OCPo and OCPr in trehalose}

\begin{figure}[h]
	\centering
	\includegraphics[scale=1.0]{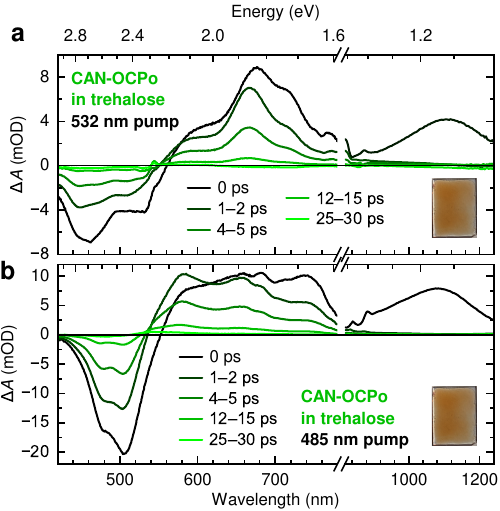}
	\caption{\textbf{Picosecond transient absorption spectra of CAN-binding OCPo in trehalose glass with pump wavelength \SI{532}{\nano\m} (a) or \SI{485}{\nano\m} (b) with visible (left) and NIR (right) probes.} Note that panel (a) is depicted in the main text (Figure \ref{fig:STTA_LTTA_OCPo_r}a), and in turn is reproduced from our previous work.\cite{Sutherland2023} Spectra have been averaged between the times indicated. Pump fluence was set to \SI{200}{\micro\J\per\square\centi\m}.}
	\label{fig:STTA_OCPo}
\end{figure} 

\begin{figure}[h]
	\centering
	\includegraphics[scale=1.0]{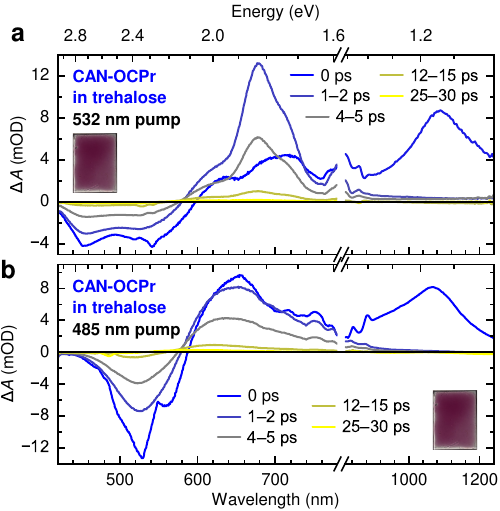}
	\caption{\textbf{Picosecond transient absorption spectra of CAN-binding OCPr in trehalose glass with pump wavelength \SI{532}{\nano\m} (a) or \SI{485}{\nano\m} (b) with visible (left) and NIR (right) probes.} A pump wavelength-dependence on the spectra is apparent. Note that panel (a) is depicted in the main text (Figure \ref{fig:STTA_LTTA_OCPo_r}b), and in turn is reproduced from our previous work.\cite{Sutherland2023} Spectra have been averaged between the times indicated. Pump fluence was set to \SI{200}{\micro\J\per\square\centi\m}.}
	\label{fig:STTA_OCPr}
\end{figure}

In this section, we highlight the results of \SI{}{\pico\s} transient absorption with visible- and NIR-region probes under two different pump (excitation) wavelengths on CAN-OCPr in trehalose glass. We note that the visible/NIR-region measurements are admittedly worse quality than the UV-vis probe transient absorption, largely due to the significant sample degradation and the strong coherent artefact (from \textit{e.g.}~cross-phase modulation\cite{Berera2009}) present due to the high pump fluence densities (\SI{200}{\micro\J\per\square\centi\m}) used. Furthermore, the set maximum time delay in the measurements was $\sim$\SI{40}{\pico\s}, leading to difficulty in resolving states with lifetimes beyond that. We therefore refrain from assigning a photophysical model to OCPr; doing this would require further \SI{}{\pico\s} transient absorption measurements with a lower fluence and further pump wavelengths.

The behaviour of OCPr seemingly matches that of OCPo. The photophysics with \SI{532}{\nano\m} pump (Figure \ref{fig:STTA_OCPr}a) is similar to that of OCPo (Figure \ref{fig:STTA_OCPo}a), although with an overall redshift, consistent with the redshift seen in the absorbance (main text Figure \ref{fig:OCP_steady_state}). We see signatures of a longer-lived S* feature around \SI{600}{\nano\m} probe when using \SI{485}{\nano\m} pump (Figure \ref{fig:STTA_OCPr}b), easily seen with the loss of the near-isosbestic point, suggesting that OCPr also has an S* feature. However, the OCPr GSB at probe wavelength $\sim$\SI{480}{\nano\m} appears relatively diminished using \SI{485}{\nano\m} pump compared to \SI{532}{\nano\m} pump. This is the opposite to the relation seen in OCPo, most easily seen in main text Figure \ref{fig:STTA_OCPo_ex_dep}. We note that the signal is noisy in the $\sim$\SI{480}{\nano\m} region due to the \SI{485}{\nano\m} pump.

\clearpage

\subsubsection{Global lifetime analysis}
\label{sec:GLA_DADS_OCPo_OCPr}

As a comparison of the visible-probe \SI{}{\pico\s} transient absorption data of OCPo (Figure \ref{fig:STTA_OCPo}) to that using a UV-vis probe (Figure \ref{fig:STTA_OCPo_ex_dep}), in addition to seeing if an S*-associated component is fit in the case for OCPr (Figure \ref{fig:STTA_OCPr}), global lifetime analysis on the data was performed. This was done using the Glotaran 1.5.1 software package (http://glotaran.org) \cite{Snellenburg2012}, a GUI for the R package TIMP\cite{Mullen2007}. Data used had already been processed with the steps outlined in the methods (Section \ref{sec:methods_TA}); in particular, a chirp correction had already been applied, so that a term to account for chirp did not need to be included in the fitting. Noisy regions in the data due to pump scatter were excluded for all times to ensure a good fit of the rest of the data. Noisy red and blue ends in the data associated with tails of the probe were also excluded, so that the fitted wavelengths were \SI{430}{\nano\m} to \SI{780}{\nano\m}. The fitting was weighted favourably at later delay times to ensure good fits; Table \ref{tab:STTA_GLA_DADS_weighting} shows the weighting applied. Due to the strong coherent artefact feature in the first \SI{0.5}{\pico\s}, only data beyond that time was fitted. Thus, in the model, terms to account for the coherent artefact and S\textsubscript{2}-like states were not included. This left a relatively simple fitted model of a number of decay-associated difference spectra (DADS) decaying exponentially in parallel.

\begin{table}[ht]
	\renewcommand*{\arraystretch}{1.2} 
	\newcolumntype{C}[1]{>{\centering\arraybackslash}m{#1}} 
	\begin{center}
		\small 
		\begin{tabular}{|C{38mm}|C{27mm}|}
			\hline
			\textbf{Time range (\SI{}{\pico\s})} & \textbf{Weighting} \\
			\hline
			0.5~--~10 & 1 \\ \hline
			10~--~20 & 2 \\ \hline
			20~--~30 & 3 \\ \hline
			30~--~35 & 4 \\ \hline
			$>$35 & 5 \\ \hline
		\end{tabular}
	\end{center}
	\vspace{-6pt}
	\caption{\textbf{Weightings applied to time-ranges of the visible \SI{}{\pico\s} transient absorption data for the global lifetime analysis.} Note that the maximum time delay in these experiments was $\sim$\SI{40}{\pico\s}.}
	\label{tab:STTA_GLA_DADS_weighting}
\end{table}

\begin{figure}[h]
	\centering
	\includegraphics[scale=1.0]{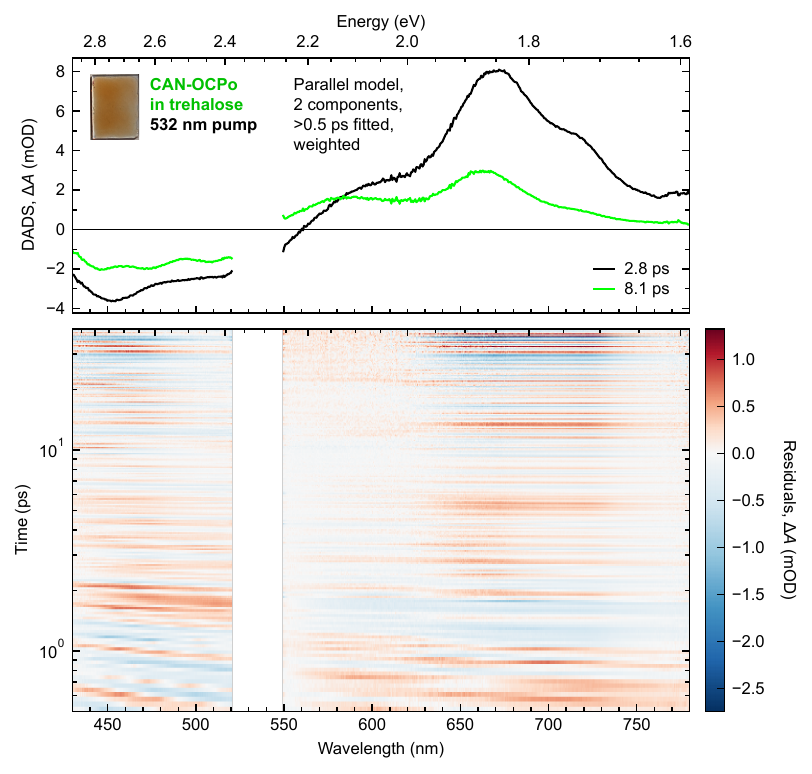}
	\caption{\textbf{Results of global lifetime analysis with a 2-component parallel model on transient absorption data of CAN-binding OCPo in trehalose with pump wavelength \SI{532}{\nano\m} and a visible probe: DADS (top) and residuals (bottom).} Only the wavelength range 430--\SI{780}{\nano\m} and times $>$\SI{0.5}{\pico\s} were fitted, and noisy data from 520.5--\SI{549.5}{\nano\m} due to significant pump scatter was excluded from the fit. DADS time constants are specified in the legend. $\textrm{Residuals} = \textrm{Data}-\textrm{Fit}$. See text for further details. Data and analysis are reproduced from our previous work.\cite{Sutherland2023}}
	\label{fig:STTA_GLA_DADS_OCPo_532nm}
\end{figure}

\begin{figure}[h]
	\centering
	\includegraphics[scale=1.0]{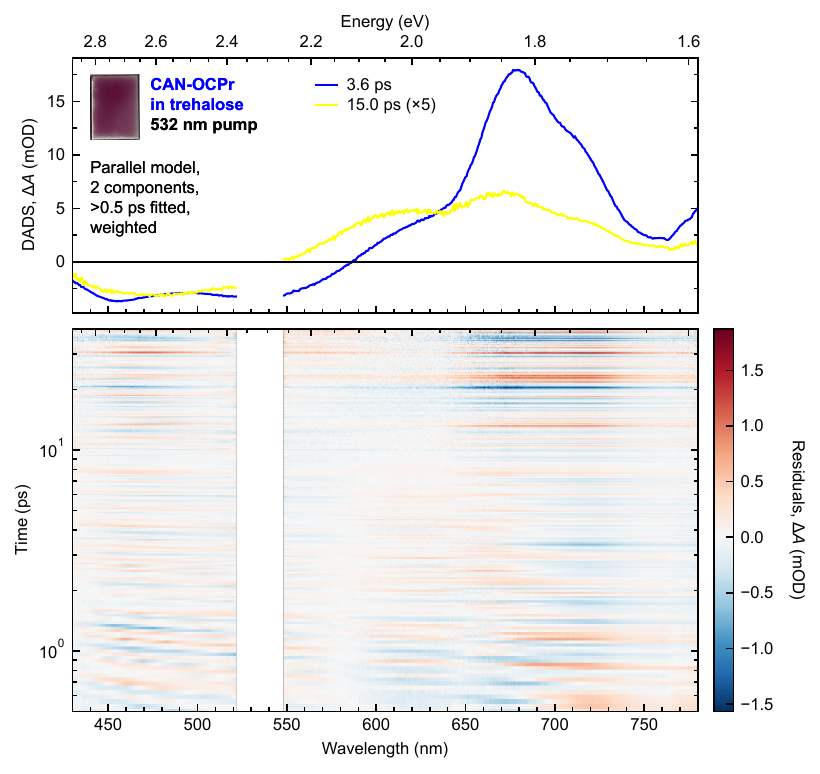}
	\caption{\textbf{Results of global lifetime analysis with a 2-component parallel model on transient absorption data of CAN-binding OCPr in trehalose with pump wavelength \SI{532}{\nano\m} and a visible probe: DADS (top) and residuals (bottom).} Only the wavelength range 430--\SI{780}{\nano\m} and times $>$\SI{0.5}{\pico\s} were fitted, and noisy data from 522--\SI{548}{\nano\m} due to significant pump scatter was excluded from the fit. DADS time constants are specified in the legend; multiplications refer to scalings applied to the DADS. $\textrm{Residuals} = \textrm{Data}-\textrm{Fit}$. See text for further details. Data and analysis are reproduced from our previous work.\cite{Sutherland2023}}
	\label{fig:STTA_GLA_DADS_OCPr_532nm}
\end{figure}

2-component global lifetime analysis of the \SI{532}{\nano\m} pump data shown in the top panels of Figure \ref{fig:STTA_OCPo} and Figure \ref{fig:STTA_OCPr} are shown in Figure \ref{fig:STTA_GLA_DADS_OCPo_532nm} and Figure \ref{fig:STTA_GLA_DADS_OCPr_532nm}. Fitting a 2-component parallel decay model in an artefact-free region of the visible-probe data beyond the initial coherent artefact and S\textsubscript{2}-associated response gives two decay-associated difference spectra (DADS) for both the OCPo data and OCPr data, with the longer time-constant DADS relatively weaker and blueshifted in both cases. A single component is not sufficient to adequately fit the region of the data, and a fitting a third component gives results with spurious DADS profiles. The results for OCPo are consistent (in the sense of the number of components required for a reasonable fit) with a 3-component global target analysis model applied to the UV-vis probe transient absorption data for pump wavelengths from \SI{495}{\nano\m} to \SI{580}{\nano\m} (and for the 2-component global lifetime analysis on the data for $\lambda_\textrm{pump}=\SI{600}{\nano\m}$); see Section \ref{sec:GTA_SADS_OCPo}.

Global lifetime analysis of the \SI{485}{\nano\m} pump data shown in the bottom panels of Figure \ref{fig:STTA_OCPo} and Figure \ref{fig:STTA_OCPr} clearly show the appearance of S*, with a third component required to adequately fit the data. The results for OCPo are consistent (in the sense of the number of components  required for a reasonable fit) with the 4-component (or 5-component) global target analysis model applied to the UV-vis probe transient absorption data for \SI{400}{\nano\m} to \SI{495}{\nano\m} pump wavelengths; see Section \ref{sec:GTA_SADS_OCPo}.

We note that sample degradation (caused by the higher pump fluences in comparison with the UV-vis transient absorption data) likely affects the fitted time constants and DADS profiles. No further states with longer lifetimes were identified from these analyses of visible-probe transient absorption data, largely due to the limited delay time range chosen in the experiments (up to \SI{40}{\pico\s}).

\begin{figure}[h]
	\centering
	\includegraphics[scale=1.0]{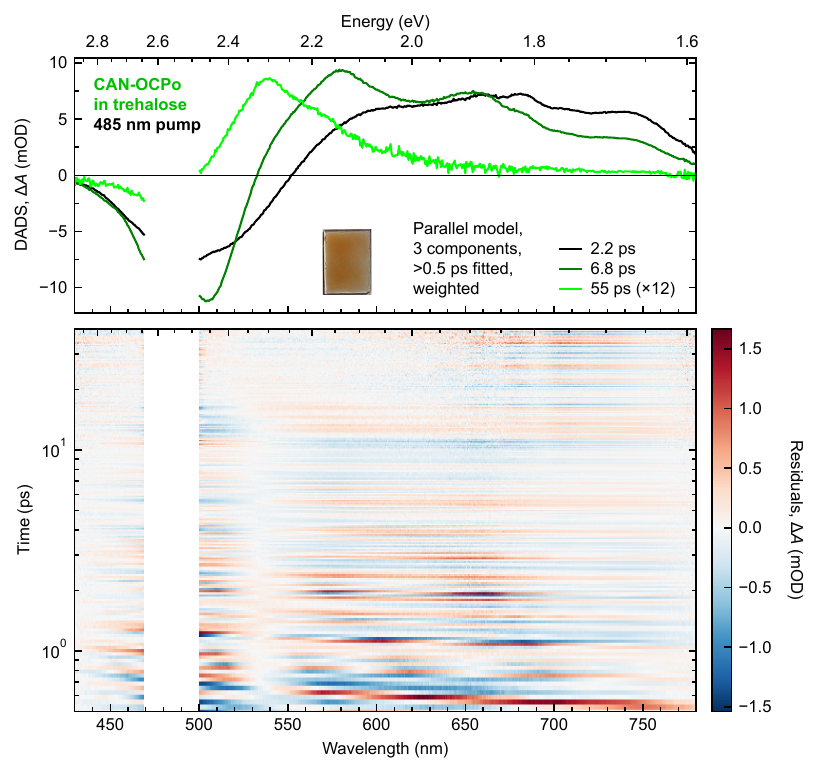}
	\caption{\textbf{Results of global lifetime analysis with a 3-component parallel model on transient absorption data of CAN-binding OCPo in trehalose with pump wavelength \SI{485}{\nano\m} and a visible probe: DADS (top) and residuals (bottom).} Only the wavelength range 430--\SI{780}{\nano\m} and times $>$\SI{0.5}{\pico\s} were fitted, and noisy data from 469.5--\SI{500}{\nano\m} due to significant pump scatter was excluded from the fit. DADS time constants are specified in the legend; multiplications refer to scalings applied to the DADS. $\textrm{Residuals} = \textrm{Data}-\textrm{Fit}$. See text for further details.}
	\label{fig:STTA_GLA_DADS_OCPo_485nm}
\end{figure}

\begin{figure}[h]
	\centering
	\includegraphics[scale=1.0]{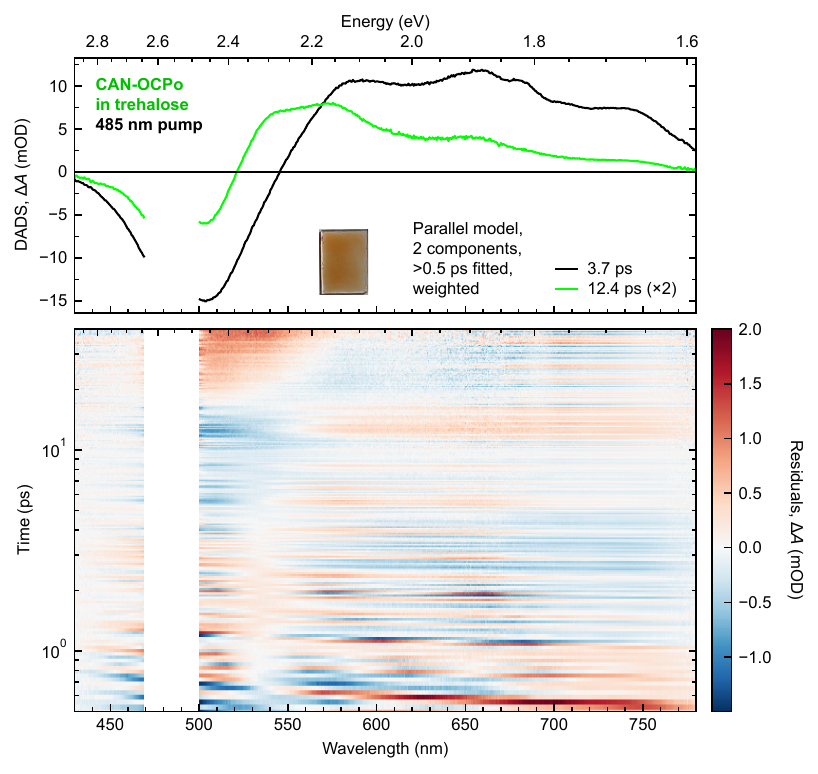}
	\caption{\textbf{Results of global lifetime analysis with a 2-component parallel model on transient absorption data of CAN-binding OCPo in trehalose with pump wavelength \SI{485}{\nano\m} and a visible probe: DADS (top) and residuals (bottom).} From the structure of the residuals, it is apparent that 2 components is not sufficient to fit this data. Only the wavelength range 430--\SI{780}{\nano\m} and times $>$\SI{0.5}{\pico\s} were fitted, and noisy data from 469.5--\SI{500}{\nano\m} due to significant pump scatter was excluded from the fit. DADS time constants are specified in the legend; multiplications refer to scalings applied to the DADS. $\textrm{Residuals} = \textrm{Data}-\textrm{Fit}$. See text for further details.}
	\label{fig:STTA_GLA_DADS_OCPo_485nm_2comp}
\end{figure}

\begin{figure}[h]
	\centering
	\includegraphics[scale=1.0]{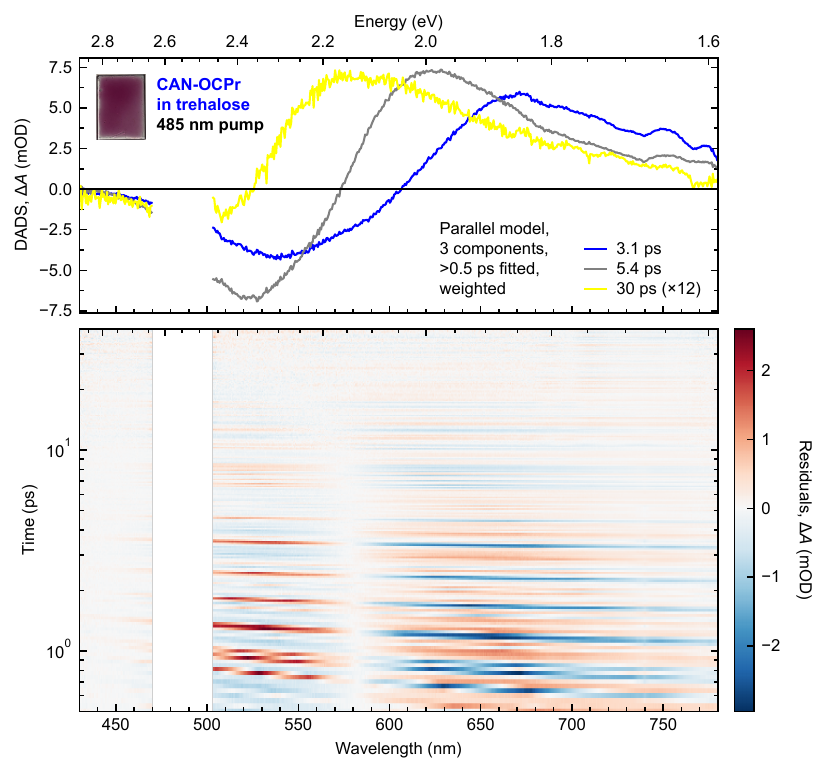}
	\caption{\textbf{Results of global lifetime analysis with a 3-component parallel model on transient absorption data of CAN-binding OCPr in trehalose with pump wavelength \SI{485}{\nano\m} and a visible probe: DADS (top) and residuals (bottom).} Only the wavelength range 430--\SI{780}{\nano\m} and times $>$\SI{0.5}{\pico\s} were fitted, and noisy data from 470--\SI{503}{\nano\m} due to significant pump scatter was excluded from the fit. DADS time constants are specified in the legend; multiplications refer to scalings applied to the DADS. $\textrm{Residuals} = \textrm{Data}-\textrm{Fit}$. See text for further details.}
	\label{fig:STTA_GLA_DADS_OCPr_485nm}
\end{figure}

\begin{figure}[h]
	\centering
	\includegraphics[scale=1.0]{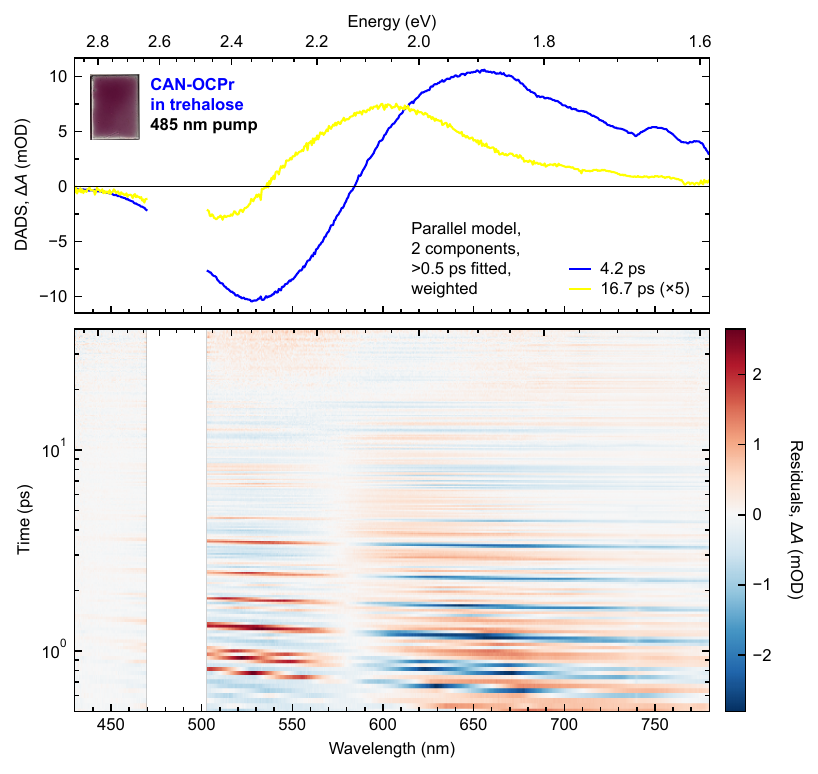}
	\caption{\textbf{Results of global lifetime analysis with a 2-component parallel model on transient absorption data of CAN-binding OCPr in trehalose with pump wavelength \SI{485}{\nano\m} and a visible probe: DADS (top) and residuals (bottom).} Only the wavelength range 430--\SI{780}{\nano\m} and times $>$\SI{0.5}{\pico\s} were fitted, and noisy data from 470--\SI{503}{\nano\m} due to significant pump scatter was excluded from the fit. DADS time constants are specified in the legend; multiplications refer to scalings applied to the DADS. $\textrm{Residuals} = \textrm{Data}-\textrm{Fit}$. See text for further details.}
	\label{fig:STTA_GLA_DADS_OCPr_485nm_2comp}
\end{figure}

\clearpage

\subsection{Supplementary visible \SI{}{\nano\s}--\SI{}{\milli\s} transient absorption materials}

\subsubsection{Transient absorption on OCPo and OCPr in trehalose}

\begin{figure}[h]
	\centering
	\includegraphics[scale=1.0]{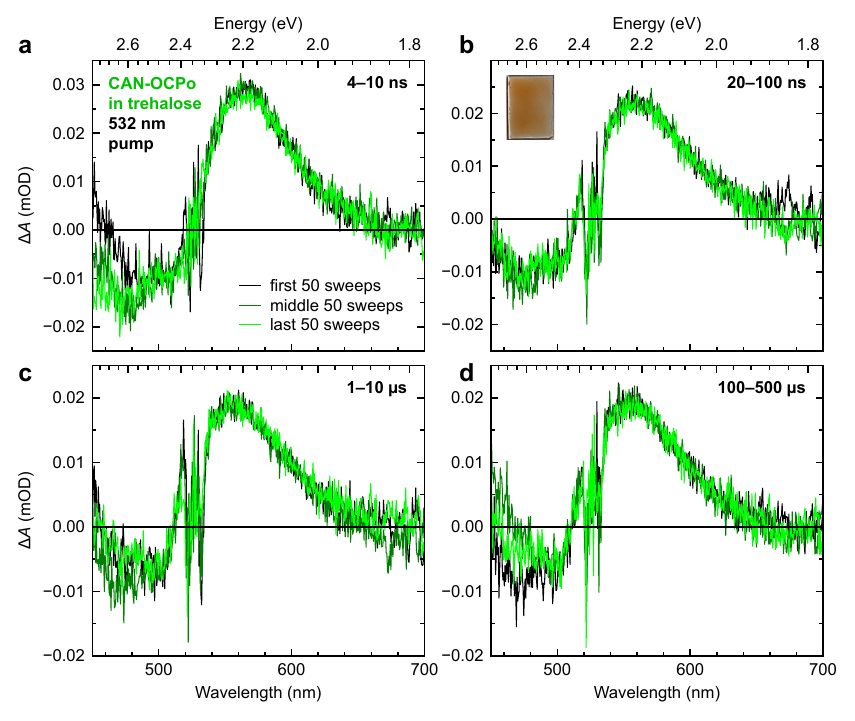}
	\caption{\textbf{Visible-probe \SI{}{\nano\s}--\SI{}{\milli\s} transient absorption spectra of CAN-binding OCPo in trehalose with pump wavelength \SI{532}{\nano\m}, with the first (black), middle (dark green), and last (light green) 50 sweeps (\textit{i.e.}~spectral replicates) averaged.} The similarity between the spectra demonstrate the lack of OCPo$\rightarrow$OCPr photoconversion in the trehalose films during pulsed-laser experiments. Variations for probe $<$\SI{500}{\nano\m} are attributable to weak, unstable white-light, while that for probe $\sim$\SI{532}{\nano\m} are due to weak probe light (by the \SI{532}{\nano\m} notch filter) and residual pump scatter. Contrast with main text Figure \ref{fig:STTA_LTTA_OCPo_r}c, where all 150 spectra of the 18-hour measurement are averaged. Panels (a--d) show different time delays (spectra are averaged between the times indicated). Pump fluence was set to \SI{200}{\micro\J\per\square\centi\m}.}
	\label{fig:LTTA_OCPo_50sweeps}
\end{figure}

\clearpage

\begin{figure}[h]
	\centering
	\includegraphics[scale=1.0]{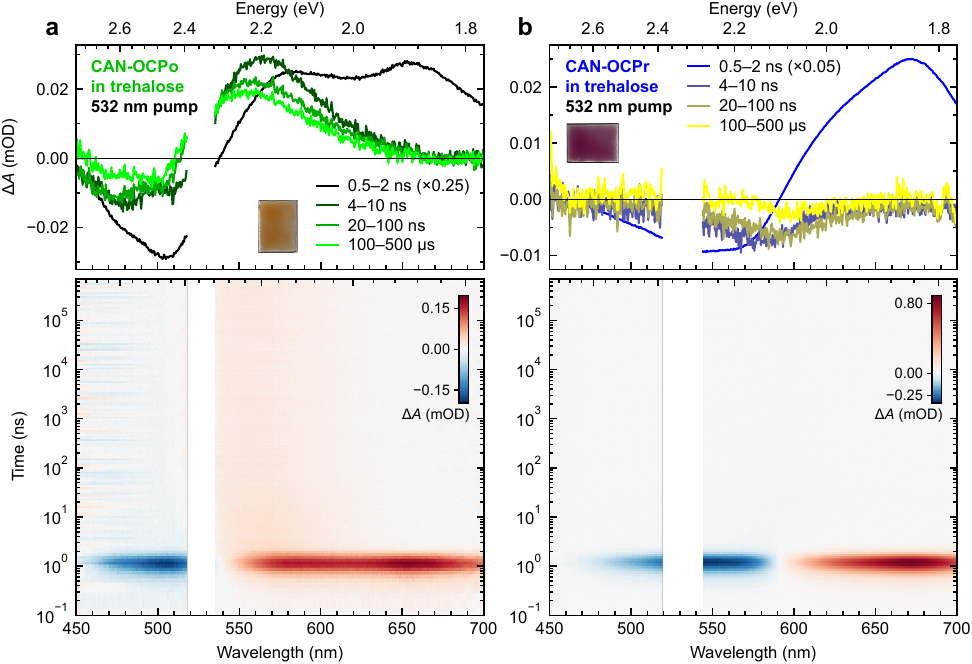}
	\caption{\textbf{Visible-probe \SI{}{\nano\s}--\SI{}{\milli\s} transient absorption spectra (top) and surfaces (bottom) of CAN-binding OCPo (a, left) and OCPr (b, right) in trehalose with pump wavelength \SI{532}{\nano\m}.} The dissimilarity between the OCPo/OCPr transient absorption data demonstrate that the lack of OCPo$\rightleftharpoons$OCPr conversion in the trehalose films during pulsed-laser experiments. Spectra have been averaged between the times indicated. Note that the 0.5--\SI{2}{\nano\s} spectra represent the IRF-limited singlet photophysics, and note the scaling factors. Noisy data from 518--\SI{535}{\nano\m} (a) and 519--\SI{544}{\nano\m} (b) due to low probe light (by \SI{532}{\nano\m} notch filter) and significant pump scatter have been excluded for clarity. Pump fluence was set to \SI{200}{\micro\J\per\square\centi\m} for both measurements.}
	\label{fig:LTTA_OCPo_r_tr}
\end{figure}

\clearpage

\subsubsection{Transient absorption dynamics of OCPo in a flow cell}

\begin{figure}[h]
	\centering
	\includegraphics[scale=1.0]{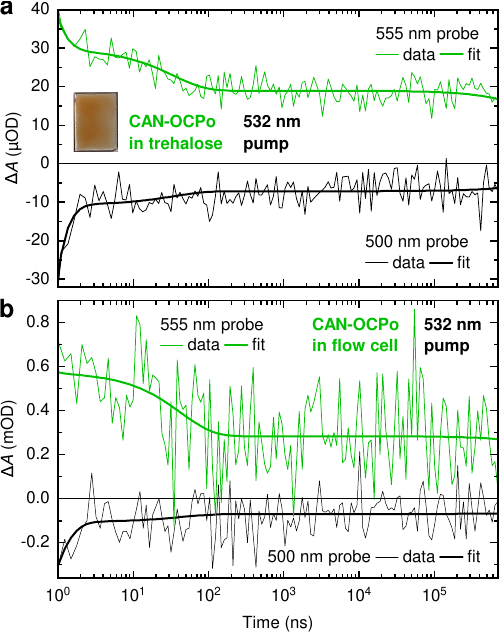}
	\caption{\textbf{Visible-probe \SI{}{\nano\s}--\SI{}{\milli\s} transient absorption dynamics of CAN-binding OCPo in trehalose (a) and in buffer (b) with pump wavelength \SI{532}{\nano\m}.} Data (thin lines) are a single probe wavelength (\textit{i.e.}~not averaged); see Figures \ref{fig:STTA_LTTA_OCPo_r}c (trehalose) and \ref{fig:STTA_LTTA_OCPo_r}d (buffer) for spectra. Fit (thick lines) are a single wavelength from the 4-component parallel global lifetime analysis on the data; see Figures \ref{fig:LTTA_GLA_EADS_OCPo_trehalose_4comp} (trehalose) and \ref{fig:LTTA_GLA_EADS_OCPo_buffer_4comp} (buffer). Note the different $y$-axis scales between each panel.}
	\label{fig:LTTA_GLA_OCPo_kinetics}
\end{figure}

Nanosecond-millisecond transient absorption dynamics of CAN-OCPo in trehalose at probe wavelengths \SI{500}{\nano\m} and \SI{555}{\nano\m} are shown in Figure \ref{fig:LTTA_GLA_OCPo_kinetics}a as thin lines. This is from the same dataset as that for the spectra of Figure \ref{fig:STTA_LTTA_OCPo_r}c. Similar transient absorption dynamics of CAN-OCPo in buffer, refreshed with a flow cell, are shown in Figure \ref{fig:LTTA_GLA_OCPo_kinetics}b as thin lines. Spectra from the same dataset are shown in Figure \ref{fig:STTA_LTTA_OCPo_r}d. As discussed in the main text, this transient absorption data in buffer shows no obvious differences to CAN-OCPo measured in trehalose or to 3$'$hECN-OCP reported elsewhere.\cite{Konold2019a}

\clearpage

\subsubsection{Global lifetime analysis}

\begin{table}[h!]
	\renewcommand*{\arraystretch}{1.2} 
	\newcolumntype{C}[1]{>{\centering\arraybackslash}m{#1}} 
	\begin{center}
		\small 
		\begin{tabular}{|C{52mm}|C{52mm}|}
		    \hline
		    \textbf{5-component} & \textbf{4-component} \\
		    \textbf{sequential model} & \textbf{sequential model} \\ \hline
			$t_1=\SI{100}{\pico\s}$ & $t_1=\SI{100}{\pico\s}$ \\
			$t_2=\SI{1}{\nano\s}$ & $t_2=\SI{1}{\nano\s}$ \\  \hline
			\rowcolor{green!15!white} 
			$t_3=\SI{50}{\nano\s}$ & $t_3=\SI{50}{\nano\s}$ \\
			\rowcolor{green!15!white} 
			$t_4=\SI{10}{\micro\s}$ & --- \\
			\rowcolor{green!15!white} 
			$t_5=\infty$ & $t_4=\infty$ \\ \hline
			$\textrm{IRF centre}=\SI{1}{\nano\s}$ (trehalose) or \SI{0.65}{\nano\s} (buffer) & $\textrm{IRF centre}=\SI{1}{\nano\s}$ (trehalose) or \SI{0.65}{\nano\s} (buffer)\\
			$\textrm{IRF FWHM}=\SI{0.35}{\nano\s}$ & $\textrm{IRF FWHM}=\SI{0.35}{\nano\s}$ \\ \hline
		\end{tabular}
	\end{center}
	\vspace{-6pt}
	\caption{\textbf{Summary of initial parametrisations used in the global lifetime analysis of the visible \SI{}{\nano\s}--\SI{}{\milli\s} transient absorption data.} The time constants $t_3$, $t_4$, and $t_5$ (light green) were chosen to match the $\geq$\SI{50}{\nano\s} EADS time constants determined by Konold \textit{et al.}~when using a higher pump power.\cite{Konold2019a} The 4-component sequential model omits the \SI{10}{\micro\s} component. Note that a different initial IRF centre is used for the trehalose and buffer data. As sequential models are being used, the decay rate constants $k_i$ are the inverse of the corresponding time constants $t_i$ ($k_i=t_i^{-1}$).}
	\label{tab:LTTA_GLA_initial}
\end{table}

Global lifetime analysis was also performed on a selection of the \SI{}{\nano\s}--\SI{}{\milli\s} transient absorption data, using similar sequential models Konold \textit{et al.}~used for transient absorption data on 3$'$hECN-OCP with a \SI{475}{\nano\m} pump\cite{Konold2019a}, to check if similar $>$\SI{1}{\nano\s} photoproducts described in their work are being formed in our samples, even when OCPo is within trehalose. This was also done using the Glotaran 1.5.1 software package (http://glotaran.org),\cite{Snellenburg2012} a GUI for the R package TIMP\cite{Mullen2007}. Data had already been processed with the steps outlined in the methods (Section \ref{sec:methods_TA}). Noisy regions in the data due to pump scatter were set to zero $\Delta A$ for all times to ensure a good fit of the rest of the data. Noisy red and blue ends in the data associated with tails of the probe were also excluded. A number of models with different numbers of components were applied to the data, with the resulting evolution-associated difference spectra (EADS) corresponding to different photoproducts that exponentially decay into one another in sequence. A summary of the initial choices for time constants (\textit{i.e.}~decay rate constants) and the instrument response function (IRF) centre and FWHM in the applied models is shown in Table \ref{tab:LTTA_GLA_initial}.

\begin{figure}[h]
	\centering
	\includegraphics[scale=1.0]{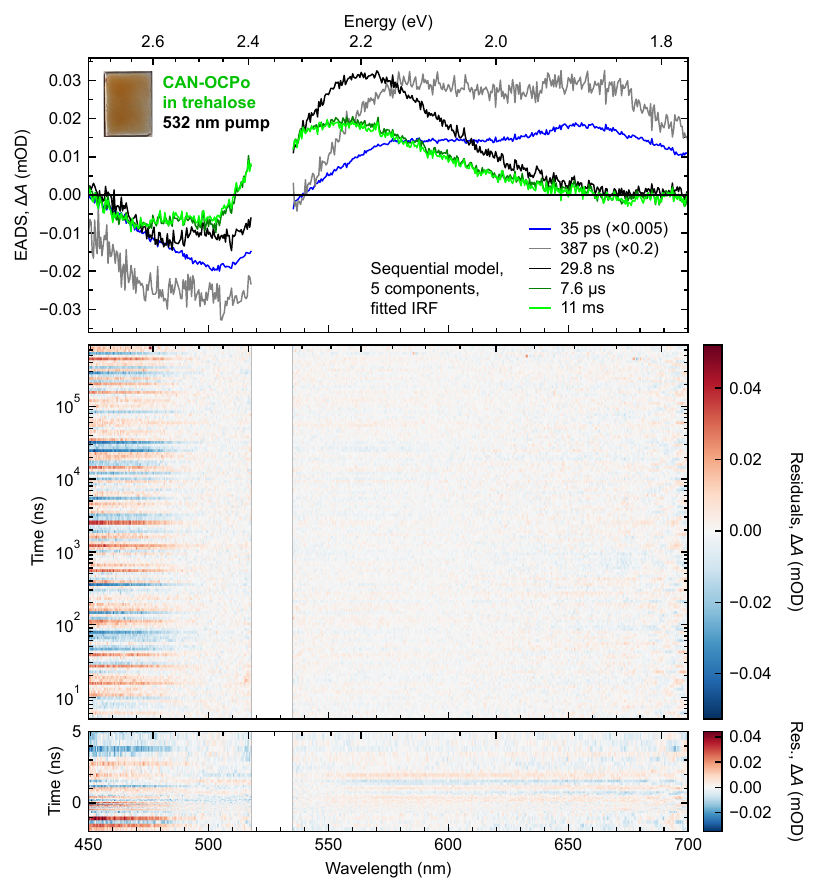}
	\caption{\textbf{Results of global lifetime analysis with a 5-component sequential model on transient absorption data of CAN-binding OCPo in trehalose with pump wavelength \SI{532}{\nano\m} and a visible probe: EADS (top) and residuals (middle, bottom).} Only the wavelength range 450--\SI{700}{\nano\m} was fitted, and noisy data from 518--\SI{535}{\nano\m} due to significant pump scatter was excluded from the fit. EADS time constants are specified in the legend; multiplications refer to scalings applied to the EADS. The fitted IRF has centre \SI{1.09}{\nano\s} and FWHM \SI{0.36}{\nano\s}. $\textrm{Residuals} = \textrm{Data}-\textrm{Fit}$; note the logarithmic time-scale in the middle panel, and the linear time-scale and different residuals scale in the bottom panel. See text for further details.}
	\label{fig:LTTA_GLA_EADS_OCPo_trehalose_5comp}
\end{figure}

\begin{figure}[h]
	\centering
	\includegraphics[scale=1.0]{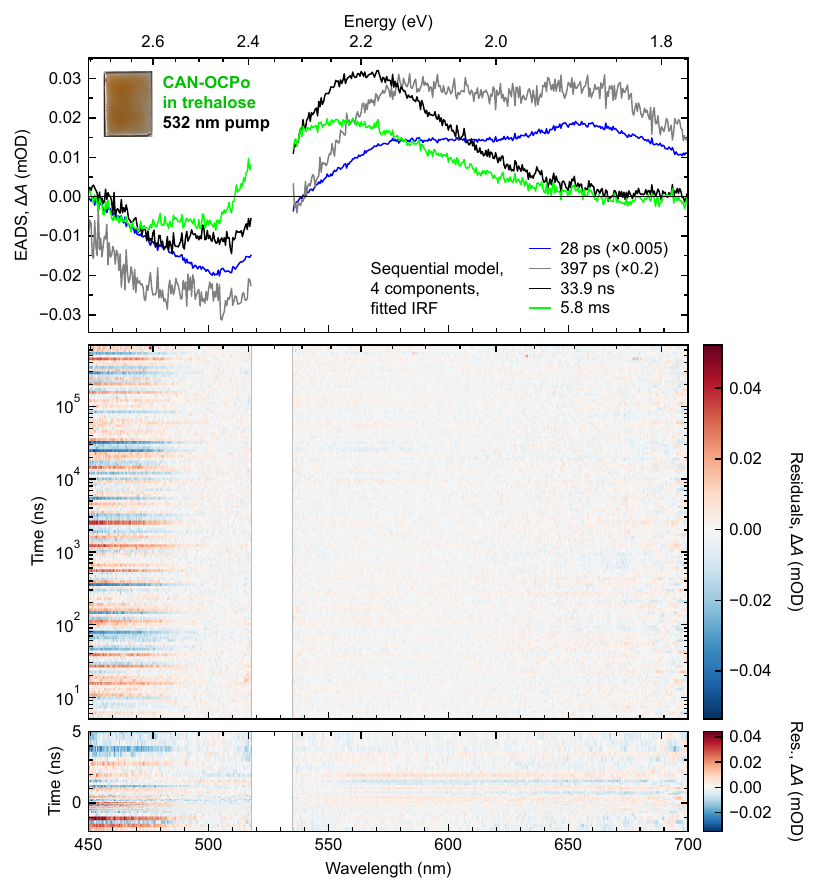}
	\caption{\textbf{Results of global lifetime analysis with a 4-component sequential model on transient absorption data of CAN-binding OCPo in trehalose with pump wavelength \SI{532}{\nano\m} and a visible probe: EADS (top) and residuals (middle, bottom).} Only the wavelength range 450--\SI{700}{\nano\m} was fitted, and noisy data from 518--\SI{535}{\nano\m} due to significant pump scatter was excluded from the fit. EADS time constants are specified in the legend; multiplications refer to scalings applied to the EADS. The fitted IRF has centre \SI{1.09}{\nano\s} and FWHM \SI{0.36}{\nano\s}. $\textrm{Residuals} = \textrm{Data}-\textrm{Fit}$; note the logarithmic time-scale in the middle panel, and the linear time-scale and different residuals scale in the bottom panel. See text for further details.}
	\label{fig:LTTA_GLA_EADS_OCPo_trehalose_4comp}
\end{figure}

\begin{figure}[h]
	\centering
	\includegraphics[scale=1.0]{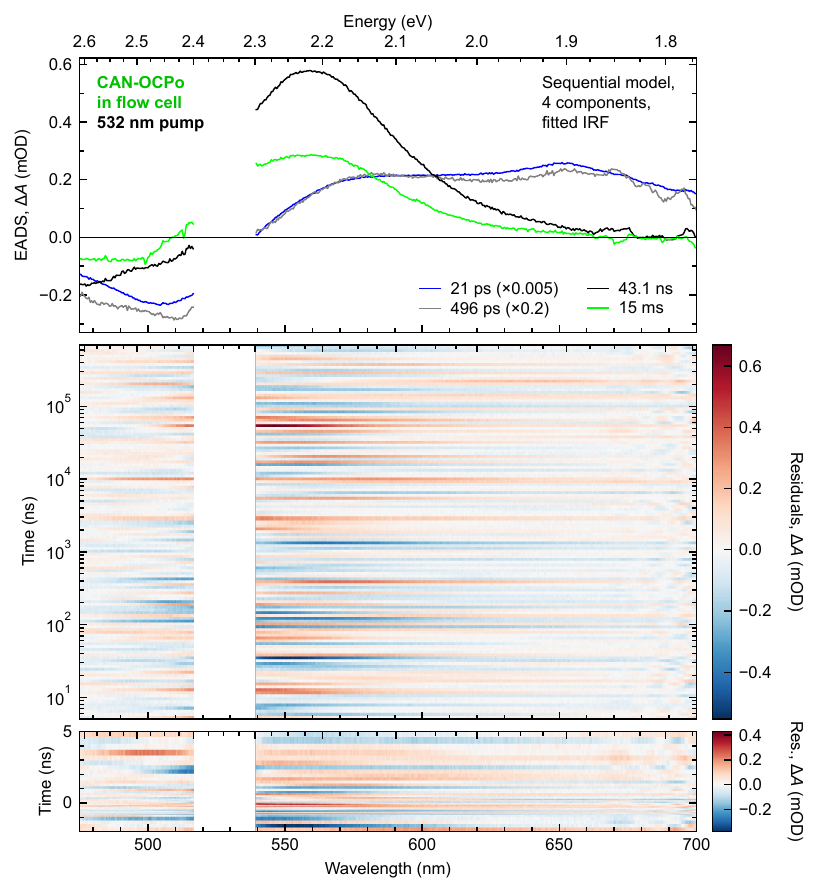}
	\caption{\textbf{Results of global lifetime analysis with a 4-component sequential model on transient absorption data of CAN-binding OCPo in buffer with pump wavelength \SI{532}{\nano\m} and a visible probe: EADS (top) and residuals (middle, bottom).} Only the wavelength range 475--\SI{700}{\nano\m} was fitted, and noisy data from 517--\SI{539}{\nano\m} due to significant pump scatter was excluded from the fit. EADS time constants are specified in the legend; multiplications refer to scalings applied to the EADS. The fitted IRF has centre \SI{0.69}{\nano\s} and FWHM \SI{0.29}{\nano\s}. $\textrm{Residuals} = \textrm{Data}-\textrm{Fit}$; note the logarithmic time-scale in the middle panel, and the linear time-scale and different residuals scale in the bottom panel. See text for further details.}
	\label{fig:LTTA_GLA_EADS_OCPo_buffer_4comp}
\end{figure}

A 5-component sequential fit to the CAN-OCPo in trehalose transient absorption data gives latter components very similar to those found by Konold \textit{et al.}, both in spectral profile and associated time constant. Figure \ref{fig:LTTA_GLA_EADS_OCPo_trehalose_5comp} shows the fit, with EADS in the top panel and time constants specified in the legend. The first two EADS do not resemble those seen in Konold \textit{et al.}, in terms of spectral profile and especially in associated time constants \cite{Konold2019a}. This is likely due to the instrument response function in our \SI{}{\nano\s}--\SI{}{\milli\s} transient absorption setup being $\sim$\SI{1}{\nano\s}, which is greater than the first three EADS time constants from a 6-component sequential fit in Konold \textit{et al.} (\SI{1.4}{\pico\s}, \SI{4.5}{\pico\s}, \SI{24}{\pico\s}), and greater than our own fitted EADS time constants (\SI{30.32}{\pico\s}, \SI{423.5}{\pico\s}) and Gaussian IRF ($\textrm{FWHM}=\SI{350.2}{\pico\s}$). Our own attempts of a 6-component sequential fit result in degenerate EADS. However, the subsequent EADS (\SI{29.8}{\nano\s}, \SI{7.6}{\micro\s}, \SI{11}{\milli\s} closely match those of Konold \textit{et al}. (\SI{50}{\nano\s}, \SI{50}{\micro\s}, infinite when using a higher pump power, and \SI{50}{\nano\s}, \SI{25}{\micro\s}, infinite when using a lower pump power), further demonstrating that we are seeing the same photoproducts as 3$'$hECN-OCP in solution, despite our CAN-OCPo being trapped in trehalose. We note that the profile of the two EADS associated with \SI{7.6}{\micro\s} and \SI{11}{\milli\s} match very closely, with a smaller amplitude decrease in the excited-state absorption (ESA) and especially in the ground-state bleach (GSB) compared to Konold \textit{et al.}'s comparable EADS. We therefore print the results of a 4-component sequential fit on the same data in Figure \ref{fig:LTTA_GLA_EADS_OCPo_buffer_4comp}; it is apparent from the structure of the residuals that omitting a $\sim$\SI{10}{\micro\s} component does not affect the fit quality. The 4-component global fit is displayed as dynamics in Figure \ref{fig:LTTA_GLA_OCPo_kinetics}a as bold lines (data are thin lines).

\clearpage

\begin{figure}[h]
	\centering
	\includegraphics[scale=1.0]{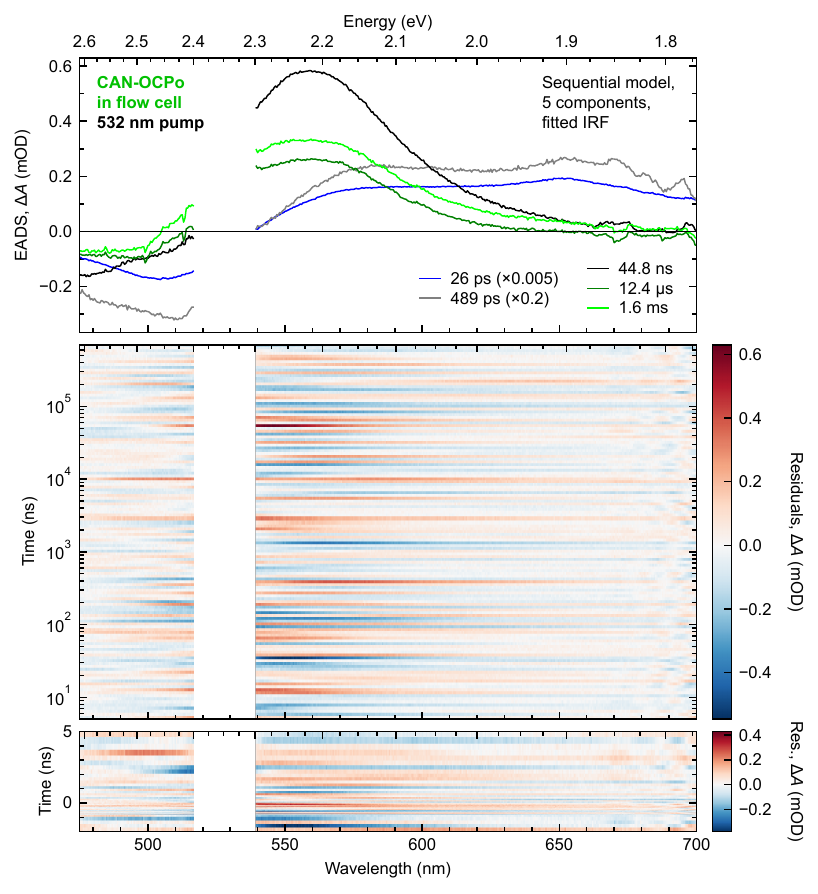}
	\caption{\textbf{Results of global lifetime analysis with a 5-component sequential model on transient absorption data of CAN-binding OCPo in buffer with pump wavelength \SI{532}{\nano\m} and a visible probe: EADS (top) and residuals (middle, bottom).} Only the wavelength range 475--\SI{700}{\nano\m} was fitted, and noisy data from 517--\SI{539}{\nano\m} due to significant pump scatter was excluded from the fit. EADS time constants are specified in the legend; multiplications refer to scalings applied to the EADS. The fitted IRF has centre \SI{0.67}{\nano\s} and FWHM \SI{0.28}{\nano\s}. $\textrm{Residuals} = \textrm{Data}-\textrm{Fit}$; note the logarithmic time-scale in the middle panel, and the linear time-scale and different residuals scale in the bottom panel. See text for further details.}
	\label{fig:LTTA_GLA_EADS_OCPo_buffer_5comp}
\end{figure}

As a final check, we undertake the same 5-component and 4-component sequential global lifetime analysis on our transient absorption data of CAN-OCPo in buffer in a flow cell, and the results are indeed consistent with both Konold \textit{et al.}'s and our own in trehalose. The 5-component model results are shown in Figure \ref{fig:LTTA_GLA_EADS_OCPo_buffer_5comp}. Here, the EADS associated with \SI{1.6}{\milli\s} appears to shift to more positive $\Delta A$ for all wavelengths, compared to the EADS for \SI{12.4}{\micro\s}. This is inconsistent with both our in-trehalose results and Konold \textit{et al.}'s results.\cite{Konold2019a} The 4-component model (dynamics in Figure \ref{fig:LTTA_GLA_OCPo_kinetics}b and EADS in Figure \ref{fig:LTTA_GLA_EADS_OCPo_buffer_4comp}) where the $\sim$\SI{10}{\micro\s} EADS is essentially omitted appears to fit the data as well as the 5-component model.

We emphasise that in these global lifetime analyses of the \SI{}{\nano\s}--\SI{}{\milli\s} transient absorption data, our first two EADS (time constants $<$\SI{1}{\nano\s}) have little physical significance. We also note that the maximum time delay in these experiments was \SI{0.7}{\milli\s}, and that the final EADS all have a fitted time constant exceeding that.

\clearpage

\subsection{Image showing no conversion in encapsulated CAN-OCPo}

\begin{figure}[h]
	\centering
	\includegraphics[width=0.82\linewidth]{"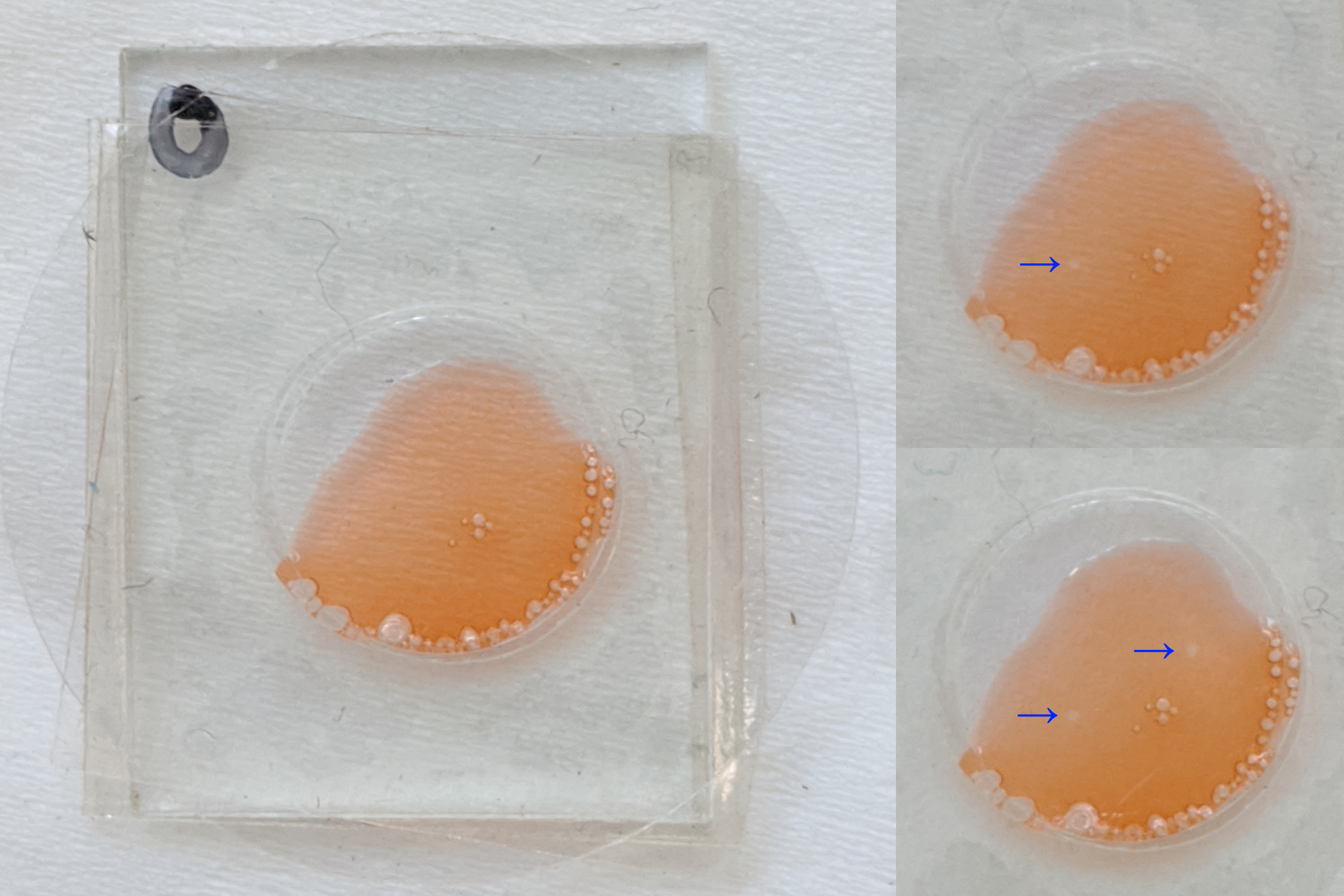"}
	\caption{\textbf{CAN-OCPo sample in trehalose-sucrose glass with cover slip encapsulation.} Images to the right are the same sample after a first (top-right) and second (bottom-right) high pump-fluence picosecond transient absorption measurement; note the bleach spots indicated with blue arrows. No conversion to OCPr is apparent. In the $\sim$\SI{30}{\minute} measurements, the film was pumped with \SI{532}{\nano\m}, \SI{3.6}{\milli\watt}, \SI{5}{\kilo\hertz} pump pulses, with \SI{465}{\micro\metre} and \SI{340}{\micro\m} the major and minor diameters ($1/e^2$). Therefore the pump fluence was \SI{580}{\micro\J\per\square\centi\m}. The majority of picosecond transient absorption measurements used significantly lower fluences and did not induce significant bleaching.}
	\label{fig:degredation_CAN-OCPo_trehalose}
\end{figure}

\clearpage

\subsection{Supplementary UV-vis \SI{}{\pico\s} transient absorption materials}

\subsubsection{Pump spectra and comments on `bands'}

For our pump wavelength-dependent \SI{}{\pico\s} transient absorption measurements on CAN-OCPo in trehalose, we obtain \SI{420}{\nano\m} to \SI{600}{\nano\m} pump pulses with an optical parametric amplifier (TOPAS Prime, Light Conversion) and subsequent frequency mixer (NirUVis, Light Conversion). Certain pump wavelength `bands' require a different TOPAS Prime and NirUVis configuration. For \SI{420}{\nano\m} to \SI{480}{\nano\m} pump pulses, a configuration we dub band (A) is used (specifically, fourth harmonic generation by twice-doubling the idler); for 480--\SI{535}{\nano\m}, band (B) is used (sum frequency generation when mixing \SI{800}{\nano\m} pump and idler); for 535--\SI{600}{\nano\m}, band (C) is used (sum frequency generation when mixing \SI{800}{\nano\m} pump and signal). Measurements using each band were completed on separate experimental days (\textit{i.e.}~one day for band (B), one later day for band (A), and so on). We note that for \SI{400}{\nano\m} pump pulses, a frequency doubler utilising \textbeta-barium borate crystals is used instead, and is denoted (TP) (see methods Section \ref{sec:methods_TA} for further experimental details).

\begin{figure}[h!]
	\centering
	\includegraphics[scale=1.0]{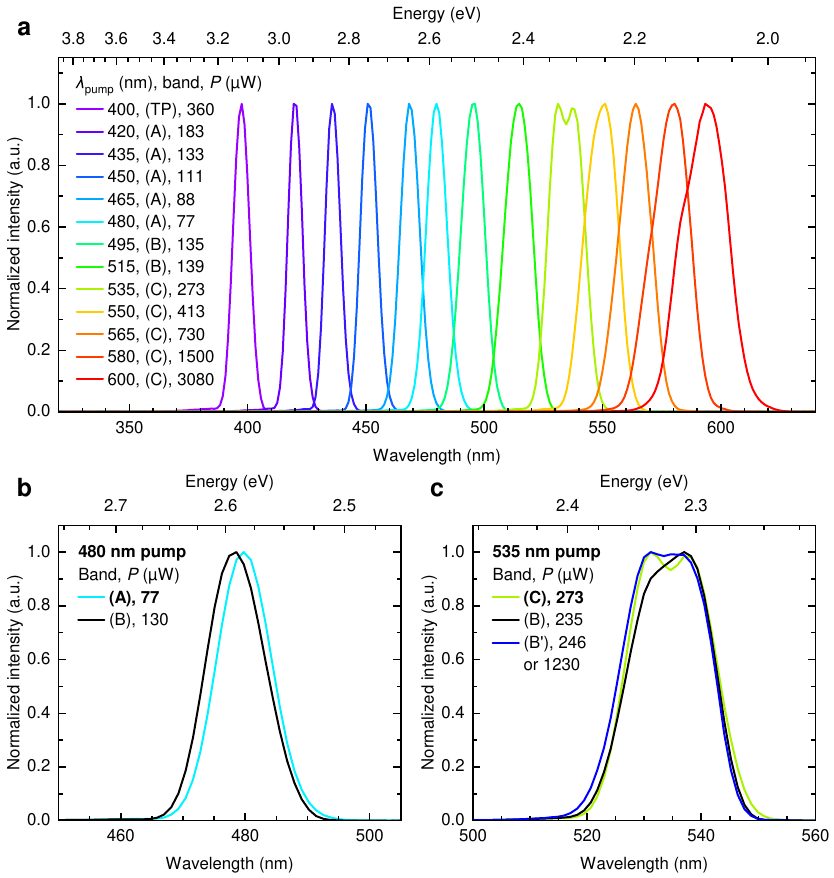}
	\caption{\textbf{Normalised intensity spectra of the pumps used in the UV-vis-probe \SI{}{\pico\s} transient absorption experiments, normalised to the maximum intensity.} Pump wavelengths (as set in WinTopas4), the TimePlate usage (here dubbed (TP)) or optical parametric amplifier and mixer configuration for a band (here dubbed (A), (B), (C)), and the measured powers (of \SI{5}{\kilo\Hz} pulses measured per-second) are specified in the legend. \textbf{(a)} Intensity spectra of the pumps used in the experiments shown in the main text (Figures \ref{fig:STTA_OCPo_ex_dep}, \ref{fig:STTA_OCPo_kinetics_superfigure}, and \ref{fig:OCP_fitted_kinetics_comparison}a). \textbf{(b)} Intensity spectra of \SI{480}{\nano\m} pumps in band (A) (main text data) and band (B) (SI data; see Figure \ref{fig:STTA_480nm_comp}). \textbf{(c)} Intensity spectra of \SI{535}{\nano\m} pumps in band (C) (main text data), band (B) (SI data; see Figure \ref{fig:STTA_535nm_comp}), and band (B') (SI data; see Figure \ref{fig:STTA_535nm_fluence_dep}). 
	The pump spot sizes at the sample position were generally not measured to save time, but are known to vary per-day and per-pump wavelength, with previous diameter ($1/e^2$) measurements typically in the range 400--\SI{800}{\micro\m}.}
	\label{fig:STTA_ex_wav_dep_pump_scatter}
\end{figure}

Figure \ref{fig:STTA_ex_wav_dep_pump_scatter}a shows normalised intensity spectra for the pump wavelengths used for transient absorption measurements depicted in Figures \ref{fig:STTA_OCPo_ex_dep}, \ref{fig:STTA_OCPo_kinetics_superfigure}, and \ref{fig:OCP_fitted_kinetics_comparison}a. The legend shows the nominal pump wavelength (as set in WinTopas4), the band used in generating that wavelength, and the measured pump powers (of the \SI{5}{\kilo\hertz} pulses measured per-second). Note that band (A) is used in generating the $\lambda_\textrm{pump}=\SI{480}{\nano\m}$ used for experiments in the main text, and that band (B) is used in generating the main text $\lambda_\textrm{pump}=\SI{535}{\nano\m}$.

\clearpage

\subsubsection{Replicates at same pump wavelengths in different bands}
\label{sec:UV-ps-TA_replicates_same_wavelength}

\begin{figure}[h!]
	\centering
	\includegraphics[scale=1.0]{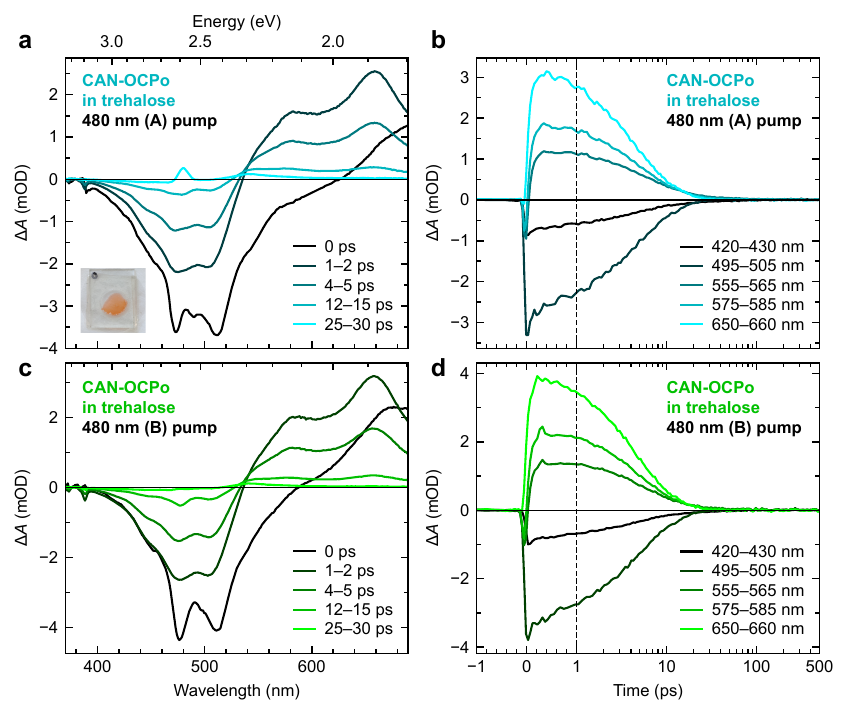}
	\caption{\textbf{Transient absorption spectra (a,c) and dynamics (b,d) on CAN-binding OCPo in trehalose using \SI{480}{\nano\m} pump wavelength in band (A) (a,b) and band (B) (c,d).} Both pumps give near-identical features, with differences attributable to small variations between the pump spectra (see Figure \ref{fig:STTA_ex_wav_dep_pump_scatter}b). We note that a different film spot was used between band (A) and band (B), and each band's measurements were done on separate experimental days. The dynamic plots (b,d) have a linear time-axis up to \SI{1}{\pico\s}, and subsequently logarithmic up to \SI{500}{\pico\s}.}
	\label{fig:STTA_480nm_comp}
\end{figure}

\begin{figure}[h!]
	\centering
	\includegraphics[scale=1.0]{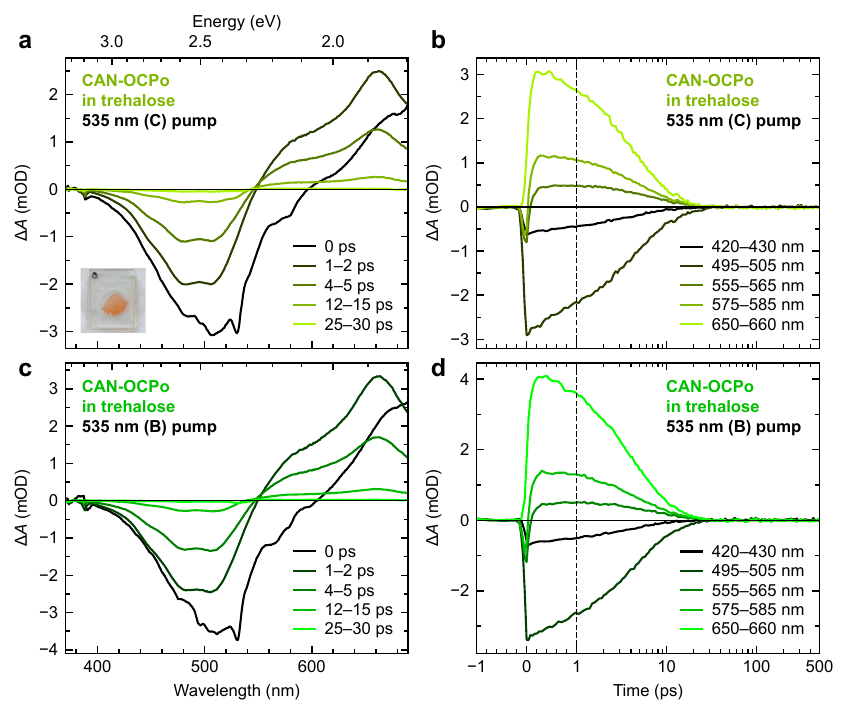}
	\caption{\textbf{Transient absorption spectra (a,c) and dynamics (b,d) on CAN-binding OCPo in trehalose using \SI{535}{\nano\m} pump wavelength in band (C) (a,b) and band (B) (c,d).} Both pumps give near-identical features, with differences attributable to small variations between the pump spectra (see Figure \ref{fig:STTA_ex_wav_dep_pump_scatter}c). We note that a different film spot was used between band (C) and band (B), and each band's measurements were done on separate experimental days. The dynamic plots (b,d) have a linear time-axis up to \SI{1}{\pico\s}, and subsequently logarithmic up to \SI{500}{\pico\s}.}
	\label{fig:STTA_535nm_comp}
\end{figure}

The use of different TOPAS Prime configurations does not have any effect on the transient absorption of CAN-OCPo in trehalose. We undertook control measurements using nominally (as set in WinTopas4) \SI{480}{\nano\m} and \SI{535}{\nano\m} pump wavelengths generated using band (B). For $\lambda_\textrm{pump}=\SI{480}{\nano\m}$, a comparison of the normalised intensity spectra between band (A) (main text) and band (B) (control) is shown in Figure \ref{fig:STTA_ex_wav_dep_pump_scatter}b, and a comparison of transient absorption spectra and dynamics is shown in Figure \ref{fig:STTA_480nm_comp}. The results are near-identical. For $\lambda_\textrm{pump}=\SI{535}{\nano\m}$, a comparison of the normalised intensity spectra between band (C) (main text) and band (B) (control) is shown in Figure \ref{fig:STTA_ex_wav_dep_pump_scatter}c, and a comparison of transient absorption spectra and dynamics is shown in Figure \ref{fig:STTA_535nm_comp}. Again, the results are near-identical. We note that a different CAN-OCPo film spot was used for each measurement, and (as stated above) measurements using each band were done on separate experimental days.

\clearpage

\subsubsection{Pump fluence independence}

\begin{figure}[h!]
	\centering
	\includegraphics[scale=1.0]{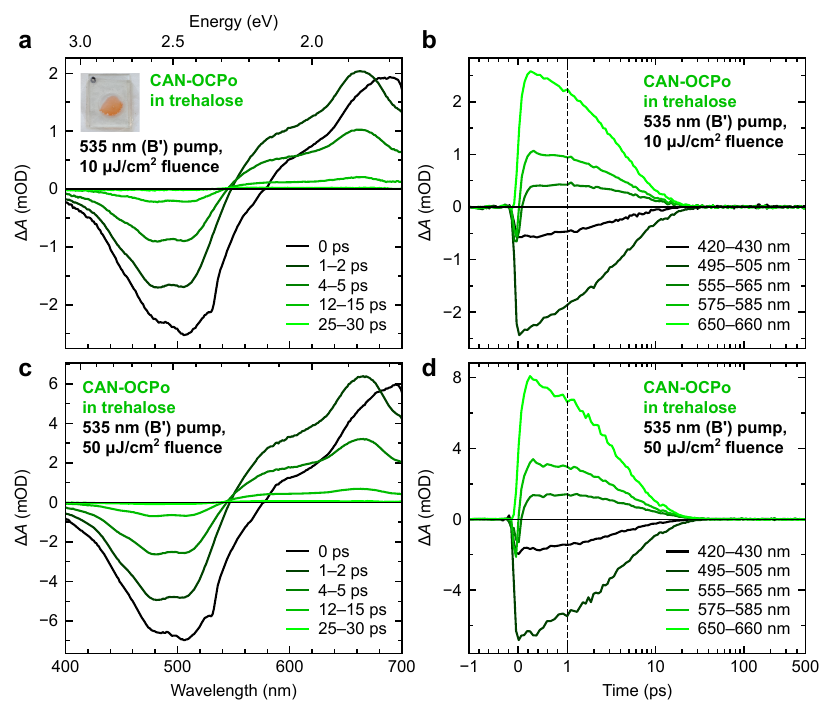}
	\caption{\textbf{Pump fluence dependence on the transient absorption spectra (a,c) and dynamics (b,d) of the same spot on CAN-binding OCPo in trehalose using \SI{535}{\nano\m} pump wavelength in band (B').} Both the \SI{10}{\micro\joule\per\square\centi\m} pump fluence (a,b) and the subsequent \SI{50}{\micro\joule\per\square\centi\m} fluence (c,d) gave near-identical spectra (a,c) and dynamics (b,d) on the same film spot, with differences attributable to sample degradation. The dynamic plots (b,d) have a linear time-axis up to \SI{1}{\pico\s}, and subsequently logarithmic up to \SI{500}{\pico\s}.}
	\label{fig:STTA_535nm_fluence_dep}
\end{figure}

A brief check shows that the pump fluence has no effect on the CAN-OCPo in trehalose transient absorption beyond increased degradation, although only a few pump fluences were used at $\lambda_{\rm{pump}}=\SI{535}{\nano\m}$ only. Figure \ref{fig:STTA_535nm_fluence_dep} shows the transient absorption spectra (panels a,c) and dynamics (b,d) using a \SI{10}{\micro\joule\per\square\centi\m} pump fluence (a,b) and then a \SI{50}{\micro\joule\per\square\centi\m} pump fluence (c,d) at the same spot on the film. Pump powers for those fluences were tuned following a measurement of the pump diameters ($1/e^2$), which were found to be \SI{934.1}{\micro\m} (major axis) and \SI{671.1}{\micro\m} (minor axis). Therefore, \SI{273}{\micro\watt} (a,b) and \SI{1230}{\micro\watt} (c,d) pump powers gave the respective pump fluences. See methods Section \ref{sec:methods_pump_fluence} for further details. A subsequent measurement using \SI{200}{\micro\joule\per\square\centi\m} pump fluence (\SI{4920}{\micro\watt} pump power) led to near-complete sample degradation (not shown). We note that band (B') was the same TOPAS Prime configuration as band (B), but the former was a different experimental day, with additional post-sample filters compared to the other UV-vis-probe measurements, along with slightly differing pump intensity spectra (see Figure \ref{fig:STTA_ex_wav_dep_pump_scatter}c).

\clearpage

\subsubsection{Global fit of dynamics parameters}

\begin{table}[ht]
	\renewcommand*{\arraystretch}{1.2} 
	\newcolumntype{C}[1]{>{\centering\arraybackslash}m{#1}} 
	\begin{center}
		\small 
		\begin{tabular}{|C{24mm}|C{26mm}|C{26mm}|C{26mm}|}
			\hline
			\boldmath{$\lambda$\textsubscript{\textbf{pump}}} \textbf{(\SI{}{\nano\m})} &  \boldmath{$\alpha_1$} & \boldmath{$\alpha_2$} & \boldmath{$\alpha_3$} \\
			\hline
			\textbf{400} & $0.243\pm0.047$ & $0.315\pm0.043$ & $0.500\pm0.010$ \\ \hline
			\textbf{420} & $0.167\pm0.054$ & $0.485\pm0.047$ & $0.430\pm0.010$ \\ \hline
			\textbf{435} & $0.238\pm0.055$ & $0.475\pm0.050$ & $0.350\pm0.009$ \\ \hline
			\textbf{450} & $0.271\pm0.061$ & $0.540\pm0.055$ & $0.267\pm0.009$ \\ \hline
			\textbf{465} & $0.333\pm0.070$ & $0.615\pm0.063$ & $0.161\pm0.008$ \\ \hline
			\textbf{480} & $0.355\pm0.072$ & $0.637\pm0.066$ & $0.090\pm0.008$ \\ \hline
			\textbf{495} & $0.512\pm0.076$ & $0.598\pm0.073$ & $0.012\pm0.007$ \\ \hline
			\textbf{515} & $0.533\pm0.077$ & $0.602\pm0.074$ & $0.006\pm0.007$ \\ \hline
			\textbf{535} & $0.591\pm0.073$ & $0.509\pm0.071$ & $0.011\pm0.007$ \\ \hline
			\textbf{550} & $0.735\pm0.074$ & $0.452\pm0.076$ & $0.000\pm0.007$ \\ \hline
			\textbf{565} & $0.783\pm0.070$ & $0.361\pm0.073$ & $0.000\pm0.007$ \\ \hline
			\textbf{580} & $0.850\pm0.068$ & $0.297\pm0.073$ & $0.004\pm0.007$ \\ \hline
			\textbf{600} & $0.909\pm0.064$ & $0.194\pm0.070$ & $0.015\pm0.007$ \\ \hline
		\end{tabular}
	\end{center}
	\vspace{-6pt}
	\caption{\textbf{Fit parameters found using a global triexponential fit to the data in Figure \ref{fig:STTA_OCPo_kinetics_superfigure}.} The triexponential equation used is $\Delta A_\textrm{fit}(t) = \alpha_1\exp(-t/\tau_1) + \alpha_2\exp(-t/\tau_2) + \alpha_3\exp(-t/\tau_3)$ with a lower bound $\alpha_i \geq 0$ applied. The fitted time constants are $\tau_1 = 2.8\pm\SI{0.2}{\pico\s}$, $\tau_2 = 6.5\pm\SI{0.4}{\pico\s}$, and $\tau_3 = 65\pm\SI{2}{\pico\s}$; note that as a global fit, these are the same for all pump wavelengths. Errors specified here are fit parameter standard errors.}
	\label{tab:STTA_OCP_fit_params}
\end{table}

\subsubsection{Global target analysis summary and discussion}
\label{sec:GTA_SADS_OCPo_summary}

It is usual, in analysis of OCP transient absorption data, to disentangle and quantify spectral contributions using analysis such as dynamic fitting with multi-exponential fits, or spectro-kinetic analysis such as global lifetime/target analysis \cite{Berera2012, Berera2013, Konold2019a, Slouf2017, Nizinski2022a, Khan2022}. However, due to the IVR and VC apparent within the S\textsubscript{1} lifetimes, these analyses are not completely suited for this system. In particular, the aforementioned multi-exponential fitting implicitly assumes well-defined species, and that first-order rate equations describing concentration changes between species may be written (their solutions being a sum of exponentials). Meanwhile, global lifetime/target analysis assumes that the spectra of the species are time-independent (\textit{i.e.}~well-defined) and that the kinetics (concentration traces) are probe wavelength-independent \cite{Ruckebusch2012, vanStokkum2004, Fernandez-Teran2022}, a condition called bilinearity. These assumptions do not hold due to the effect of IVR and VC\cite{Balevicius2015, Balevicius2016, Balevicius2018, Balevicius2019, Sebelik2022}.

\begin{figure}[h!]
	\centering
	\includegraphics[scale=1.0]{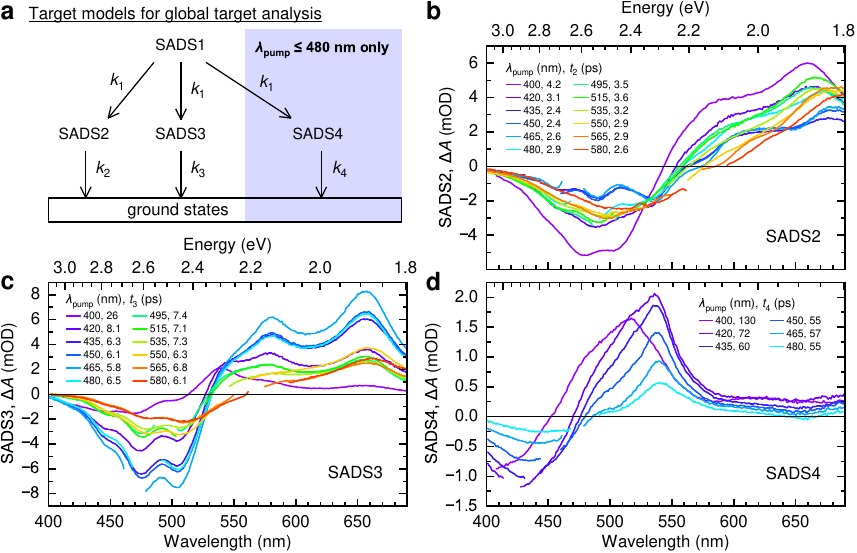}
	\caption{\textbf{Sample of the results of global target analysis performed on the transient absorption data.} \textbf{(a)} Target model used in the global target analysis performed, with resulting species-associated difference spectra (SADS) shown in the other panels: \textbf{(b)} SADS2, \textbf{(c)} SADS3, \textbf{(d)} SADS4; note that SADS4 is fitted only with pump wavelengths $\leq$\SI{480}{\nano\m}. Full results and additional global target analysis are detailed in Section \ref{sec:GTA_SADS_OCPo}.}
	\label{fig:STTA_OCPo_SADS_superfigure}
\end{figure}

Nevertheless, to compare with literature, depict a sample of our global target analysis models in Figure \ref{fig:STTA_OCPo_SADS_superfigure}a, and a selection of the results in Figure \ref{fig:STTA_OCPo_SADS_superfigure}b--d, with further details below. Here we apply models involving 3 or 4 components depending on the pump wavelength, hence resulting in 3 or 4 species-associated difference spectra (SADS). The 3-component model (Figure \ref{fig:STTA_OCPo_SADS_superfigure}a, excluding blue box) sufficiently fits transient absorption data with pumps from \SI{495}{\nano\m} to \SI{580}{\nano\m}, while a fourth component (Figure \ref{fig:STTA_OCPo_SADS_superfigure}a, SADS4-inclusive) is required to fit the data with pumps from \SI{400}{\nano\m} to \SI{480}{\nano\m}, once again consistent with the presence of long-lived forms of OCPo and the associated S* for $\lambda_\textrm{pump}\leq\SI{480}{\nano\m}$. We note that the profile of all SADS and their fitted time constant $t_i$ vary with excitation wavelength, suggesting that the model is not valid with this dataset; SADS4 varies particularly markedly depending on the pump wavelength used. We discuss the global lifetime analysis further, give plots of all SADS, and show additional models below.

\subsubsection{Global target analysis extended results}
\label{sec:GTA_SADS_OCPo}

Global target analysis on the UV-vis-probe \SI{}{\pico\s} transient absorption datasets was done using the Glotaran 1.5.1 software package (http://glotaran.org) \cite{Snellenburg2012}, a GUI for the R package TIMP\cite{Mullen2007}. Since a target model is applied, we term the extracted $\Delta A$-profiles species-associated difference spectra (SADS). Data used had already been processed with the steps outlined in  the methods (Section \ref{sec:methods_TA}); in particular, a chirp correction had already been applied, so that a term to account for chirp did not need to be included in the fitting. Noisy regions in the data due to pump scatter were excluded for all times to ensure a good fit of the rest of the data. Noisy red and blue ends in the data associated with tails of the probe were also excluded, so that the fitted wavelengths were typically \SI{400}{\nano\m} to \SI{690}{\nano\m}. Terms to account for the instrument response function (IRF) and the coherent artefact were included. For the IRF, a Gaussian is convoluted with the exponential decays and parametrised within the model, with its centre and full-width half-maximum (FWHM) fitted. For the coherent artefact, an additional component with the time profile of the IRF is fitted; we do not show this component in our figures.

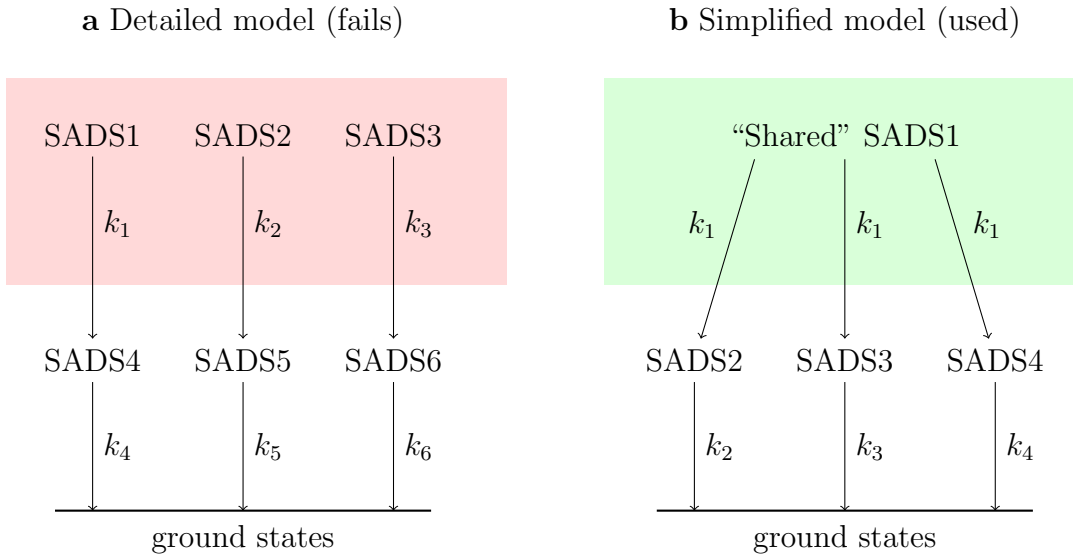
\begin{figure}[h]
\centering
\begin{tikzpicture}[scale=1]



\fill[red!15!white] (-3.15,-1) rectangle (3.5,1.75);

\node at (0,2.5) {\textbf{a} Detailed model (fails)};

\node (S2*) at (0,1) {SADS2};
\node (S1*) at (0,-2) {SADS5};
\draw [->] (S2*) -- node [anchor=south west] {$k_2$} (S1*);
\draw [->] (S1*) -- node [anchor=west] {$k_5$} (0,-4);

\draw[-, thick] (-2.5,-4) -- (2.5, -4);
\node (S0*) at (0,-4.375) {ground states};

\node (S2) at (-2,1) {SADS1};
\node (S1) at (-2,-2) {SADS4};
\draw [->] (S2) -- node [anchor=south west] {$k_1$} (S1);
\draw [->] (S1) -- node [anchor=west] {$k_4$} (-2,-4);

\node (S2t) at (2,1) {SADS3};
\node (S1t) at (2,-2) {SADS6};
\draw [->] (S2t) -- node [anchor=south west] {$k_3$} (S1t);
\draw [->] (S1t) -- node [anchor=west] {$k_6$} (2,-4);


\def\xshift{8}

\node at (\xshift,2.5) {\textbf{b} Simplified model (used)};

\fill[green!15!white] (\xshift-3.2,-1) rectangle (\xshift+3.2,1.75);

\node (sS2) at (\xshift,1) {`Shared' SADS1};
\node (sS1*) at (\xshift,-2) {SADS3};
\draw [->] (\xshift,0.675) -- node [anchor=south west] {$k_1$} (sS1*);
\draw [->] (sS1*) -- node [anchor=west] {$k_3$} (\xshift,-4);

\draw[-, thick] (\xshift-2.5,-4) -- (\xshift+2.5, -4);
\node (sS0*) at (\xshift,-4.375) {ground states};

\node (sS1) at (\xshift-2,-2) {SADS2};
\draw [->] (\xshift-1.2,0.675) -- node [anchor=south east] {$k_1$} (sS1);
\draw [->] (sS1) -- node [anchor=west] {$k_2$} (\xshift-2,-4);

\node (sS1t) at (\xshift+2,-2) {SADS4};
\draw [->] (\xshift+1.2,0.675) -- node [anchor=south west] {$k_1$} (sS1t);
\draw [->] (sS1t) -- node [anchor=west] {$k_4$} (\xshift+2,-4);

\end{tikzpicture}
\vspace{5pt}
\caption{\textbf{Diagrams of representative global target analysis models applied to the UV-vis \SI{}{\pico\s} transient absorption data. (a)} The model considering distinct S\textsubscript{2}-like states from ground-state heterogeneities that fails to converge or give reasonable results. \textbf{(b)} The simplified model that was used instead, with a shared S\textsubscript{2}-like state and shared decay rate $k_1$ to the distinct SADS. The instrument response function and coherent artefact are accounted for in the model. Note that in some applications of the model, the number of SADS is adjusted (\textit{e.g.}~SADS4 may be removed, or a SADS5 added). See SI text for full details.}
\label{fig:STTA_GTA_models}
\end{figure}

We generally use a model depicted in Figure \ref{fig:STTA_GTA_models}b where we fit a SADS corresponding to a single initially populated S\textsubscript{2}-like state that subsequently decays with an equal (shared) time constant into each of the fitted SADS, with those subsequently decaying with different time constants into the ground state. In the main text, we eventually assign different ground-state heterogeneities within OCPo; a target model possibly more accurate to this is depicted in Figure \ref{fig:STTA_GTA_models}a. It includes multiple SADS corresponding to multiple S\textsubscript{2}-like states that decay individually with different time constants into a paired SADS, subsequently decaying with different time constants into a ground state. However, the $\sim$\SI{100}{\femto\s} time resolution, the relatively temporal sampling around time zero, and the coherent artefact obscure distinct states within the first $\sim$\SI{200}{\femto\s}, so attempts applying the more detailed model failed to converge or give reasonable results. We consequently use the aforementioned model with a single, shared, fast-decaying SADS. Regardless, we believe that this does not remove support for a multiple ground-state assignment.

We find that the 4-SADS model is needed to fit transient absorption datasets with pump wavelengths from \SI{400}{\nano\m} to \SI{480}{\nano\m}, but the 3-SADS model is sufficient to fit those with pump wavelengths from \SI{495}{\nano\m} to \SI{600}{\nano\m}. The associated fitted decay time constants vary but are somewhat consistent with those extracted from the global triexponential fit of the normalised dynamics (main text Figure \ref{fig:STTA_OCPo_kinetics_superfigure} and SI Table \ref{tab:STTA_OCP_fit_params}). The requirement that a fourth SADS is needed to fit transient absorption datasets for $\lambda_\textrm{pump}\leq\SI{480}{\nano\m}$ is once again consistent with the presence of long-lived forms of OCPo and associated S* features for those pump wavelengths.

\begin{figure}[h]
	\centering
	\includegraphics[scale=1.0]{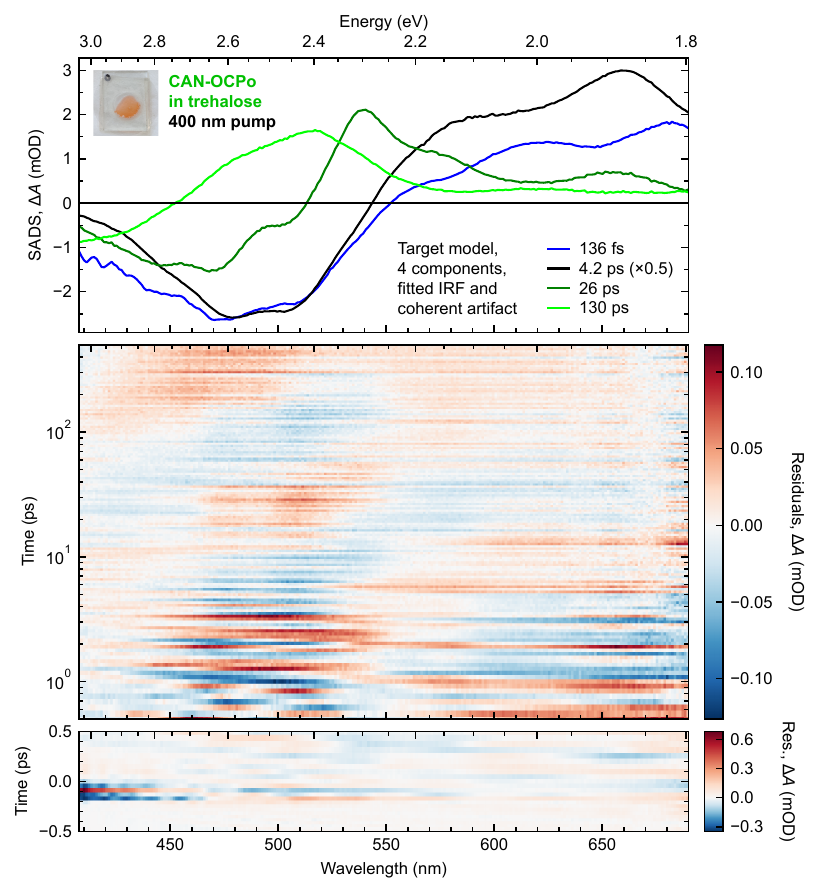}
	\caption{\textbf{Results of global target analysis with a 4-component target model on transient absorption data of CAN-binding OCPo in trehalose with pump wavelength \SI{400}{\nano\m} and a UV-vis probe: SADS (top) and residuals (middle, bottom).} Only the wavelength range 407--\SI{690}{\nano\m} was fitted to exclude noisy data due to low probe light and significant pump scatter. SADS time constants are specified in the legend; multiplications refer to scalings applied to the SADS. The fitted IRF has centre \SI{-78}{\femto\s} and FWHM \SI{64}{\femto\s}. A coherent artefact with the concentration profile of the IRF was fitted but not shown here. $\textrm{Residuals} = \textrm{Data}-\textrm{Fit}$; note the logarithmic time-scale in the middle panel, and the linear time-scale and different residuals scale in the bottom panel. See text for further details.}
	\label{fig:STTA_GTA_3_TP-400nm_ex-crop_plot}
\end{figure}

\begin{figure}[h]
	\centering
	\includegraphics[scale=1.0]{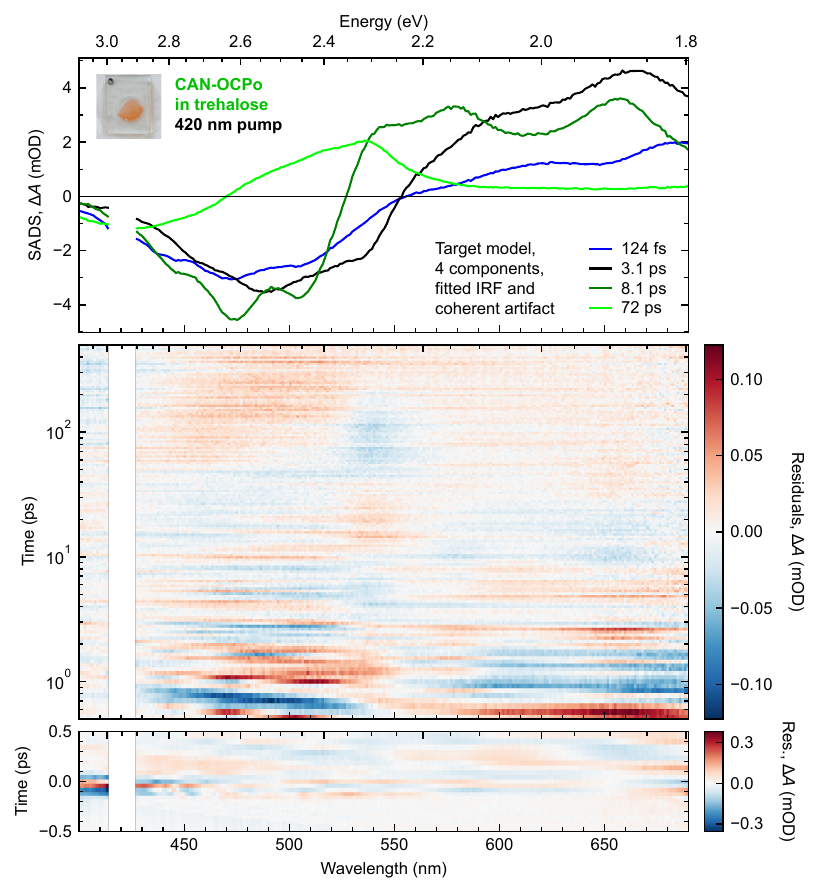}
	\caption{\textbf{Results of global target analysis with a 4-component target model on transient absorption data of CAN-binding OCPo in trehalose with pump wavelength \SI{420}{\nano\m} and a UV-vis probe: SADS (top) and residuals (middle, bottom).} Only the wavelength range 400--\SI{690}{\nano\m} was fitted, and noisy data from 414--\SI{427}{\nano\m} due to significant pump scatter was excluded from the fit. SADS time constants are specified in the legend. The fitted IRF has centre \SI{-50}{\femto\s} and FWHM \SI{47}{\femto\s}. A coherent artefact with the concentration profile of the IRF was fitted but not shown here. $\textrm{Residuals} = \textrm{Data}-\textrm{Fit}$; note the logarithmic time-scale in the middle panel, and the linear time-scale and different residuals scale in the bottom panel. See text for further details.}
	\label{fig:STTA_GTA_3_A-420nm_ex-cut_crop_plot}
\end{figure}

\begin{figure}[h]
	\centering
	\includegraphics[scale=1.0]{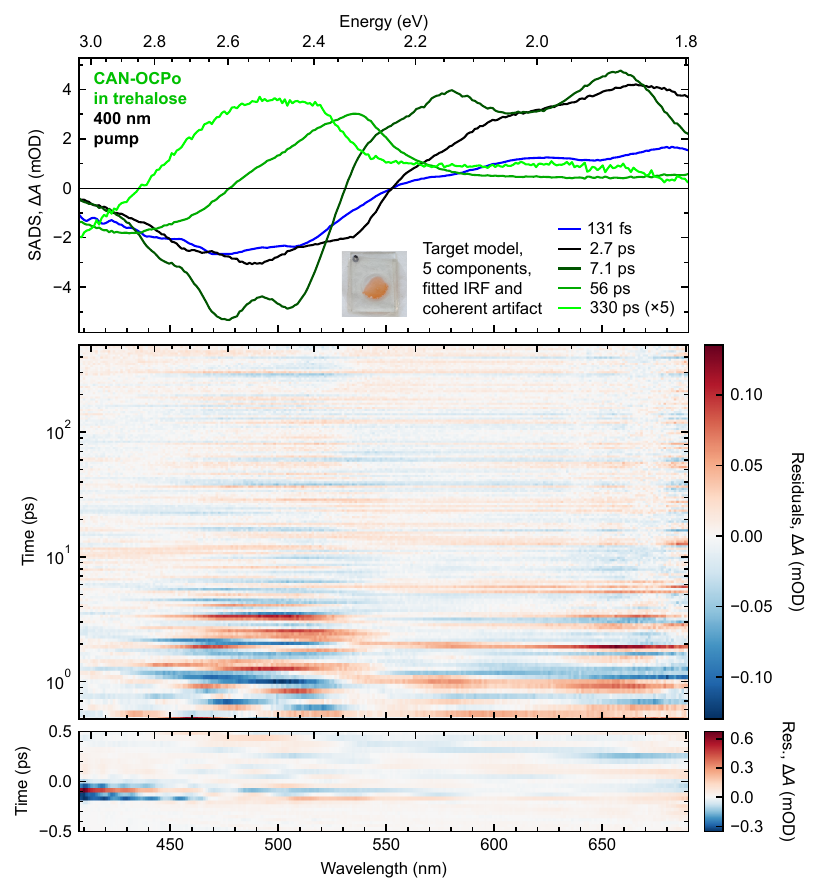}
	\caption{\textbf{Results of global target analysis with a 5-component target model on transient absorption data of CAN-binding OCPo in trehalose with pump wavelength \SI{400}{\nano\m} and a UV-vis probe: SADS (top) and residuals (middle, bottom).} Only the wavelength range 407--\SI{690}{\nano\m} was fitted to exclude noisy data due to low probe light and significant pump scatter. SADS time constants are specified in the legend; multiplications refer to scalings applied to the SADS. The fitted IRF has centre \SI{-88}{\femto\s} and FWHM \SI{59}{\femto\s}. A coherent artefact with the concentration profile of the IRF was fitted but not shown here. $\textrm{Residuals} = \textrm{Data}-\textrm{Fit}$; note the logarithmic time-scale in the middle panel, and the linear time-scale and different residuals scale in the bottom panel. See text for further details.}
	\label{fig:STTA_GTA_4_TP-400nm_ex-crop_plot}
\end{figure}

\begin{figure}[h]
	\centering
	\includegraphics[scale=1.0]{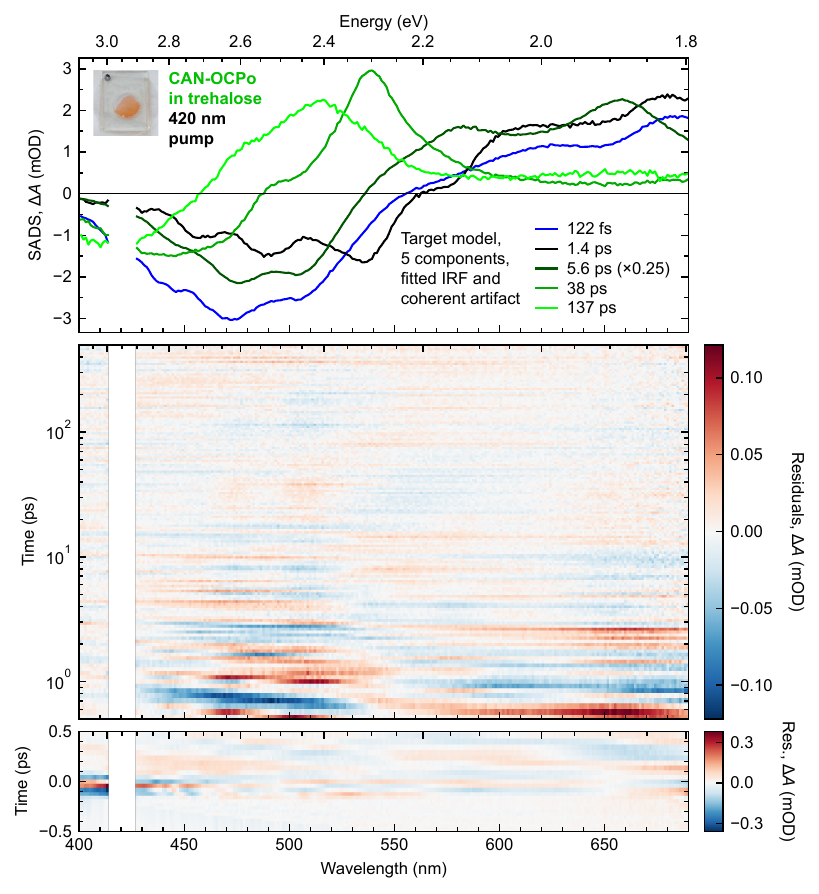}
	\caption{\textbf{Results of global target analysis with a 5-component target model on transient absorption data of CAN-binding OCPo in trehalose with pump wavelength \SI{420}{\nano\m} and a UV-vis probe: SADS (top) and residuals (middle, bottom).} Only the wavelength range 400--\SI{690}{\nano\m} was fitted, and noisy data from 414--\SI{427}{\nano\m} due to significant pump scatter was excluded from the fit. SADS time constants are specified in the legend; multiplications refer to scalings applied to the SADS. The fitted IRF has centre \SI{-50}{\femto\s} and FWHM \SI{46}{\femto\s}. A coherent artefact with the concentration profile of the IRF was fitted but not shown here. $\textrm{Residuals} = \textrm{Data}-\textrm{Fit}$; note the logarithmic time-scale in the middle panel, and the linear time-scale and different residuals scale in the bottom panel. See text for further details.}
	\label{fig:STTA_GTA_4_A-420nm_ex-crop_plot}
\end{figure}

We note that the four-SADS model somewhat underfits the transient absorption data for the pump wavelength \SI{400}{\nano\m} (Figure \ref{fig:STTA_GTA_3_TP-400nm_ex-crop_plot}) and possibly \SI{420}{\nano\m} (Figure \ref{fig:STTA_GTA_3_A-420nm_ex-cut_crop_plot}) with an observable structure in the residuals. The result of respective five-SADS models are shown in Figures \ref{fig:STTA_GTA_4_TP-400nm_ex-crop_plot} and \ref{fig:STTA_GTA_4_A-420nm_ex-crop_plot}. The requirement of SADS5 for a better fit is indicative of further ground-state heterogeneity.

\begin{figure}[h]
	\centering
	\includegraphics[scale=1.0]{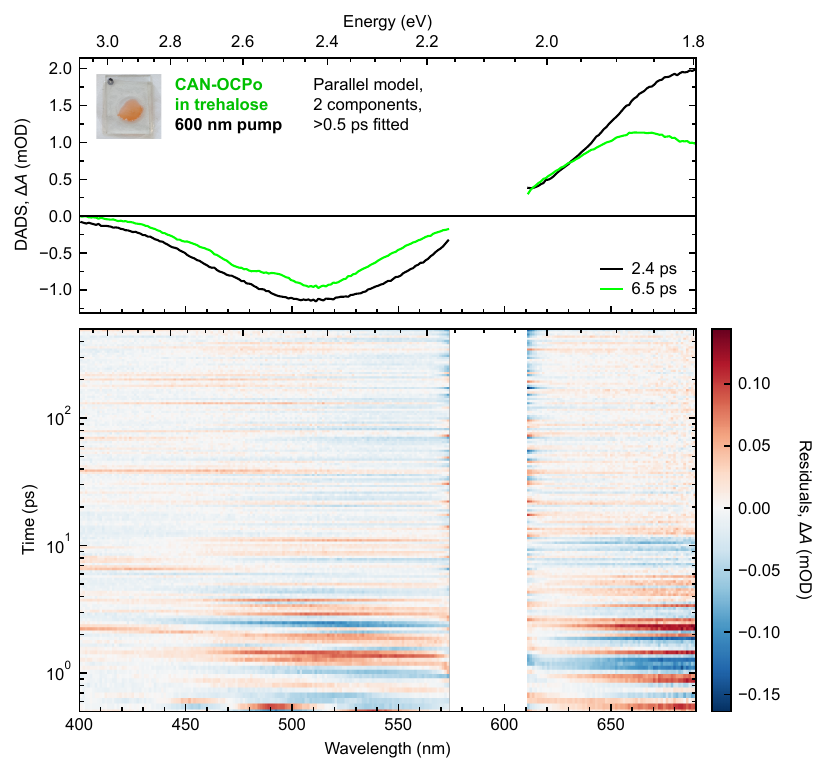}
	\caption{\textbf{Results of global lifetime analysis with a 2-component parallel model on transient absorption data of CAN-binding OCPo in trehalose with pump wavelength \SI{600}{\nano\m} and a UV-vis: DADS (top) and residuals (bottom).} Only the wavelength range 400--\SI{690}{\nano\m} and times $>$\SI{0.5}{\pico\s} were fitted, and noisy data from 574--\SI{611}{\nano\m} due to significant pump scatter was excluded from the fit. DADS time constants are specified in the legend. $\textrm{Residuals} = \textrm{Data}-\textrm{Fit}$. See text for further details.}
	\label{fig:STTA_GLA_ii_C-600nm_ex-cut_crop2_plot}
\end{figure}

An exception to the `target' fitting procedure specified above is the analysis of the \SI{600}{\nano\m} pump wavelength transient absorption data (Figure \ref{fig:STTA_GLA_ii_C-600nm_ex-cut_crop2_plot}). As a relatively high pump power is used (see legend in Figure \ref{fig:STTA_ex_wav_dep_pump_scatter}), the coherent artefact is relatively strong, so the early-time dynamics cannot be easily fitted with the target model. We therefore fitted times beyond \SI{0.5}{\pico\s} with a 2-component parallel global lifetime analysis model with no IRF or coherent artefact terms, very similar to the model applied to the visible-probe transient absorption data (Section \ref{sec:GLA_DADS_OCPo_OCPr}), although with no weighting of the data here due to greater maximum delay used ($\sim$\SI{500}{\pico\s}). This corresponds to two decay-associated difference spectra (DADS) decaying in parallel after initial population from the unmodelled (cropped out) S\textsubscript{2}-like states. DADS1 and DADS2 are comparable to SADS2 and SADS3 respectively.

The variations in the SADS profiles and associated decay time constants per-pump wavelength (particularly for SADS2 and SADS3) are indicative that global target (and lifetime) analysis is not valid with this dataset. This is because of the affects of internal vibrational redistribution (IVR) and vibrational cooling (VC). The assumption of bilinearity (separable spectral and kinetic contributions) in using global lifetime/target analysis\cite{Ruckebusch2012, vanStokkum2004, Fernandez-Teran2022} does not hold, with the IVR/VC essentially giving the spectra some time-dependence\cite{Balevicius2015, Balevicius2016, Balevicius2018, Balevicius2019}. Indeed, SADS2 and SADS3 have a mixed and unknown physical correspondence, likely mixed S\textsubscript{1} decay, IVR, and VC between the different ground-state forms of CAN-OCPo. Likewise, SADS1 may correspond to S\textsubscript{2}/IVR/VC for multiple CAN-OCPo forms (as discussed above; see Figure \ref{fig:STTA_GTA_models}), in addition to a potential coherent artefact contribution (the additional IRF-time profile component may not always account for it fully). Meanwhile, SADS4 (and SADS5) can more safely be associated with the S*-yielding form of OCPo due to the 1-order difference between their associated decay time constants compared to those for SADS1, SADS2, and SADS3.

\begin{figure}[h]
	\centering
	\includegraphics[scale=1.0]{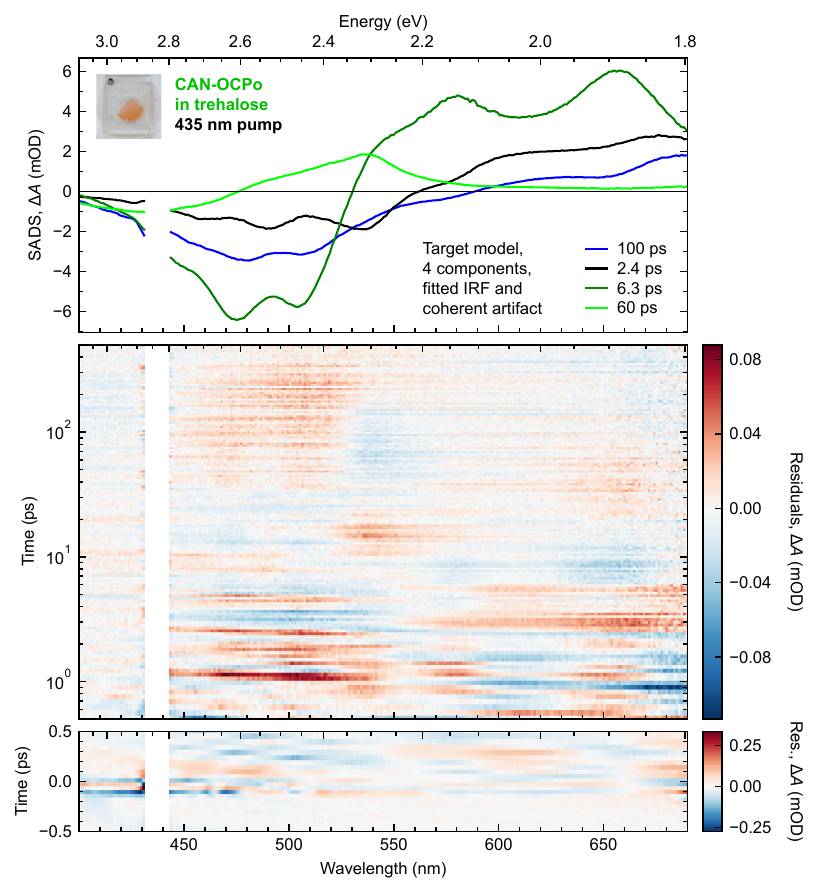}
	\caption{\textbf{Results of global target analysis with a 4-component target model on transient absorption data of CAN-binding OCPo in trehalose with pump wavelength \SI{435}{\nano\m} and a UV-vis probe: SADS (top) and residuals (middle, bottom).} Only the wavelength range 400--\SI{690}{\nano\m} was fitted, and noisy data from 432--\SI{443}{\nano\m} due to significant pump scatter was excluded from the fit. SADS time constants are specified in the legend. The fitted IRF has centre \SI{-55}{\femto\s} and FWHM \SI{36}{\femto\s}. A coherent artefact with the concentration profile of the IRF was fitted but not shown here. $\textrm{Residuals} = \textrm{Data}-\textrm{Fit}$; note the logarithmic time-scale in the middle panel, and the linear time-scale and different residuals scale in the bottom panel. See text for further details.}
	\label{fig:STTA_GTA_3_A-435nm_ex-cut_crop_plot}
\end{figure}

\begin{figure}[h]
	\centering
	\includegraphics[scale=1.0]{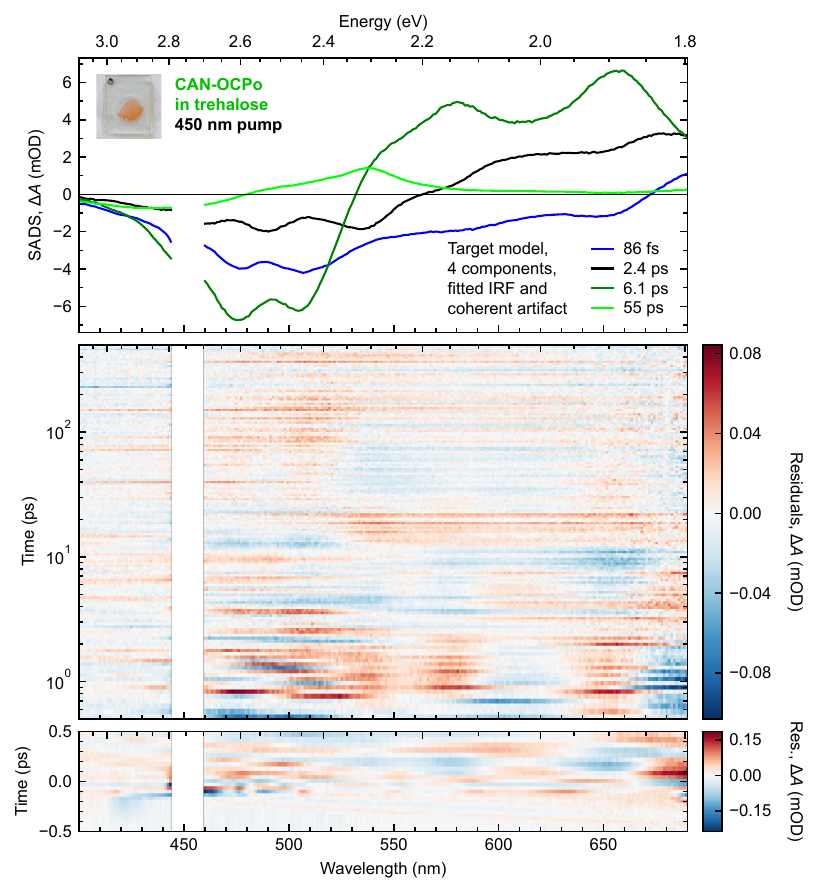}
	\caption{\textbf{Results of global target analysis with a 4-component target model on transient absorption data of CAN-binding OCPo in trehalose with pump wavelength \SI{450}{\nano\m} and a UV-vis probe: SADS (top) and residuals (middle, bottom).} Only the wavelength range 400--\SI{690}{\nano\m} was fitted, and noisy data from 444--\SI{459}{\nano\m} due to significant pump scatter was excluded from the fit. SADS time constants are specified in the legend. The fitted IRF has centre \SI{-66}{\femto\s} and FWHM \SI{30}{\femto\s}. A coherent artefact with the concentration profile of the IRF was fitted but not shown here. $\textrm{Residuals} = \textrm{Data}-\textrm{Fit}$; note the logarithmic time-scale in the middle panel, and the linear time-scale and different residuals scale in the bottom panel. See text for further details.}
	\label{fig:STTA_GTA_3_A-450nm_ex-cut_crop_plot}
\end{figure}

\begin{figure}[h]
	\centering
	\includegraphics[scale=1.0]{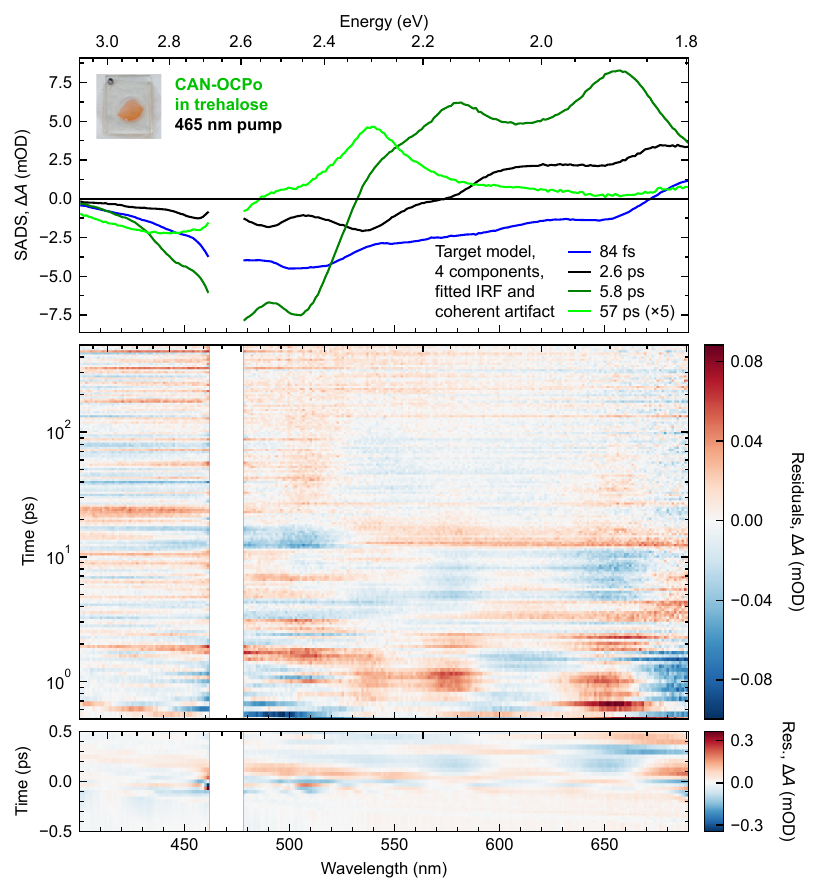}
	\caption{\textbf{Results of global target analysis with a 4-component target model on transient absorption data of CAN-binding OCPo in trehalose with pump wavelength \SI{465}{\nano\m} and a UV-vis probe: SADS (top) and residuals (middle, bottom).} Only the wavelength range 400--\SI{690}{\nano\m} was fitted, and noisy data from 462--\SI{478}{\nano\m} due to significant pump scatter was excluded from the fit. SADS time constants are specified in the legend; multiplications refer to scalings applied to the SADS. The fitted IRF has centre \SI{-53}{\femto\s} and FWHM \SI{26}{\femto\s}. A coherent artefact with the concentration profile of the IRF was fitted but not shown here. $\textrm{Residuals} = \textrm{Data}-\textrm{Fit}$; note the logarithmic time-scale in the middle panel, and the linear time-scale and different residuals scale in the bottom panel. See text for further details.}
	\label{fig:STTA_GTA_3_A-465nm_ex-cut_crop_plot}
\end{figure}

\begin{figure}[h]
	\centering
	\includegraphics[scale=1.0]{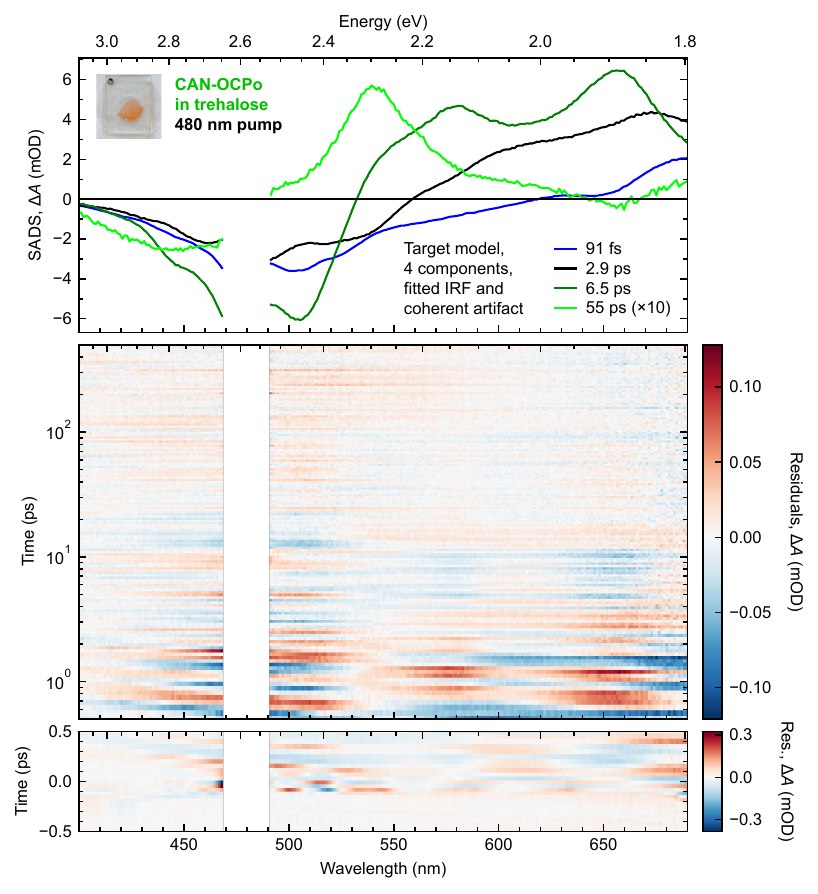}
	\caption{\textbf{Results of global target analysis with a 4-component target model on transient absorption data of CAN-binding OCPo in trehalose with pump wavelength \SI{480}{\nano\m} and a UV-vis probe: SADS (top) and residuals (middle, bottom).} Only the wavelength range 400--\SI{690}{\nano\m} was fitted, and noisy data from 469--\SI{491}{\nano\m} due to significant pump scatter was excluded from the fit. SADS time constants are specified in the legend; multiplications refer to scalings applied to the SADS. The fitted IRF has centre \SI{-29}{\femto\s} and FWHM \SI{31}{\femto\s}. A coherent artefact with the concentration profile of the IRF was fitted but not shown here. $\textrm{Residuals} = \textrm{Data}-\textrm{Fit}$; note the logarithmic time-scale in the middle panel, and the linear time-scale and different residuals scale in the bottom panel. See text for further details.}
	\label{fig:STTA_GTA_3_A-480nm_ex-cut_crop_plot}
\end{figure}

\begin{figure}[h]
	\centering
	\includegraphics[scale=1.0]{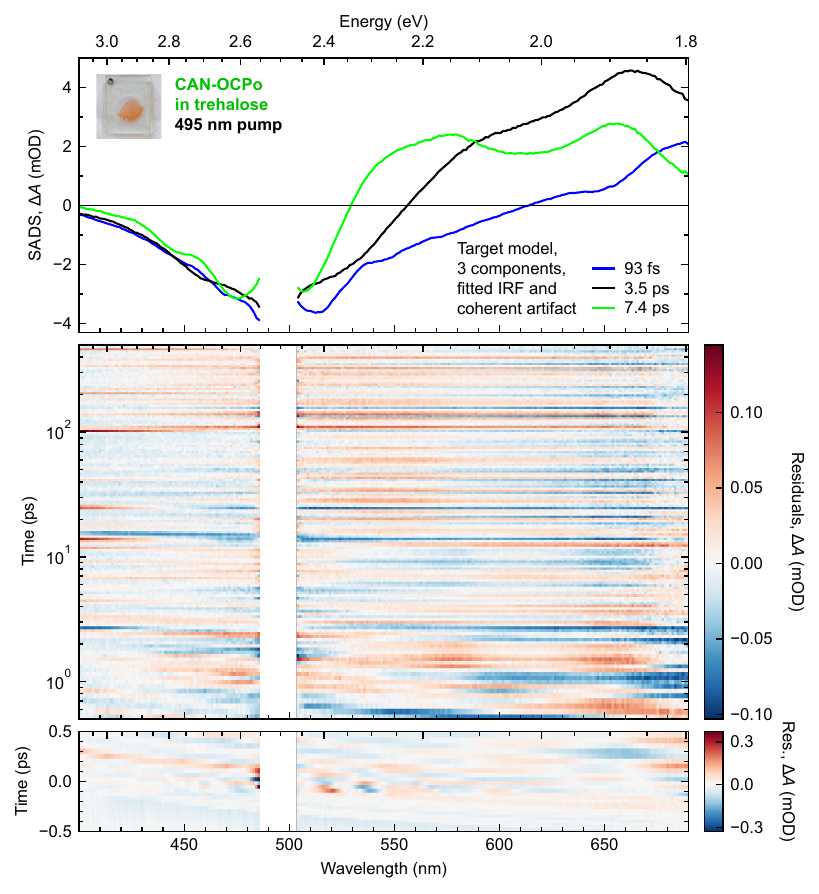}
	\caption{\textbf{Results of global target analysis with a 3-component target model on transient absorption data of CAN-binding OCPo in trehalose with pump wavelength \SI{495}{\nano\m} and a UV-vis probe: SADS (top) and residuals (middle, bottom).} Only the wavelength range 400--\SI{690}{\nano\m} was fitted, and noisy data from 486--\SI{504}{\nano\m} due to significant pump scatter was excluded from the fit. SADS time constants are specified in the legend. The fitted IRF has centre \SI{-25}{\femto\s} and FWHM \SI{38}{\femto\s}. A coherent artefact with the concentration profile of the IRF was fitted but not shown here. $\textrm{Residuals} = \textrm{Data}-\textrm{Fit}$; note the logarithmic time-scale in the middle panel, and the linear time-scale and different residuals scale in the bottom panel. See text for further details.}
	\label{fig:STTA_GTA_2_B-495nm_ex-cut_crop_plot}
\end{figure}

\begin{figure}[h]
	\centering
	\includegraphics[scale=1.0]{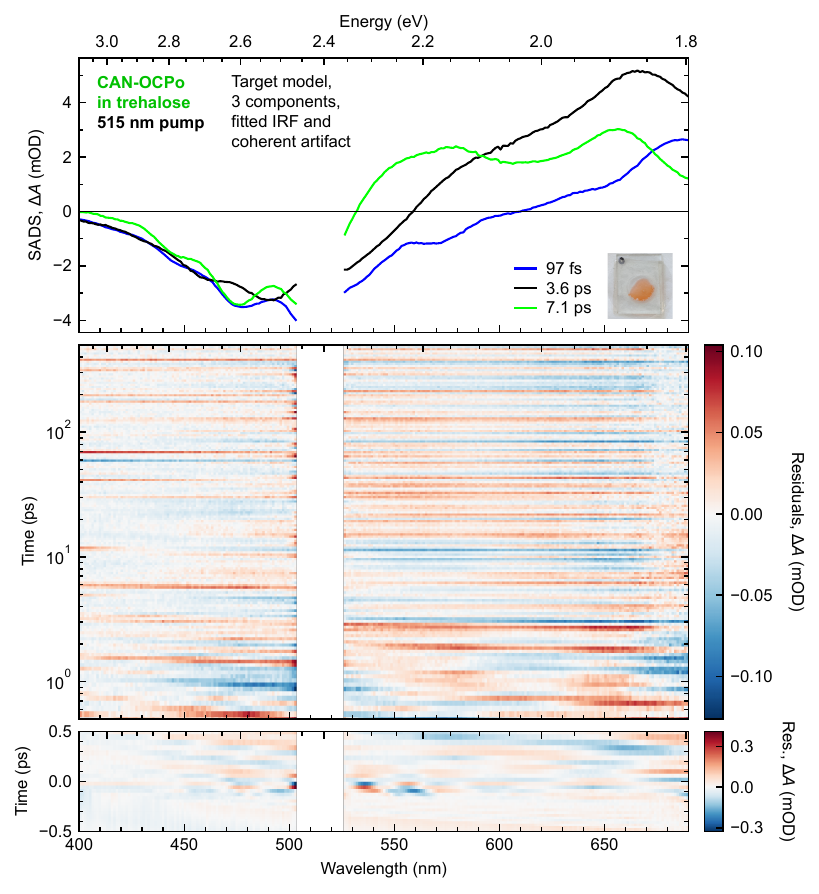}
	\caption{\textbf{Results of global target analysis with a 3-component target model on transient absorption data of CAN-binding OCPo in trehalose with pump wavelength \SI{515}{\nano\m} and a UV-vis probe: SADS (top) and residuals (middle, bottom).} Only the wavelength range 400--\SI{690}{\nano\m} was fitted, and noisy data from 504--\SI{526}{\nano\m} due to significant pump scatter was excluded from the fit. SADS time constants are specified in the legend. The fitted IRF has centre \SI{-33}{\femto\s} and FWHM \SI{33}{\femto\s}. A coherent artefact with the concentration profile of the IRF was fitted but not shown here. $\textrm{Residuals} = \textrm{Data}-\textrm{Fit}$; note the logarithmic time-scale in the middle panel, and the linear time-scale and different residuals scale in the bottom panel. See text for further details.}
	\label{fig:STTA_GTA_2_B-515nm_ex-cut_crop_plot}
\end{figure}

\begin{figure}[h]
	\centering
	\includegraphics[scale=1.0]{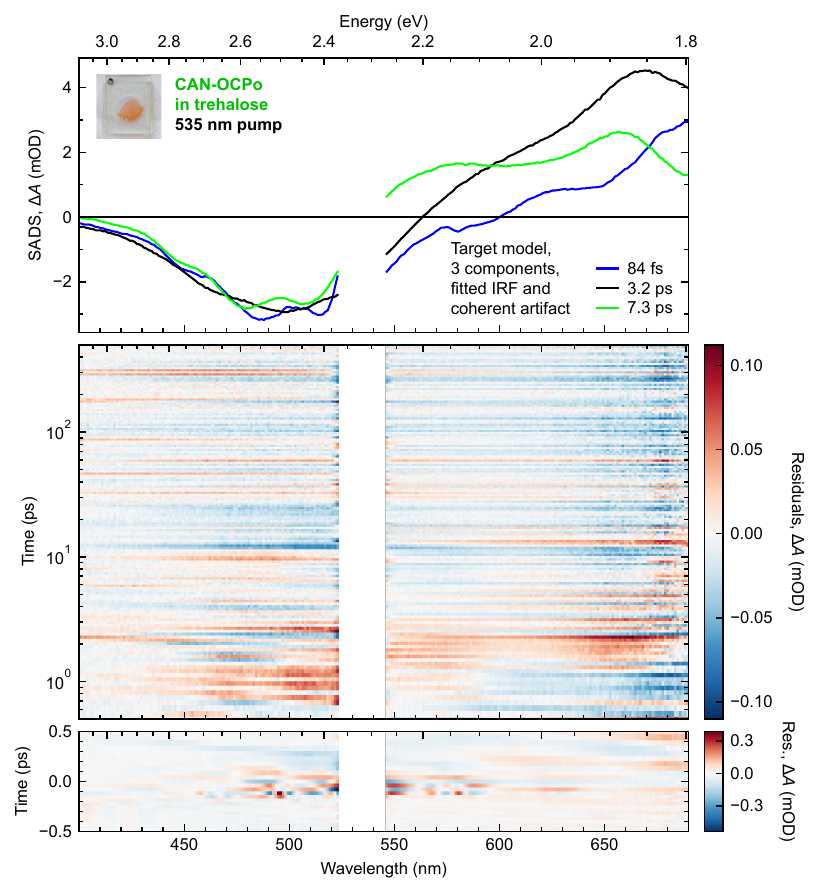}
	\caption{\textbf{Results of global target analysis with a 3-component target model on transient absorption data of CAN-binding OCPo in trehalose with pump wavelength \SI{535}{\nano\m} and a UV-vis probe: SADS (top) and residuals (middle, bottom).} Only the wavelength range 400--\SI{690}{\nano\m} was fitted, and noisy data from 524--\SI{546}{\nano\m} due to significant pump scatter was excluded from the fit. SADS time constants are specified in the legend. The fitted IRF has centre \SI{-30}{\femto\s} and FWHM \SI{53}{\femto\s}. A coherent artefact with the concentration profile of the IRF was fitted but not shown here. $\textrm{Residuals} = \textrm{Data}-\textrm{Fit}$; note the logarithmic time-scale in the middle panel, and the linear time-scale and different residuals scale in the bottom panel. See text for further details.}
	\label{fig:STTA_GTA_2_C-535nm_ex-cut_crop_plot}
\end{figure}

\begin{figure}[h]
	\centering
	\includegraphics[scale=1.0]{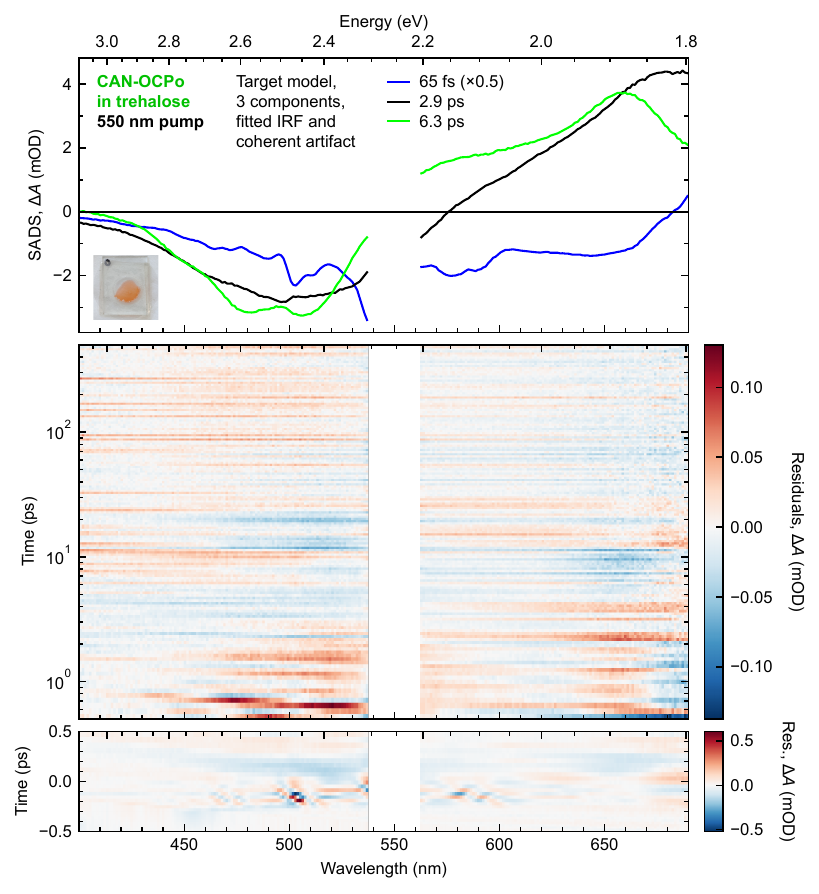}
	\caption{\textbf{Results of global target analysis with a 3-component target model on transient absorption data of CAN-binding OCPo in trehalose with pump wavelength \SI{550}{\nano\m} and a UV-vis probe: SADS (top) and residuals (middle, bottom).} Only the wavelength range 400--\SI{690}{\nano\m} was fitted, and noisy data from 538--\SI{562}{\nano\m} due to significant pump scatter was excluded from the fit. SADS time constants are specified in the legend; multiplications refer to scalings applied to the SADS. The fitted IRF has centre \SI{-136}{\femto\s} and FWHM \SI{51}{\femto\s}. A coherent artefact with the concentration profile of the IRF was fitted but not shown here. $\textrm{Residuals} = \textrm{Data}-\textrm{Fit}$; note the logarithmic time-scale in the middle panel, and the linear time-scale and different residuals scale in the bottom panel. See text for further details.}
	\label{fig:STTA_GTA_2_C-550nm_ex-cut_crop_plot}
\end{figure}

\begin{figure}[h]
	\centering
	\includegraphics[scale=1.0]{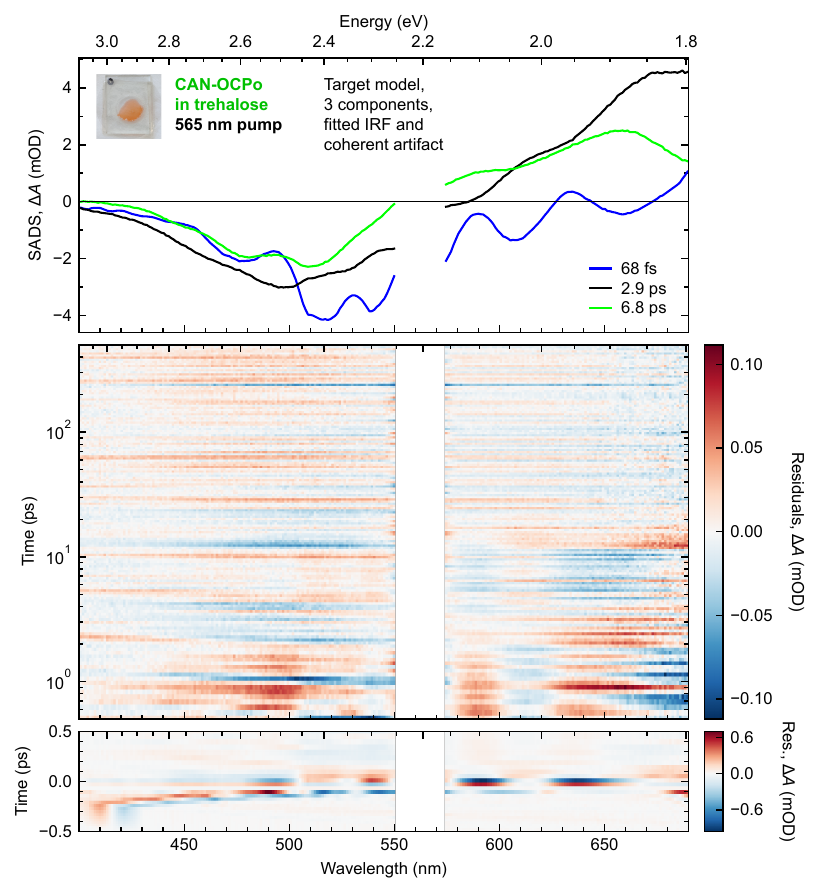}
	\caption{\textbf{Results of global target analysis with a 3-component target model on transient absorption data of CAN-binding OCPo in trehalose with pump wavelength \SI{565}{\nano\m} and a UV-vis probe: SADS (top) and residuals (middle, bottom).} Only the wavelength range 400--\SI{690}{\nano\m} was fitted, and noisy data from 551--\SI{574}{\nano\m} due to significant pump scatter was excluded from the fit. SADS time constants are specified in the legend. The fitted IRF has centre \SI{-49}{\femto\s} and FWHM \SI{5}{\femto\s}. A coherent artefact with the concentration profile of the IRF was fitted but not shown here. $\textrm{Residuals} = \textrm{Data}-\textrm{Fit}$; note the logarithmic time-scale in the middle panel, and the linear time-scale and different residuals scale in the bottom panel. See text for further details.}
	\label{fig:STTA_GTA_2_C-565nm_ex-cut_crop_plot}
\end{figure}

\begin{figure}[h]
	\centering
	\includegraphics[scale=1.0]{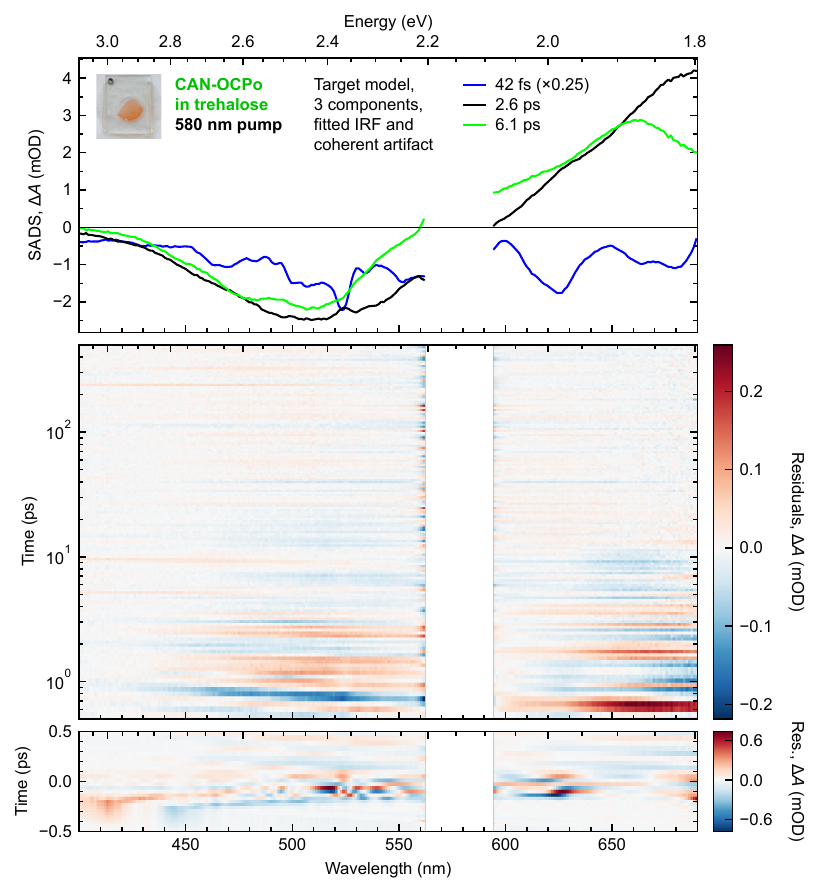}
	\caption{\textbf{Results of global target analysis with a 3-component target model on transient absorption data of CAN-binding OCPo in trehalose with pump wavelength \SI{580}{\nano\m} and a UV-vis probe: SADS (top) and residuals (middle, bottom).} Only the wavelength range 400--\SI{690}{\nano\m} was fitted, and noisy data from 562--\SI{594}{\nano\m} due to significant pump scatter was excluded from the fit. SADS time constants are specified in the legend; multiplications refer to scalings applied to the SADS. The fitted IRF has centre \SI{-72}{\femto\s} and FWHM \SI{43}{\femto\s}. A coherent artefact with the concentration profile of the IRF was fitted but not shown here. $\textrm{Residuals} = \textrm{Data}-\textrm{Fit}$; note the logarithmic time-scale in the middle panel, and the linear time-scale and different residuals scale in the bottom panel. See text for further details.}
	\label{fig:STTA_GTA_2_C-580nm_ex-cut_crop_plot}
\end{figure}

\clearpage

\subsection{Supplementary pump wavelength-dependent photoconversion materials}

\subsubsection{Dark-adapted absorbance}

\begin{figure}[h]
	\centering
	\includegraphics[scale=1.0]{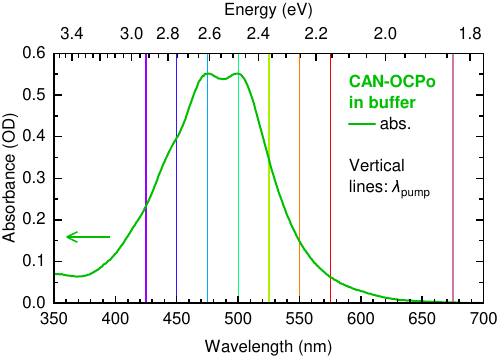}
	\caption{\textbf{Absorbance of CAN-binding OCPo used in the photoconversion dependence on pump wavelength experimentation.} The green line shows the dark-adapted OCP absorbance in a \SI{2}{\milli\metre} cuvette with normal incidence. Vertical lines indicate the nominal pump wavelengths ($\pm$\SI{5}{\nano\m} nominal centre wavelength accuracy, $<$\SI{10}{\nano\m} nominal FWHM) used in the experiment.}
	\label{fig:PC_OCP_absorbance}
\end{figure}

\subsubsection{Experimental oversights discussion}
\label{sec:PC_OCP_oversights}

The number of pump photons incident per second (photon rate) at the power meter position is given by
\begin{equation}
\dot{N}_\textrm{\textgamma} = \frac{P}{E_\textrm{\textgamma}} = P\frac{\lambda_{\textrm{pump}}}{hc}
\label{eqn:photon_rate}
\end{equation}
where $P$ is the steady-state power of the pump, $E_\textrm{\textgamma} = hc/\lambda_{\textrm{pump}}$ is the per-photon energy, $\lambda_{\textrm{pump}}$ is the pump wavelength, $h$ is Planck's constant, and $c$ is the speed of light in a vacuum. The initial fraction of pump photons absorbed by the OCPo, assuming a zero angle of incidence and negligible scattering from the quartz cuvette, is
\begin{equation}
F_\textrm{dark} = 1-10^{-A_{\textrm{dark}}}
\end{equation}
where $A_{\textrm{dark}}$ is the absorbance of the dark-adapted OCP in buffer measured in a separate UV-vis spectrometer at zero angle of incidence (Figure \ref{fig:PC_OCP_absorbance}). Therefore, overlooking the later-discussed power losses and angle dependence, the initial number of pump-induced OCPo excitations per second (excitation rate) is given by
\begin{equation}
\dot{N}_\textrm{pump} = \dot{N}_\textrm{\textgamma}F_\textrm{dark} = P\frac{\lambda_{\textrm{pump}}}{hc}\left(1-10^{-A_{\textrm{dark}}}\right)
\end{equation}
which was controlled to a value of $\dot{N}_\textrm{pump}=\SI{2.53e16}{\per\s}$ in the experiment for most pump wavelengths. The exception to this control was for \SI{675}{\nano\metre} pump, where instead the photon rate (Equation \ref{eqn:photon_rate}) was controlled to $\dot{N}_\textrm{\textgamma}=\SI{8.76e16}{\per\s}$, the same as that for $\lambda=\SI{550}{\nano\metre}$; this is because OCPo is non-absorbing at \SI{675}{\nano\metre}, and the tuneable filter is unable to reach the higher photon rates (\textit{i.e.}~$\dot{N}_\textrm{\textgamma}=\SI{1.89e17}{\per\s}$ at $\lambda_{\textrm{pump}}=\SI{575}{\nano\metre}$) when set to \SI{675}{\nano\metre}.

However, as noted in the main text, the experimental oversights require some scrutiny in response to the significant pump wavelength-dependence on the OCPo$\rightarrow$OCPr yield. We are able to assess in post the effect of two of these oversights and show that they do not yield the trends observed. First, the beam path after the pump power measurement position involves reflection by three UV-enhanced Al mirrors and focusing by a lens. This results in a considerable power (photon rate) loss, defined here by a fraction $R_P$ that has some pump-wavelength dependence. $R_P$ was determined once all measurement runs (three experimental replicates) were complete by taking comparative pump power measurements at the usual position as well as the sample position (doing this during the main experiment was impossible due to spatial limitations, and would also cause pre-measurement photoconversion). The resulting correction to the photon rate is
\begin{equation}
\dot{N}_\textrm{\textgamma}^\textrm{corr.} = R_P \dot{N}_\textrm{\textgamma} = R_P P\frac{\lambda_{\textrm{pump}}}{hc}
\end{equation}
which we note remains incident on the cuvette rather than on the buffer. The second oversight was that the pump light was not close to normal incidence to the cuvette; it was at an angle of incidence $\theta \sim \SI{40}{\degree}$, resulting in the pump path length within the buffer being extended compared to that at normal incidence by a factor $1/\cos\theta$. Hence, since absorbance is directly proportional to path length in the sample (from the Beer-Lambert law\cite{LakowiczBook}), the corrected absorbance is $A_{\textrm{dark}}/\cos\theta$, and therefore the correct fraction of pump photons absorbed is
\begin{equation}
F_\textrm{dark}^\textrm{corr.} = 1-10^{-A_{\textrm{dark}}/\cos\theta}
\end{equation}
where we recall that $A_{\textrm{dark}}$ is determined at normal incidence in a separate UV-vis spectrometer. We note we are continuing to assume negligible scattering of the pump off the high-quality cuvette, when for a significantly off-normal angle of incidence this may fail to be the case. Combining these corrections, we obtain
\begin{equation}
\dot{N}_\textrm{pump}^\textrm{corr.} = \dot{N}_\textrm{\textgamma}^\textrm{corr.}F_\textrm{dark}^\textrm{corr.} = R_P P\frac{\lambda_{\textrm{pump}}}{hc}\left(1-10^{-A_{\textrm{dark}}/\cos\theta}\right)
\label{eqn:ex_rate_corr}
\end{equation}
which is the corrected pump-induced excitation rate in OCPo.

\begin{table}[ht]
	\renewcommand*{\arraystretch}{1.2} 
	\newcolumntype{C}[1]{>{\centering\arraybackslash}m{#1}} 
	\begin{center}
		\footnotesize 
		\begin{tabular}{|C{11mm}|C{10mm}C{9mm}C{10mm}C{13mm}C{13mm}|C{10mm}C{8mm}C{13mm}C{13mm}|}
			\hline
			\boldmath{$\lambda_{\textbf{pump}}$} \textbf{(nm)} & \boldmath{$A_{\textbf{dark}}$ \textbf{(\SI{}{\OD})}} & \boldmath{$F_\textbf{dark}$} & \boldmath{$P$} \textbf{(\SI{}{\micro\watt})} & \boldmath{$\dot{N}_\textbf{\textgamma}$} \boldmath{$\times 10^{-16}$} & \boldmath{$\dot{N}_\textbf{pump}$} \boldmath{$\times 10^{-16}$} & \boldmath{$F_\textbf{dark}^\textbf{corr.}$} & \boldmath{$R_P$} & \boldmath{$\dot{N}_\textbf{\textgamma}^{\textbf{corr.}}$} \boldmath{$\times 10^{-16}$} & \boldmath{$\dot{N}_\textbf{pump}^\textbf{corr.}$} \boldmath{$\times 10^{-16}$}\\
			\hline
			\textbf{425} & 0.233 & 0.415 & 285 & 6.10 &2.53& 0.503 & 0.55 & 3.35 & 1.69\\ \hline
			\textbf{450} & 0.397 & 0.599 & 187 & 4.23 &2.53& 0.697 & 0.54 & 2.29 & 1.60\\ \hline
			\textbf{475} & 0.551 & 0.719 & 147 & 3.52 &2.53& 0.809 & 0.54 & 1.89 & 1.53\\ \hline
			\textbf{500} & 0.552 & 0.719 & 140 & 3.52 &2.53& 0.809 & 0.52 & 1.83 & 1.48\\ \hline
			\textbf{525} & 0.345 & 0.548 & 175 & 4.62 &2.53& 0.646 & 0.51 & 2.38 & 1.53\\ \hline
			\textbf{550} & 0.148 & 0.289 & 316 & 8.76 &2.53& 0.359 & 0.49 & 4.26 & 1.53\\ \hline
			\textbf{575} & 0.062 & 0.134 & 654 & 18.9 &2.53& 0.171 & 0.48 & 9.06 & 1.55\\ \hline
		    \textbf{675} & 0 & 0 & 256 & 8.76 &0& 0 & 0.43 & 3.72 & 0 \\ \hline
		\end{tabular}
	\end{center}
	\vspace{-6pt}
	\caption{\textbf{Pump wavelengths and powers used in experiments with photoconverting OCP in buffer, with derived quantities also printed}. Corrected quantities were calculated using unrounded values and $\theta=\SI{40}{\degree}$. See Equations \ref{eqn:photon_rate} to \ref{eqn:ex_rate_corr} and SI text for definitions of the quantities.}
	\label{tab:PC_OCP_pump_powers_calc}
\end{table}

We print Table \ref{tab:PC_OCP_pump_powers_calc} with pump wavelength and the measured or calculated quantities defined in Equations \ref{eqn:photon_rate} to \ref{eqn:ex_rate_corr}, assuming an angle of incidence of $\theta=\SI{40}{\degree}$. We highlight that the corrected pump-induced excitation rate ($\dot{N}_\textrm{pump}^\textrm{corr.}$) is a minimum for $\lambda_{\textrm{pump}}=\SI{500}{\nano\m}$, and generally increasing from there in either $\lambda_{\textrm{pump}}$-direction. This does not match the dependence of pump wavelength on the photoconversion yield (\textit{i.e.}~the wavelength dependence on the dynamics in main text Figure \ref{fig:PC_OCP_Aratio}b). So we conclude that these two oversights are not the determinant for the trend in OCPo$\rightarrow$OCPr photoconversion.

We note that some significant sources of error here are due to the pump power fluctuations, the intrinsic error of the Thorlabs S120VC power meter (note that $R_P$ is calculated from two pump power measurements per pump wavelength, distinct from the single $P$ pump power measurement per pump wavelength per experimental replicate), and the visual estimation (rather than a measurement) of the pump angle of incidence $\theta$.

A significant oversight that we cannot account for post-experiment is the poor pump-probe overlap; the pump spot diameter ($\sim$\SI{30}{\micro\m}) is significantly smaller than the white-light spot diameter ($\sim$\SI{1}{\milli\metre}). This is likely to result in spectral biasing; if \textit{e.g.} the white-light spot is relatively blue within the overlap region, then the product OCPr absorbance will appear artificially blue. Further, the pump spot diameter and position potentially varies per-wavelength due to some dispersive optics used (in particular, the focusing lens).

The shapes of the dynamics (main text Figure \ref{fig:PC_OCP_Aratio}b) should be treated with caution; we have controlled for the same initial rate of OCPo excitations (see above), but the excitation rate for $t>\SI{0}{\s}$ is muddled by the presence of OCPr within the pump spot and the diffusion of OCPo and OCPr in and out of the pump spot, the probe spot, and the dark-adapted bulk in the rest of the cuvette. There will also constantly be some back-conversion from OCPr to OCPo. These dynamics should therefore not be treated the same way as chemical kinetics, or `normal' transient absorption dynamic traces, and is why we have refrained from further analysis and assignment in this study.

\end{document}